\documentclass[11pt]{amsart}

\usepackage{amsfonts}
\usepackage{amsmath}
\usepackage{color}
\usepackage{hyperref}

\usepackage{pict2e}
\usepackage{graphicx}
\numberwithin{equation}{section}
\allowdisplaybreaks[4]

\newcommand{\sect}[1]{\setcounter{equation}{0}\section{#1}}

\DeclareMathOperator{\diag}{diag}

\newcommand{\Tr}{\mathop{\mathrm{Tr}}}
\newcommand{\Pf}{\mathop{\mathrm{Pf}}}
\newcommand{\sign}{\mathop{\mathrm{sign}}}
\newcommand{\erf}{\mathop{\mathrm{erf}}}
\newcommand{\erfc}{\mathop{\mathrm{erfc}}}
\newcommand{\hc}{^\dagger}
\newcommand{\nn}{\nonumber}

\newcommand{\eins}{\leavevmode\hbox{\small1\kern-3.8pt\normalsize1}}

\begin{document}
\title[Symmetry Transition for Real Gaussian Random Matrices]
 {\bf{
Preserving topology while breaking chirality: 
From chiral orthogonal to  anti-symmetric Hermitian ensemble
 }}

\author{Gernot Akemann, Mario Kieburg, Adam Mielke, and Pedro Vidal}
\address{Faculty of Physics, Bielefeld University,  P.O. Box 100131, D-33501 Bielefeld, Germany} \email{akemann@physik.uni-bielefeld.de, mkieburg@physik.uni-bielefeld.de,\newline amielke@math.uni-bielefeld.de}

\begin{abstract}
We consider a parameter dependent ensemble of two real random matrices with Gaussian distribution. It describes the transition between the symmetry class of 
the chiral Gaussian orthogonal ensemble (Cartan class B$|$DI) and the ensemble of antisymmetric Hermitian random matrices (Cartan class B$|$D).
It enjoys the special feature that, depending on the matrix dimension $N$, it has exactly $\nu=0$ $(1)$ zero-mode for $N$ even (odd), throughout the symmetry transition. This ``topological protection" is reminiscent of properties of topological insulators. We show that our ensemble represents a Pfaffian point process 
which is typical for such transition ensembles. 
On a technical level, our results follow from the applicability of the Harish-Chandra integral over the orthogonal group. The matrix-valued kernel determining all eigenvalue correlation functions is explicitly constructed in terms of skew-orthogonal polynomials, depending on the topological index $\nu=0,1$. These polynomials interpolate between Laguerre and even (odd) Hermite polynomials for $\nu=0$ $(1)$, in terms of which the two limiting symmetry classes can be solved.
Numerical simulations illustrate our analytical results for the spectral density and an expansion for the distribution of the smallest eigenvalue at finite $N$.

\end{abstract}


\date{\today}
\maketitle

\newpage
\section{Introduction}
\label{sec:intro}

In its regime of applicability random matrix theory (RMT) allows one to describe the universal local spectral statistics of a given physical system once the appropriate symmetry class is identified. Such applications include electrons in disordered systems, quantum chaos, or Quantum Chromodynamics (QCD) to name just a few, and we refer to \cite{book} for a recent collection of many modern applications.

Moreover, RMT is versatile enough to describe the transition between {\it different} symmetry classes and therefore the effect of symmetry breaking. Two classical random two-matrix models were introduced and solved by Mehta and Pandey \cite{PandeyMehta,MehtaPandey}, describing the effect of time-reversal symmetry breaking.
They considered the transition between the  Gaussian unitary ensemble (GUE), relevant for systems without time-reversal invariance, and the 
Gaussian orthogonal ensemble (GOE) \cite{PandeyMehta} as well as the Gaussian symplectic ensemble (GSE) \cite {MehtaPandey} for systems with time-reversal symmetry and integer or half-integer spin, respectively.

In the present work we study a transition ensemble where topology, that is the number of the generic zero-eigenvalues, is preserved, while chirality is broken. One realisation of the corresponding random matrix is defined as
\begin{align}\label{eq:Yaltdef.1}
  J=i\left(\begin{array}{cc}
 a A & \widetilde{W}\\
  -\widetilde{W}^{T} & a B
 \end{array}
\right),
\end{align}
where $A$ and $B$ are two real antisymmetric matrices of sizes $n\times n$ and $(n+\nu)\times (n+\nu)$, respectively, $\widetilde{W}$ is an $n\times(n+\nu)$ real matrix with $\nu=0,1$, and $a$ is a positive real coupling constant. This model exhibits four different scenarios depending on whether $A$ and $B$ are odd- or even-dimensional since the size determines the number of generic zero modes of an antisymmetric matrix.

Let us recall the arguments from the existing literature why such a symmetry transition is worth studying in the context of topological insulators. These have been classified according to dimension and global symmetries~\cite{Kitaev, Ludwig,Chiu,Slager1,Slager2}, see also~\cite{Beenakker,HK,Elliot} for reviews. As explained for example in~\cite{Beenakker}, chiral symmetry appears in such systems due to a combination of time-reversal and particle-hole symmetry. Furthermore, the authors of~\cite{BA} pointed out that in the presence of disorder Majorana modes in quasi one-dimensional quantum wires with spin-orbit coupling lead to the symmetry class of Hermitian antisymmetric random matrices. The simplest choice of the distribution is a Gaussian. In view of this, together with their antisymmetry and their invariance under the orthogonal group we denote this ensemble by GAOE, a notation proposed in~\cite{MarioTim}.  Subsequently to~\cite{BA}, it was suggested in \cite{Dumitrescu}
to study the transition between the chiral Gaussian orthogonal ensemble (chGOE) and the GAOE; the transition ensemble~\eqref{eq:Yaltdef.1} is one possible choice. Several other transitions have been suggested for the corresponding Bogoliubov--de Gennes Hamiltonian, including the chGUE. We will not repeat the arguments for the respective symmetry transitions, see e.g.~\cite{Neven}, starting from the Hamiltonian picture~\cite{Hamiltonian}, and defer a more profound analysis to future work.

We expect that the matrix model \eqref{eq:Yaltdef.1} will be able to capture the statistical behaviour of such a topological superconductor as described above, because the limit of large matrix size often gives way to universality results based only on the symmetries. So as long as the identified symmetry classes are the same, the eigenvalue density correlation functions will agree as well. Especially the regime $a\gg1$ with $n$ odd and $\nu=0$ of our model may be relevant for topological superconductors in the gapless phase.
In this phase, the zero-modes of antisymmetric origin at $a=\infty$ in \eqref{eq:Yaltdef.1} can be identified as a pair of Majorana modes in a quantum wire~\cite{BA,Dumitrescu,Neven}. From the Kitaev model~\cite{Kitaev-Majorana} we know that these two modes lie on the opposite ends of the wire and show themselves as zero modes of the Boguliubov--de Gennes equation. Additionally, the Hamiltonian ideally degenerates into two independent copies of an antisymmetric Hamiltonian~\cite{BA,Dumitrescu,Neven}, when closing the gap with an appropriate magnetic field and neglecting interactions with higher energetical band structures. Each Hamiltonian has only one zero mode and hence can be associated with the subsystem containing only one of the two Majorana modes. This is reflected by the matrix $J$ which is dominated by $A$ and $B$ for $a\gg1$. Taking perturbations of the system into account caused by impurities, thermal fluctuations, or inaccuracies in the experimental setting, the two systems start to couple, at first only weakly. This coupling is designed by the matrix $\widetilde{W}$ and the parameter $1/a$ in the model~\eqref{eq:Yaltdef.1}. As a result, the zero modes are broadened and are no longer exact. To render the model analytically feasible we decided to choose $A$ and $B$ independently in~\eqref{eq:Yaltdef.1}. Hence, they are not given by the same matrix as it is the case in the physical situation, but we expect that we nevertheless keep the most important features of the system in this way. Another motivation for this particular choice~\eqref{eq:Yaltdef.1} has been the analogous structure of the Wilson-Dirac random matrix model~\cite{DSV,ADSV,AN,KieburgMixing} that has applications in lattice QCD.
Also the other three cases of $n$ and $\nu$ in the model~\eqref{eq:Yaltdef.1} may be of interest although they describe only the presence or absence of Majorana modes (which may be unpaired), and not the paired Majorana zero modes (that result from electron-hole pairs).

Depending on the dimension of the random matrix $J$ being even or odd, the ensemble~\eqref{eq:Yaltdef.1} corresponds to different symmetry classes, namely to the Cartan classes $D$ or $B$ see~\cite{AlexMartin}, respectively. In total ten different symmetry classes of random matrices exist~\cite{AlexMartin}, the three Dyson symmetry classes GO/U/SE~\cite{Dyson}, their chiral partners chGU/O/SE~\cite{Jac3fold}, as well as 4 further classes, the antisymmetric or anti-self-dual GAOE and GASE, and the two so-called Bogoliubov--de Gennes types GBOE and GBSE\footnote{We follow here the nomenclature of \cite{MarioTim} rather than the Cartan classes.}~\cite{AlexMartin}.
All of these can be solved in terms of orthogonal or skew-orthogonal Hermite or Laguerre polynomials~\cite{book,Mehta}. 
Many of the transition ensembles between one unitary and one non-unitary symmetry class have been formulated, see e.g. \cite{FHN,NF99}, including those of the four so-called non-standard classes, such as  the transition between GBOE and GASE \cite{KTNK,KatoriTanemura}.  An important tool used here is the representation of these ensembles in terms of Brownian motion, and we refer to~\cite{KTNK,KatoriTanemura} and references therein for this approach. Typically the kernel for all density correlation functions can be determined~\cite{PandeyMehta,MehtaPandey}.  Regarding the distribution of individual eigenvalues such as the smallest eigenvalue distribution, much less is known, even in the case of transitions between ensembles with unitary symmetry; see~\cite{Edelman,DN} for one of the few examples with explicit analytic results for a single ensemble. We refer to \cite{AIp1} and \cite{Nishigaki} where this question was addressed for the random two-matrix model describing the chGUE-GUE and chGUE-chGSE transition, respectively. 
The former is an example that describes the breaking of chiral symmetry by discretisation effects, using Wilson Fermions in Lattice QCD as discussed in~\cite{DSV,ADSV}. The corresponding two-matrix ensemble is a transition from the chGUE with chiral symmetry to the GUE without it~\cite{AN,KieburgMixing}. Its real analogue, valid for for 2-colour QCD, is much more involved and was studied in~\cite{MarioJacWilson}. For a transition ensemble related to the complex Wilson-Dirac random matrix model, which, however, preserves chiral symmetry, see~\cite{KimAdam,MarioTakuya}. The latter enjoys also applications in QCD.

We want to emphasise that our model ~\eqref{eq:Yaltdef.1} in principle allows for an arbitrary number of zero-modes at $a=0$, corresponding to the chGOE, depending on the rectangularity $\nu\in\mathbb{N}$ of the random matrix $\widetilde{W}$ of size $n\times(n+\nu)$ \cite{VerbNc2}. Although this general framework is possible, it would require a mix of orthogonal- and skew-orthogonal polynomials as developed in \cite{KieburgMixing}, and we therefore restrict ourselves to the two cases $\nu=0,1$ to keep the discussion comprehensible and the interpretation of topological protection intact. 

Looking at the model~\eqref{eq:Yaltdef.1} from a mathematical angle, we will see that its spectral statistics satisfies a  Pfaffian point process, see~\cite[Chapter~11.10]{book} for its definition. This was found in all of the above transition ensembles. 
Let us highlight one peculiarity that is distinct from the other models, which is its corresponding symmetry group. Whereas most models, e.g., in~\cite{PandeyMehta,MehtaPandey,DSV,ADSV,AN,KieburgMixing,FHN,NF99,KTNK,KatoriTanemura,AIp1,Nishigaki,MarioTakuya}, are usually invariant with respect to a unitary group in one or another limit, our model always satisfies an orthogonal symmetry, regardless of the value of $a$, including infinity. This difference is remarkable because of the group integral that has to be solved to obtain the joint probability density function (jpdf) of the eigenvalues. In our case the group integral is the Harish-Chandra integral~\cite{HC} of the orthogonal group which is explicitly known in terms of standard functions~\cite{FEFZ}. This knowledge is at the heart of why the model~\eqref{eq:Yaltdef.1} is analytically tractable. This group integral should not be confused with the real Itzykson--Zuber integral~\cite{IZ,KohlerGuhr,BrezinHikami} that obstructs e.g. the calculation of the jpdf of the real Wilson-Dirac operator. For this reason the authors of~ \cite{MarioJacWilson} only calculated the microscopic level density in the limit of large matrices. The Itzykson--Zuber integral~\cite{IZ,KohlerGuhr,BrezinHikami} and the Harish-Chandra integral~\cite{HC,FEFZ} are intimately related, yet they only agree when one integrates over the unitary group. The difference between these two integrals is subtle and originates in their two matrix arguments. For the Harish-Chandra integral the matrices are elements in the Lie-algebra corresponding to the group over which one integrates. In the case of the Itzykson--Zuber integral the matrices lie in the symmetric space dual to this Lie-algebra. This explains why only for the unitary group these two integrals agree.

The paper is organised as follows.
In Section~\ref{sec:MainResults} we introduce the details of the random matrix model~\eqref{eq:Yaltdef.1} including an alternative representation and state our main results. The dependence on the matrix size is presented in Subsections~\ref{sec:Even} and \ref{sec:Odd} and illustrated by numerical simulations of the spectral density compared to our analytical results. In Section~\ref{sec:jpdf} we first prove the equivalence between two alternative matrix representations with the main part being devoted to the derivation of the jpdf of the eigenvalues as a Pfaffian point process. Some details of these derivations are shown in Appendix~\ref{sec:AppendixA}. The construction of our skew-orthogonal polynomials is performed in Section~\ref{sec:sOP}, where we use the supersymmetry method and bosonisation. Additionally, we follow a non-standard approach proposed in~\cite{KieburgMixing} and briefly summarised and applied in~\cite{MarioTakuya} when defining the skew-orthogonal polynomials for an odd number of eigenvalues. It differs from Mehta's approach~\cite[Chapter~5.5]{Mehta}. The advantage is that one can readily use Heine-like formulas as in~\cite{MarioTakuya,AKP,Eynard01} for these polynomials. The validity of this approach here is shown in Appendix~\ref{sec:HeineFormulas}. Several equivalent representations including expressions in terms of the classical Hermite or Laguerre polynomials are also derived in Section~\ref{sec:sOP}, with foresight of the limits $a\to0,1$, and $\infty$ studied in Appendix~ \ref{sec:Limits}. Those limits serve as analytical checks of our results. A discussion of the influence of the interpolation parameter $a$ on the spectral density and on the smallest eigenvalue is presented in Section~\ref{rhop1} with the additional aid of Monte Carlo simulations. Our conclusions are drawn in Section~\ref{sec:Conc}.

\section{Symmetry Transition Ensemble and Main Results}
\label{sec:MainResults}

\subsection{Real Random Two-Matrix Model}
\label{sec:model}

Let us introduce the ensemble of two Gaussian real random matrices that allows us to describe a symmetry transition. This model slightly deviates from~\eqref{eq:Yaltdef.1} though it is equivalent as we will see.
We are interested in the statistics of the non-zero eigenvalues  of the sum of two purely imaginary antisymmetric random matrices 
\begin{align}
\label{Jdef}
J=Y+X\ .
\end{align}
Its individual matrix elements are distributed according to the following normalised density,
\begin{align}
 P(Y,X)=\left(\frac{\pi a^2}{2}\right)^{-N(N-1)/4}
 \left(\frac{\pi (1-a^2)}{2}\right)^{-n(n+\nu)/2}
 \exp\left[-\frac{1}{a^{2}}\text{Tr}\,Y^{2}
 -\frac{1}{1-a^{2}}\text{Tr}  \,X^{2}\right],
 \label{eq:ZNdef}
\end{align}
where $a\in (0,1)$ is a real parameter and $N=2n+\nu$. The first matrix $Y$ is an $N\times N$ antisymmetric Hermitian matrix, implying that it can be written as 
\begin{align}
Y=iH\ ,
\label{YH}
\end{align}
where $H=-H^T$ is real antisymmetric.
The second matrix $X$ is a chiral antisymmetric Hermitian matrix of the same dimension as $Y$,
\begin{align}
  X=\left(\begin{array}{cc}
  0 & iW\\
  -iW^{T} & 0
 \end{array}
\right),
\label{eq:Xdef}
\end{align}
with $W$ an $n\times(n+\nu)$ real matrix without further symmetries. The parameter $\nu=0,1$ takes two values and indicates whether the total matrix dimension $N$ is even ($\nu=0$) or odd ($\nu=1$). Thence, $\nu$ counts the number of exact zero eigenvalues of the matrix $J$, independently of the parameter $a$  
that drives the symmetry transition. For that reason we call $\nu$ the preserved topology. 
Furthermore, we equip the two random matrices with flat Lebesgue measures for all independent matrix elements, $[dY]=\prod_{i=1}^{N}\prod_{i<j=2}^N dH_{i,j}$ and $[dX]=\prod_{i=1}^{n}\prod_{j=1}^{n+\nu} dW_{i,j}$ on the real numbers. This is equivalent to the above stated normalisation in~\eqref{eq:ZNdef}, i.e.
\begin{align}
\label{Pnorm}
\int [dX] [dY] P(Y,X)=1\ .
\end{align}

In the two limits $a\to0$ and $a\to1$ of the density~\eqref{eq:ZNdef},  we obtain the two classical ensembles between which the transition interpolates.
For $a\to0$ we obtain the chGOE also called real Wishart or Laguerre Orthogonal Ensemble. 
The chGOE yields a Pfaffian point process that can be described in terms of skew-orthogonal Laguerre polynomials \cite{VerbNc2}. This ensemble can be defined for an arbitrary number of zero eigenvalues $\nu\geq0$, choosing the matrix $W$ in \eqref{eq:Xdef} to be of size $n\times(n+\nu)$, see~\cite{VerbNc2}. 
In this general case, we would have to employ a combination of skew-orthogonal and orthogonal polynomials, following the ideas pursued in~\cite{KieburgMixing,MarioTakuya}.
However, we focus on the cases $\nu=0,1$ here, as only then the number of zero eigenvalues of $X$, $Y$ and $J$ agree. 
For $\nu>1$, either one or no zero eigenvalue would be preserved, whereas the remaining ones would broaden when increasing $a>0$, see \cite{AN,KieburgMixing} for a similar phenomenon.

In the limit $a\to1$ we obtain the Gaussian ensemble of antisymmetric Hermitian matrices (GAOE). It represents a determinantal point process \cite{Mehta} and can be solved in terms of only even (odd) Hermite polynomials, for $\nu=0$ ($\nu=1$), respectively.

In subsection~\ref{sec:3matrix} we derive an equivalent representation of the random matrix ensemble~\eqref{Jdef}-\eqref{eq:ZNdef}, which is given by the following rescaled random matrix
\begin{align}\label{eq:Yaltdef}
  J=i\left(\begin{array}{cc}
 a A & \widetilde{W}\\
  -\widetilde{W}^{T} & a B
 \end{array}
\right).
\end{align}
Its three individual matrices are distributed according to the normalised density
\begin{align}
 P(A,B,\widetilde{W})=\left(\frac{\pi}{2}\right)^{-N(N-1)/4}
 \exp\left[\text{Tr}\,AA^T+\text{Tr}\,BB^T
 -2\text{Tr}\,  \widetilde{W}\widetilde{W}^T\right].
 \label{eq:ZNdef-b}
\end{align}
This time we have two real antisymmetric matrices $A=-A^T$ and $B=-B^T$, and a matrix $\widetilde{W}$ that is a rectangular matrix like $W$. All independent matrix elements are again equipped with the flat Lebesgue measure. The benefit of this matrix model is its pellucid interpretation when $a$ takes certain values. In the limit $a\to0$ we recover the chGOE and in the limit $a\to1$ the GAOE of size $N=2n+\nu$. Moreover, the representation~\eqref{eq:Yaltdef} also allows us to choose values for $a>1$, as the density \eqref{eq:ZNdef-b} is still integrable in contrast to~\eqref{eq:ZNdef}. In particular, this representation makes it possible to also take the limit $a\to\infty$, in which the two diagonal blocks $A$ and $B$ dominate, and hence the ensemble~\eqref{eq:ZNdef-b} separates into a direct sum of two GAOE's, see \cite{KimAdam} for a similar mechanism in the case of two coupled chGOEs. Exactly this limit, in combination with the choices of $\nu=0$ and $n$ being odd, is expected to correspond to the physical situation of creating two Majorana modes, one at each end of the quantum wire, see~\cite{BA,Neven,Dumitrescu,Kitaev-Majorana}.

As a side remark, the matrix $J$ in~\eqref{eq:Yaltdef} is reminiscent of the Hermitian Wilson Dirac operator $D_5$ with two colours in the fundamental representation ~\cite{MarioJacWilson}. However, there is one important difference here. The Hermitian Wilson Dirac operator $D_5$ contains two real {\it symmetric} matrices in contrast to the antisymmetric matrices $A$ and $B$ in~\eqref{eq:Yaltdef}. This property makes a crucial difference not only in the global symmetry, but also when it comes to compute the jpdf of the eigenvalues of $J$, namely only here the group integral corresponding to the orthogonal degrees of freedom is known explicitly~\cite{HC,FEFZ}. One immediate consequence from the antisymmetry of the matrix is that the eigenvalues come in ``chiral pairs" $(\lambda,-\lambda)$. This holds true for any value of the transition parameter $a$. Let us mention that this behaviour is not the case for the Hermitian Wilson Dirac operator for two colours~\cite{MarioJacWilson}, where the random matrix is real as well, albeit symmetric instead of antisymmetric. This matrix model has only been evaluated in the mean field limit so far.

\subsection{Pfaffian Point Process}
\label{sec:GeneralResults}

We first give the general structure of the spectral statistics for the eigenvalues of the random matrix~\eqref{eq:Yaltdef}, or equivalently~\eqref{Jdef}, before giving the details for specific dimensions; indeed the explicit results strongly depend on the matrix dimensions $n$ and $\nu$. 

The first main result is the jpdf of the eigenvalues $(\pm \lambda_1,\ldots,\pm \lambda_n)$, with $\lambda_j\geq 0$ the singular values of the random matrix $J$ distributed according to~\eqref{eq:ZNdef}. It  is given by the following product of a Vandermonde determinant and a Pfaffian determinant,
\begin{align}
P^{(\nu)}_n(\lambda_1,\ldots,\lambda_n)=C_{n,\nu}\,
\Delta_n\left(\{\lambda^2\}\right)
\left\{
\begin{array}{ll}
\text{Pf}\left[\ \ G_{\nu}(\lambda_j,\lambda_k) \ \ \right]_{j,k= 1,\ldots,n}\quad , &\ \  \mbox{for} \ \ n=2m\ ,\\
& \\
\text{Pf}\left[
\begin{array}{cc}
G_{\nu}(\lambda_{j},\lambda_{k}) & {g}_\nu(\lambda_{j}) \\
-{g}_\nu(\lambda_{k}) & 0 
\end{array}
\right]_{j,k= 1,\ldots,n}, &\ \  \mbox{for} \ \ n=2m+1.
\end{array}
\right.
\label{eq:jpdfdef}
\end{align}
The antisymmetric two-point weight function $G_\nu(x,y)=-G_\nu(y,x)$ is explicitly given by (see Appendix~\ref{sec:AppendixA})
\begin{align}
\begin{split}
G_\nu(x,y)=&
\frac{\pi a^2(1-a^2)}{8}
(xy)^\nu e^{-2(x^2+y^2)}\Bigg(\erf\left[\gamma(y-x)\right] \erf\left[\gamma(x+y)\right]\\
&\quad\quad\quad\quad\quad\quad\quad\quad\quad\quad\quad\quad
-\delta_{\nu,1}\frac{2}{\sqrt{\pi}}
\int_{\sqrt{2}\gamma x}^{\sqrt{2}\gamma y} du \erf\left[\sqrt{2}\gamma(x+y)-u\right]e^{-u^2}
\Bigg)
\end{split}
\label{eq:G01}
\end{align}
with $\gamma = \sqrt{(1-a^2)/a^2}$, and the one-point weight function can be written as 
\begin{align}
g_{\nu}(y) = \sqrt{\frac{\pi a^2(1-a^2)}{8}}\exp\left[-2y^2\right]\left(y \erf \left[\sqrt{2}\gamma y\right]\right)^{\nu}.
\label{eq:g01}
\end{align}
In~\eqref{eq:jpdfdef} as well as below we employ the following convention of the Vandermonde determinant,
\begin{align}
\Delta_n(\{\lambda^2\})=\prod_{1\leq a<b\leq n} (\lambda_b^2-\lambda_a^2) =\det\left[ \lambda_{i}^{2j-2}\right]_{i,j=1\ldots,n}.
\label{eq:Vander}
\end{align}
The normalisation constant reads 
\begin{align}
C_{n,\nu}=\frac{2^{\frac{n}{2}(3+n+\nu)} }{a^n (1-a^2)^{\frac{n}{2}(n+\nu)}}
\prod_{j=0}^{n-1}\frac{1}{\Gamma\left(\frac{j+3}{2}\right)\Gamma\left(\frac{j+\nu+1}{2}\right)}\ ,
\label{jpdf-const}
\end{align}
such that the jpdf~\eqref{eq:jpdfdef} is normalised to unity,
\begin{align}
\prod_{j=1}^n \int_0^\infty d\lambda_j\ P^\nu_n(\lambda_1,\ldots,\lambda_n) = 1\ .
\end{align}
Note that the singular values $\lambda_j$ are not ordered in our entire work.

From the definitions of $G_\nu(x,y)$ and $g_\nu(x)$ in~\eqref{eq:Gnudef} and \eqref{eq:gnudef}, respectively, it becomes obvious that both functions are even functions in their arguments $G_\nu(-x,y)=G_\nu(x,y)=G_\nu(x,-y)$, and $g_\nu(-x)=g_\nu(x)$. For that reason the jpdf \eqref{eq:jpdfdef} only depends on the squared eigenvalues $\lambda_j^2$ for all $j$. Moreover, the weights~\eqref{eq:G01} and~\eqref{eq:g01} and the constant~\eqref{jpdf-const} can be readily analytically continued to $a>1$ when choosing the positive root with the negative real axis as the cut. The parameter $\gamma=\sqrt{1/a^2-1}=i\sqrt{1-1/a^2}$ becomes imaginary for $a>1$ such that we use the function ${\rm erfi}(x)=\erf(i x)/i$ instead of the error-function ``$\erf$".

The $k$-point correlation functions of the jpdf~\eqref{eq:jpdfdef} are defined in the standard way~\cite{book,Mehta,PfStructure},
\begin{align}
R^\nu_k(\lambda_1,\ldots,\lambda_k)= 
\frac{n!}{(n-k)!} \int_0^\infty d\lambda_{k+1}\ldots  \int_0^\infty d\lambda_n P^\nu_n(\lambda_1,\ldots,\lambda_n)\ .
\label{eq:Rkdef}
\end{align}
They can be expressed as follows in terms of three kernels $I_n^\nu$, $S_n^\nu$ and $D_n^\nu$, which depend on the corresponding skew-orthogonal polynomials and their integral transforms, 
\begin{align}
R^\nu_k(\lambda_1,\ldots,\lambda_k)= \text{Pf}
\left[\left(
\begin{array}{rr} 
I^\nu_n(\lambda_i,\lambda_j)& S^\nu_n(\lambda_i,\lambda_j)\\
-S^\nu_n(\lambda_i,\lambda_j)& D^\nu_n(\lambda_i,\lambda_j)\\
\end{array}
\right)\right]_{i,j=1,\ldots,k}.
\label{eq:RkPf}
\end{align}
The inner bracket reflects the fact that we consider a $2\times2$ matrix-valued kernel in the Pfaffian determinant.
The expression~\eqref{eq:RkPf} is the standard form of a Pfaffian point processes, see~\cite{book,Mehta,PfStructure}.

The explicit expressions for  the three kernels differ for even and odd $n$ and will be given below. Similarly, 
the corresponding skew-orthogonal polynomials (sOP) depend on this matrix dimension, though both parities of $n$ share the very same Heine-like formulas, 
\begin{align}
\begin{split}
p_{j}^{(\nu)}(x) =& x^{-\nu}\left<\det(x\mathbf{1}_{2j+\nu}-J)\right>_{j,\nu}\ ,\\
q_{j}^{(\nu)}(x) =& x^{-\nu}\left<\det(x\mathbf{1}_{2j+\nu}-J)\left(x^2+\frac{1}{2}\Tr\,J^2 + c_j^{(\nu)}(a)\right)\right>_{j,\nu}.\label{eq:sOPHeine}
\end{split}
\end{align}
Here, $\left<\ldots\right>_{j,\nu}$ is the average over a matrix $J$ with dimensions $(n,\nu)\to(j,\nu)$ in the random matrix ensemble~\eqref{Jdef}-\eqref{eq:ZNdef}. 
The same relations were derived for one-matrix models in~\cite{Eynard01}, see also~\cite{MarioTakuya,AKP}.
The constants $c_j^{(\nu)}(a)$ are arbitrary, since sOP are not uniquely defined \cite{Mehta} and will be chosen conveniently later.

The fact that the formulas
\eqref{eq:sOPHeine} hold for general $j=0,1,2,\ldots$ is 
derived in Appendix~\ref{sec:HeineFormulas}. 
For fixed and given $j$ and $\nu$, the final result for the sOP  in~\eqref{eq:sOPHeine} reads
\begin{align}
\begin{split}
p_j^{(\nu)}(x) =& x^{-\nu} \frac{4}{\pi\sqrt{1-a^4}} \int_{-\infty}^{\infty}dy \int_{-\infty}^\infty d\lambda\, e^{-\frac{4}{1+a^2}y^2-\frac{4}{1-a^2}\lambda^2} (iy + \lambda + x)^j (iy - \lambda + x)^{j+\nu}\ ,\\
\ \ \ \ q_{j}^{(\nu)}(x) =& x^{-\nu}\left[x^2 - \frac{1}{16}\partial_x^2 + \frac{a^4-1}{2}\partial_{a^2} + \tilde{c}^{(\nu)}_j(a)\right]\left(x^{\nu}p_{j}^{(\nu)}(x)\right)\ .\label{eq:sOPMain}\
\end{split}
\end{align}
Here, the constants $\tilde{c}^{(\nu)}_j(a)$ differ from $c_j^{(\nu)}(a)$ by a shift. We want to emphasize that the result~\eqref{eq:sOPMain} is only valid for $0<a<1$, otherwise the existence of the integral is not guaranteed. When going to $a\geq1$ one has to use other equivalent representations which are derived in Section~\ref{sec:sOP}. 

We note that $p_j^{(\nu)}(x)$ and $q_j^{(\nu)}(x)$ are monic polynomials of degree $j$ and $j+1$ in the variable $x^2$, respectively. Therefore the index should not be confused with the order of the polynomials in $x$. 

In the classical ensembles of random matrices that are given by Pfaffian point processes, like the chGOE, 
the two sets of polynomials defined in~\eqref{eq:sOPHeine} yield the even and odd polynomials. In the chGOE~\cite{VerbNc2} for example, the polynomials $p_j^{(\nu)}(x)$ are given by the Laguerre polynomials (in monic normalisation) of even degree, whereas the polynomials $q_j^{(\nu)}(x)$ are given by linear combinations of Laguerre polynomials of odd degree, see also Appendix~\ref{sec:Limits}. In our model, the two polynomials play different roles, depending on $n$ being even or odd.  For $n=2m$ even, we only need the polynomials $p_{2k}^{(\nu)}(x)$ of even degree $2k$ in $x^2$ and the polynomials $q_{2k}^{(\nu)}(x)$ of odd degree $2k+1$ in $x^2$, for $k=0,1,2,\ldots$ In the other case where $n=2m+1$ is odd, only the polynomials $p_{2k-1}^{(\nu)}(x)$ of odd degree $2k-1$ in $x^2$ and the polynomials $q_{2k-1}^{(\nu)}(x)$ of even degree $2k$ in $x^2$, for $k=1,2,\ldots$, are of use.

\begin{center}
	\begin{figure}[t]
		\includegraphics[width=0.49\linewidth,angle=0]{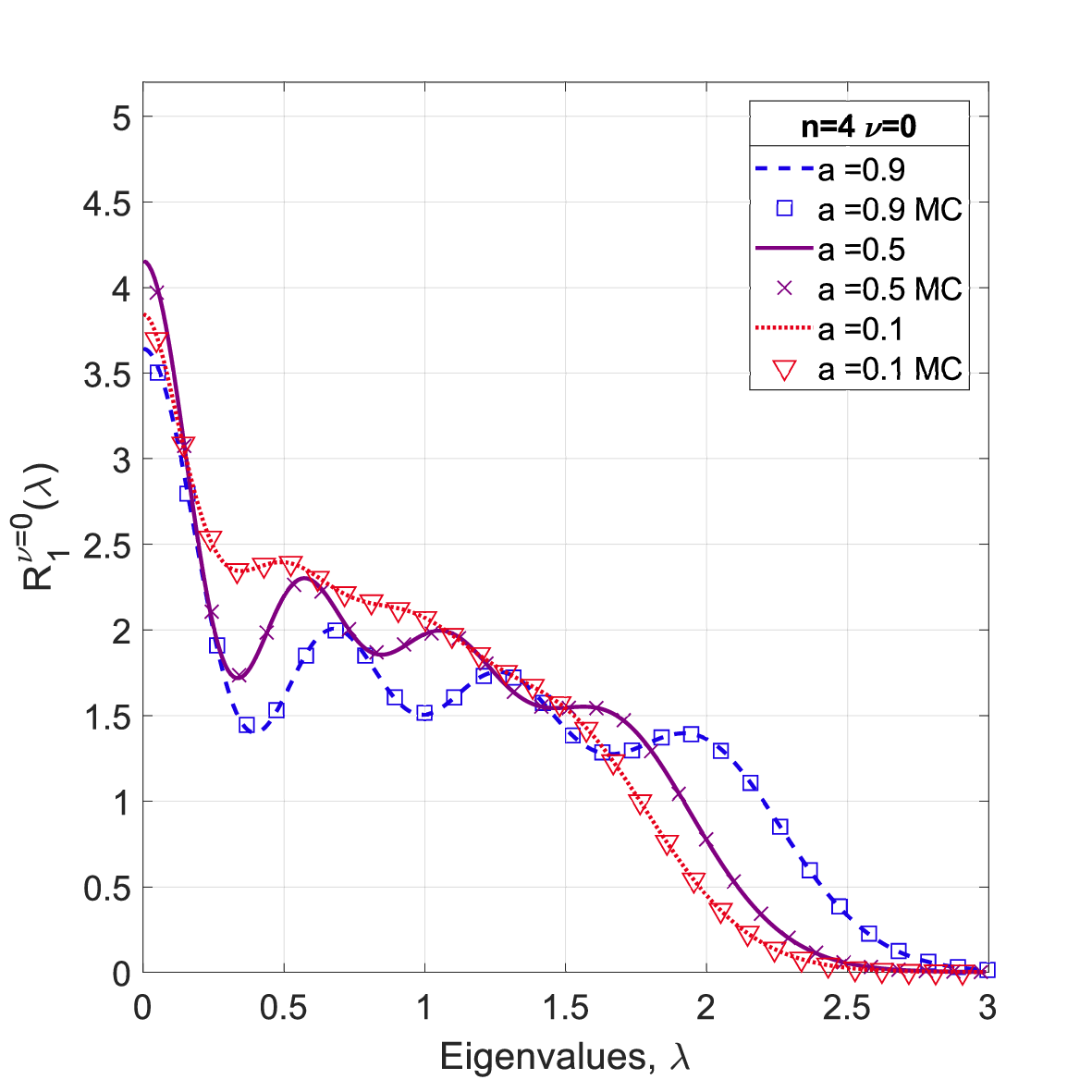}
		\includegraphics[width=0.49\linewidth,angle=0]{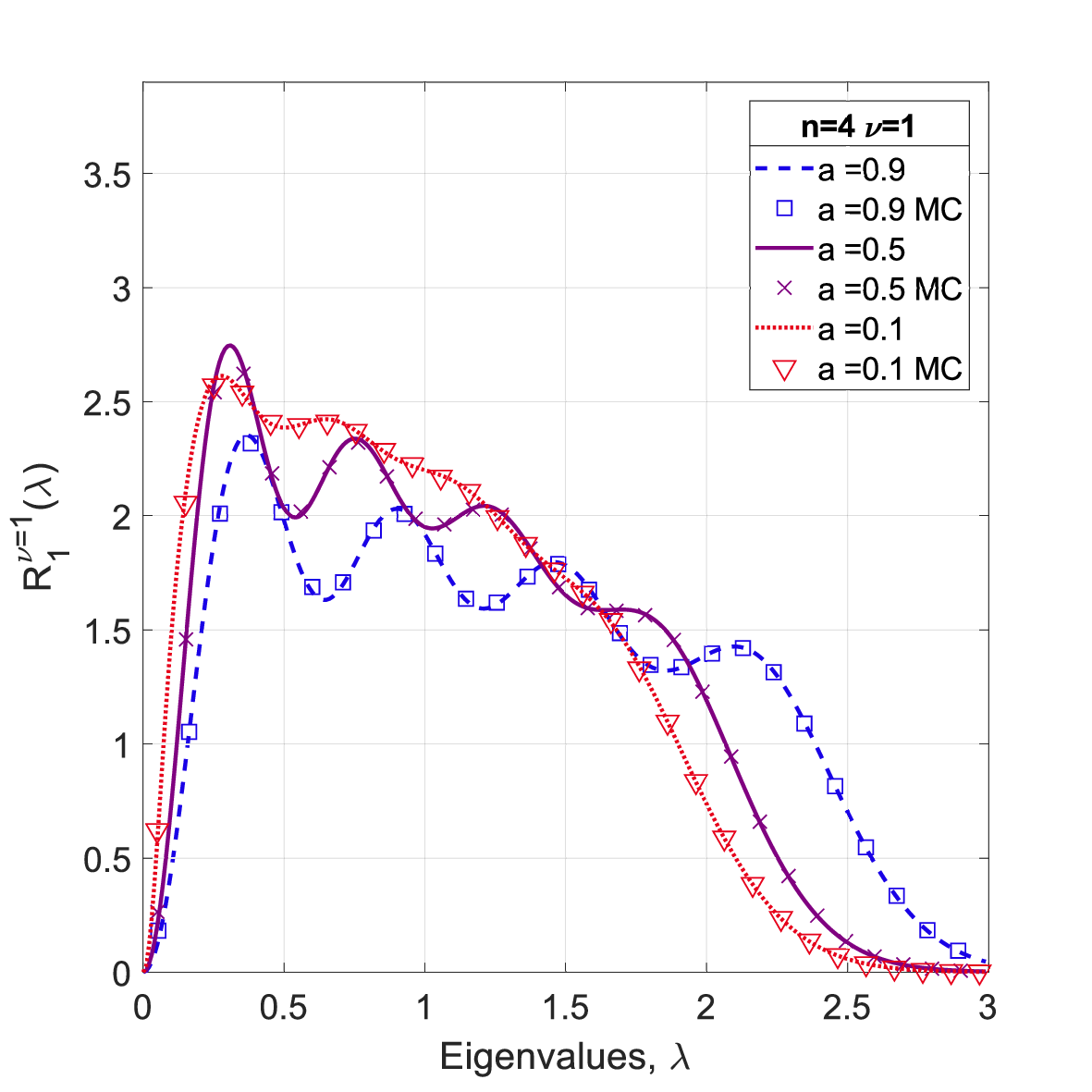}
		\caption{
			The spectral density $R_1^\nu(\lambda)$ taken from the analytical result~\eqref{eq:DensEven} (solid curves) is compared to Monte-Carlo (MC) simulations (symbols) for $n=4$ even with $\nu=0$ (left) and $\nu=1$ (right),
			at three different values of $a=0.1$ (triangles), $0.5$ (crosses), and $0.9$ (squares). The ensemble has consisted of $10^{6}$ matrices and the bin size was chosen to be approximately $0.1$. 
		}
		\label{fig:even}
	\end{figure}
\end{center}

\subsection{Kernels for Even Dimension $n=2m$}\label{sec:Even}

Let us recall that for even $n=2m$, $m=1,2,\ldots$, the jpdf \eqref{eq:jpdfdef} takes the form
\begin{align}
P^{(\nu)}_n(\lambda_1,\ldots,\lambda_n)=C_{n,\nu}\,
\Delta_n\left(\{\lambda^2\}\right)
\text{Pf}\left[G_{\nu}(\lambda_j,\lambda_k)\right]_{j,k= 1,\ldots,n}\ .
\label{eq:jpdfeven}
\end{align}
In this case we can follow the standard approach of \cite{Mehta} and define the following skew-symmetric product, labelled by the subscript ``$e$", which is based on the antisymmetric two-point weight 
$G_\nu(x,y)$ from~\eqref{eq:G01},
\begin{eqnarray}
\langle f_1,f_2\rangle_{e} = -\langle f_2,f_1\rangle_{e} =\int_{0}^{\infty}dx \int_0^\infty dy \,f_1(x)f_2(y)G_\nu(x,y) .\label{eq:sOPprodEven}
\end{eqnarray}
The goal is to find those sOP which satisfy the following skew-orthogonality relations,
\begin{eqnarray}
\langle p_{2j}^{(\nu)},p_{2k}^{(\nu)}\rangle_e=\langle q_{2j}^{(\nu)},q_{2k}^{(\nu)}\rangle_e =0\quad {\rm and}\quad \langle p_{2j}^{(\nu)},q_{2k}^{(\nu)}\rangle_e = h_{2j}^{(\nu)}\delta_{jk},\quad k,l=0,\dots,m-1.\label{eq:sOPProdEven}
\end{eqnarray}
The polynomials~\eqref{eq:sOPHeine} are these and the corresponding normalisation constants $h_{2j}^{(\nu)}$ are given by 
\begin{align}
h_{2j}^{(\nu)}
= \frac{\pi a^2 (1-a^2)^{4j+2+\nu}}{2^{8j+2\nu+7}} (2j)!(2j+\nu)!\ .
\end{align}
The three kernels that determine  the $k$-point correlation functions \eqref{eq:RkPf} can be expressed in terms of these quantities as
\begin{align}
\begin{split}
S_{2m}^\nu(x,y)=& 
\sum_{j=0}^{m-1}\frac{p_{2j}^{(\nu)}(x) \bar{q}_{2j}^{(\nu)}(y) - q_{2j}^{(\nu)}(x) \bar{p}_{2j}^{(\nu)}(y)}{h_{2j}^{(\nu)}}\ ,\\
D_{2m}^\nu(x,y)=& \sum_{j=0}^{m-1}  \frac{\bar{q}_{2j}^{(\nu)}(x) \bar{p}_{2j}^{(\nu)}(y) - \bar{p}_{2j}^{(\nu)}(x)\bar{q}_{2j}^{(\nu)}(y)}{h_{2j}^{(\nu)}} + G_\nu(x,y)\ ,\\
I_{2m}^\nu(x,y)=& \sum_{j=0}^{m-1} \frac{q_{2j}^{(\nu)}(x) p_{2j}^{(\nu)}(y) - p_{2j}^{(\nu)}(x)q_{2j}^{(\nu)}(y)}{h_{2j}^{(\nu)}}\ ,
\label{eq:3kernels}
\end{split}
\end{align}
following \cite{book,Mehta,PfStructure}. Here, we introduce
 the following integral transforms of the polynomials: 
\begin{eqnarray}
\bar{p}_{2j}^{(\nu)}(x) = \int_{0}^{\infty}dy\, p_{2j}^{(\nu)}(y) G_\nu(x,y)\quad{\rm and}\quad \bar{q}_{2j}^{(\nu)}(x) = \int_{0}^{\infty}dy\, q_{2j}^{(\nu)}(y) G_\nu(x,y).
\label{bpq-def}
\end{eqnarray}

As an example, the spectral density or 1-point function is given by~\eqref{eq:RkPf} for $k=1$,
\begin{align}
R_1^\nu(\lambda) = S_{2m}^\nu(\lambda,\lambda)=
\sum_{j=0}^{m-1}\frac{p_{2j}^{(\nu)}(\lambda) \bar{q}_{2j}^{(\nu)}(\lambda) - q_{2j}^{(\nu)}(\lambda) \bar{p}_{2j}^{(\nu)}(\lambda)}{h_{2j}^{(\nu)}}\ . \label{eq:DensEven}
\end{align}
For illustration it is compared to Monte-Carlo simulations in Figure \ref{fig:even} for different values of $a$. For a detailed discussion of the effect of the interpolation parameter $a$ we refer to Section \ref{rhop1}.

\begin{center}
	\begin{figure}
		\includegraphics[width=0.49\linewidth,angle=0]{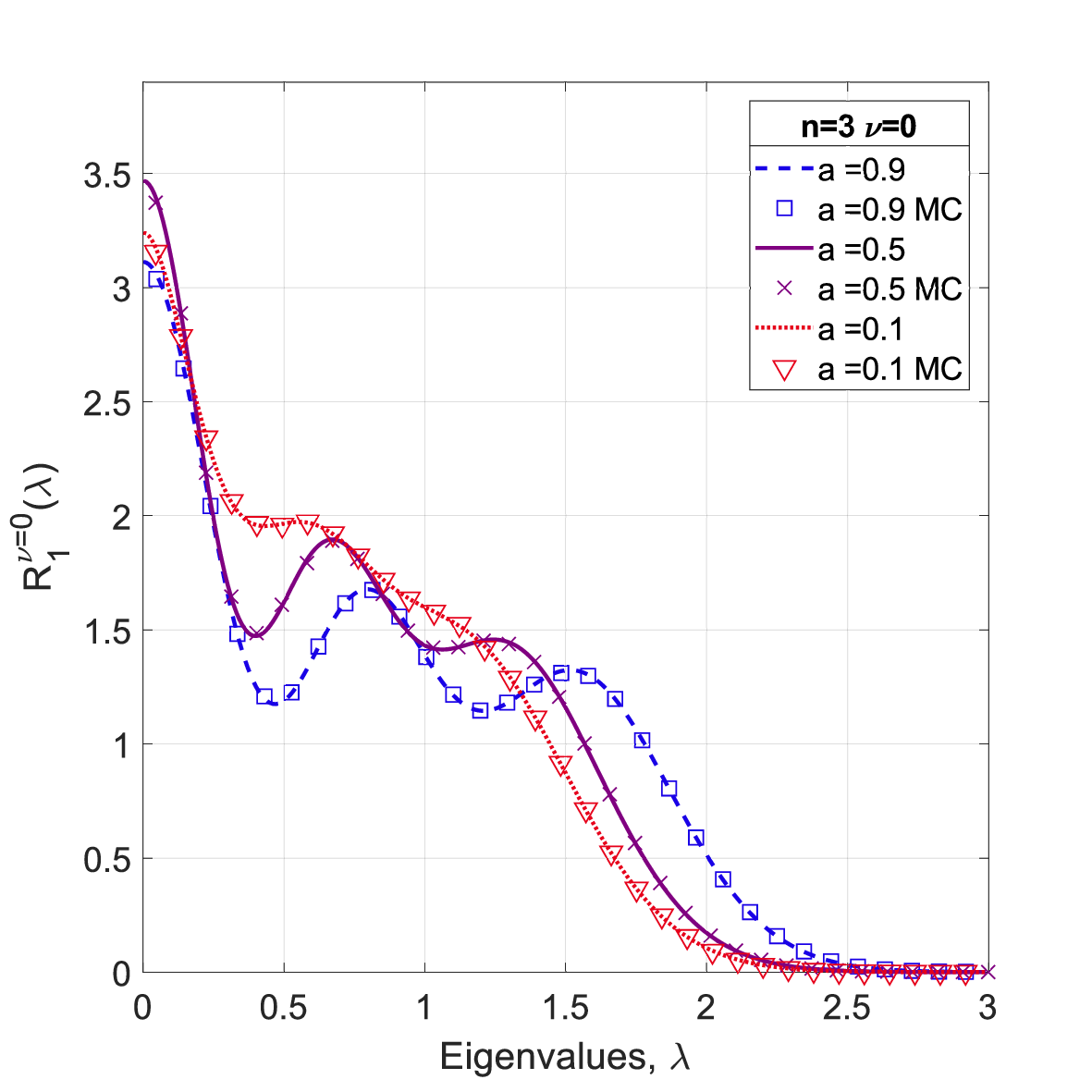}
		\includegraphics[width=0.49\linewidth,angle=0]{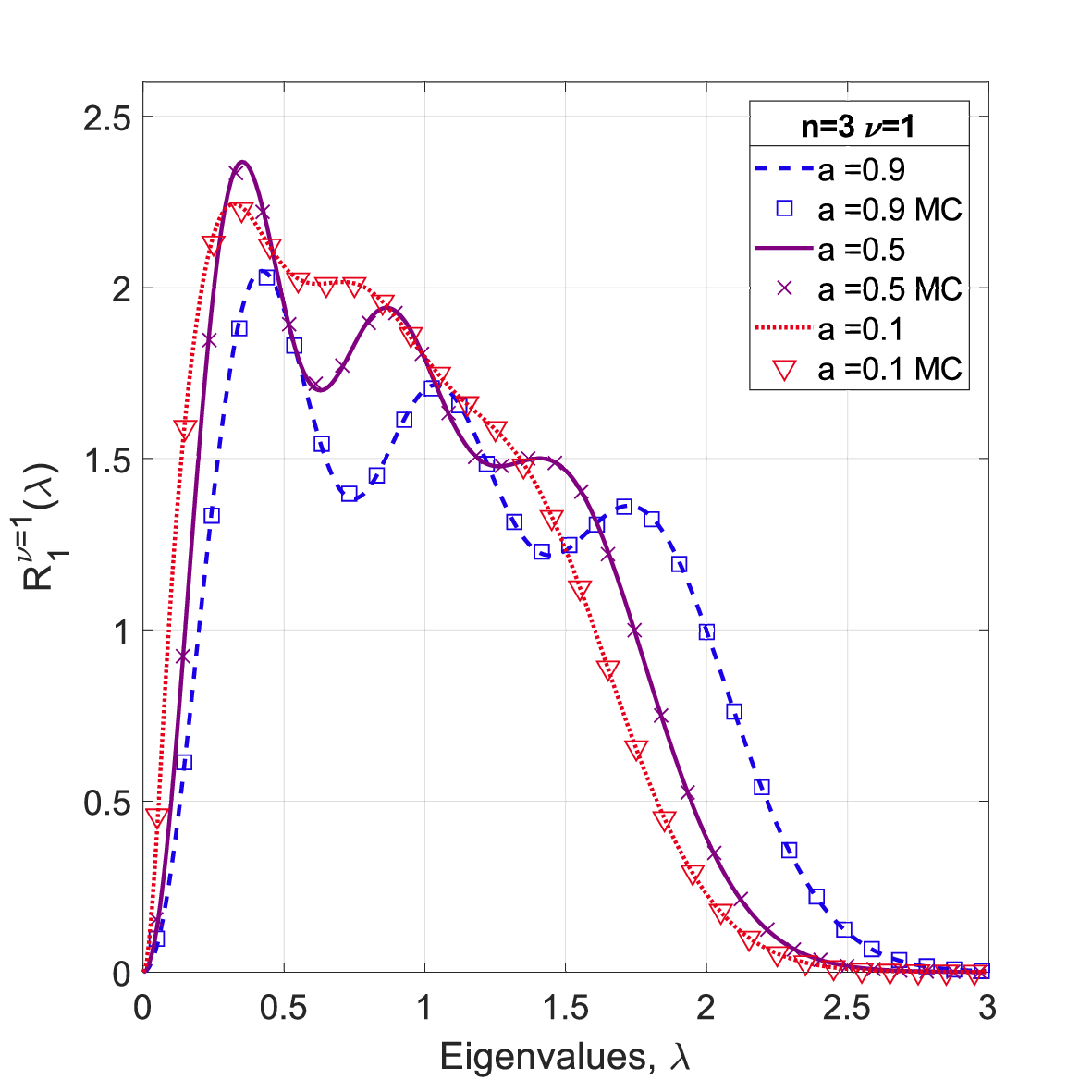}
		\caption{The analytical expression~ \eqref{eq:DensOdd} for the spectral density $R_1^\nu(\lambda)$ (solid curve) is compared to Monte-Carlo 
			(MC) simulations (symbols). We have generated $10^6$ matrices with the dimensions $n=3$ and $\nu=0$ (left) and $\nu=1$ (right). The coupling constant was chosen as in Fig.~\ref{fig:even}, namely $a=0.1$ (triangles), $0.5$ (crosses), and $0.9$ (squares), and the bin size has been set to approximately $0.1$. 
		}
		\label{fig:odd}
	\end{figure}
\end{center}

\subsection{Kernels for Odd Dimension $n=2m-1$}\label{sec:Odd}

We now turn to the odd dimensional case, $n=2m-1$ with $m=1,2,\ldots$, where 
the Pfaffian determinant in~\eqref{eq:jpdfdef} has one extra row and column containing the one-point weight function $g_\nu(x)$ from \eqref{eq:g01}, i.e.
\begin{align}
P^{(\nu)}_n(\lambda_1,\ldots,\lambda_n)=C_{n,\nu}\,
\Delta_n\left(\{\lambda^2\}\right)
\text{Pf}\left[
\begin{array}{cc}
G_{\nu}(\lambda_{j},\lambda_{k}) & {g}_\nu(\lambda_{j}) \\
-{g}_\nu(\lambda_{k}) & 0 
\end{array}
\right]_{j,k= 1,\ldots,n}.
\label{eq:jpdfodd}
\end{align}
This can also be obtained from~\eqref{eq:jpdfdef} for even $n=2m$ when sending one of its singular values, say $\lambda_{2m}$, to infinity, following the ideas of~\cite{ForrMays}. This procedure leads to the following relation $\lim_{y\gg1}G_\nu(x,y) = g_\nu(x) y^\nu g_0(y)$ that is derived as an additional check at the end of Appendix~\ref{sec:AppendixA}.

A standard approach to Pfaffian point processes with odd $n$ is to modify all polynomials from the case of $n$ even,  in order to obtain a skew-orthogonality relation for the polynomials with respect to the one-point weight. We pursue the ideas in~\cite{KieburgMixing,MarioTakuya} instead and modify the skew-symmetric product, while keeping the same polynomials 
$p_j^{(\nu)}(x)$ and $q_j^{(\nu)}(x)$.
For odd $n=2m-1$, the skew-symmetric product, denoted by the subscript ``$o$", is chosen to be
\begin{align}
\langle f_1,f_2\rangle_o=-  \langle f_2,f_1\rangle_o = \int_{0}^{\infty}dx \int_0^\infty dy\, f_1(x)f_2(y)H_\nu(x,y)
\label{eq:sOPprodOdd}
\end{align}
with the weight
\begin{eqnarray}
H_\nu(x,y) = G_\nu(x,y) - \frac{g_\nu(x)}{\bar{g}_\nu} \int_{0}^{\infty}dx'G_\nu(x',y) - \frac{g_\nu(y)}{\bar{g}_\nu} \int_{0}^{\infty}dy'G_\nu(x,y')\ 
\label{Hdef}
\end{eqnarray}
and the constant
\begin{align}
\bar{g}_\nu = \int_0^\infty dx\, g_\nu(x) = \sqrt{\frac{\pi^3 a^2}{32}}\left(\frac{1-a^2}{2\pi}\right)^{(\nu+1)/2}\ .
\label{bargdef}
\end{align}
The integral $\int_{0}^{\infty}dx'G_\nu(x',y)$ is computed in Appendix~\ref{sec:AppendixA}, see~\eqref{Gbar0.res} and~\eqref{Gbar1.res}. The jpdf~\eqref{eq:jpdfodd} does not change under replacing $G_\nu(x,y)$ by $H_\nu(x,y)$, which is still antisymmetric. The reason for this lies in the skew-symmetry of the Pfaffian determinant; we just add multiples of the last row and column to the other rows and columns without changing its value. The redefinition~\eqref{eq:sOPprodOdd} of the skew-product immediately implies that 
\begin{align}
\langle p_{2j-1}^{(\nu)},1\rangle_o\ =\ \langle q_{2j-1}^{(\nu)},1\rangle_o = 0,\qquad{\rm for}\ j=1,2,\ldots
\label{skew1}
\end{align}
Hence any polynomial is skew-orthogonal to the monomial of zeroth order.
As before, the remaining sOP starting from degree $1$ in $x^2$ onwards then satisfy
\begin{eqnarray}
\langle p_{2j-1}^{(\nu)},p_{2k-1}^{(\nu)}\rangle_o=\langle q_{2j-1}^{(\nu)},q_{2k-1}^{(\nu)}\rangle_o =0\quad{\rm and}\quad\langle p_{2j-1}^{(\nu)},q_{2k-1}^{(\nu)}\rangle_o = h_{2j-1}^{(\nu)}\delta_{jk}\ ,
\label{eq:sOPProdOddI}
\end{eqnarray}
for $j,k=1,\dots,m$, with respect to the new skew-symmetric product~\eqref{eq:sOPprodOdd}. The normalisation constants are now given by 
\begin{align}
h_{2j-1}^{(\nu)}
= \frac{\pi a^2 (1-a^2)^{4j+\nu}}{2^{8j+2\nu+3}} (2j-1)! (2j+\nu-1)!\ .
\end{align}
In addition, we also have 
\begin{eqnarray}
\int_{0}^{\infty}dx\, p_{2j-1}^{(\nu)}(x)g_\nu(x)= \int_{0}^{\infty}dx\,q_{2j-1}^{(\nu)}(x)g_\nu(x) =0\ ,\label{eq:sOPProdOdd}
\end{eqnarray}
where $j,k=1,\dots,m$.
The kernels of the $k$-point correlation function \eqref{eq:RkPf} then take a slightly different form\footnote{In slight abuse of notation we use the same names for the three kernels and the normalisation constants. Only their subscript indicates if we are in the even or odd $n$ case.}
\begin{align}
\begin{split}
S_{2m-1}^\nu(x,y)=& 
\sum_{j=1}^{m-1}\frac{p_{2j-1}^{(\nu)}(x) \tilde{q}_{2j-1}^{(\nu)}(y) - q_{2j-1}^{(\nu)}(x) \tilde{p}_{2j-1}^{(\nu)}(y)}{h_{2j-1}^{(\nu)}} + \frac{g_\nu(x)}{\bar{g}_\nu}\ ,\\
D_{2m-1}^\nu(x,y)=& \sum_{j=1}^{m-1}  \frac{\tilde{q}_{2j-1}^{(\nu)}(x) \tilde{p}_{2j-1}^{(\nu)}(y) - \tilde{p}_{2j-1}^{(\nu)}(x)\tilde{q}_{2j-1}^{(\nu)}(y)}{h_{2j-1}^{(\nu)}} + H_\nu(x,y)\ ,\\
I_{2m-1}^\nu(x,y)=& \sum_{j=1}^{m-1} \frac{q_{2j-1}^{(\nu)}(x) p_{2j-1}^{(\nu)}(y) - p_{2j-1}^{(\nu)}(x)q_{2j-1}^{(\nu)}(y)}{h_{2j-1}^{(\nu)}}\ .
\end{split}
\label{eq:3kernelsOdd}
\end{align}
Here the transformed polynomials are integrated with respect to the new two-point weight~\eqref{Hdef}
\begin{eqnarray}
\tilde{p}_{2j-1}^{(\nu)}(x) =\int_{0}^{\infty}dy\, p_{2j-1}^{(\nu)}(y) H_\nu(x,y) \quad{\rm and}\quad\tilde{q}_{2j-1}^{(\nu)}(x) = \int_{0}^{\infty}dy\, q_{2j-1}^{(\nu)}(y) H_\nu(x,y),
\label{pqtrafo-odd}
\end{eqnarray}
for $ j=1,2,\ldots$

Let us again consider the example of the spectral density. Due to the additional row in the Pfaffian and that we have to deal with the monomials of zeroth order differently, the spectral density now reads
\begin{align}
R_1^\nu(\lambda) = S_{2m-1}^\nu(\lambda,\lambda)=\sum_{j=1}^{m-1}\frac{p_{2j-1}^{(\nu)}(\lambda) \tilde{q}_{2j-1}^{(\nu)}(\lambda) - q_{2j-1}^{(\nu)}(\lambda) \tilde{p}_{2j-1}^{(\nu)}(\lambda)}{h_{2j-1}^{(\nu)}} + \frac{g_\nu(\lambda)}{\bar{g}_\nu}\ .
\label{eq:DensOdd}
\end{align}
The new term $g_\nu(\lambda)/\bar{g}_\nu$, compared to the density \eqref{eq:DensEven}, originates from this particularity for odd $n=2m-1$. It essentially describes the distribution of the smallest singular value of the random matrix $J$ since it is the only term left for $m=1$. Adding $m$ new singular values of $J$, represented as new peaks, arise only on the right-hand side of the maximum of this distribution. The new terms in the sum~\eqref{eq:DensOdd} also contribute corrections to the individual distribution of the smallest eigenvalue due to the level repulsion caused by the other singular values. The identification of the term $g_\nu(\lambda)/\bar{g}_\nu$ with the smallest eigenvalue is therefore not exact, but a good approximation.

A comparison of~\eqref{eq:DensOdd} to Monte-Carlo simulations 
is shown in Figure~\ref{fig:odd}, and we again refer to Section~\ref{rhop1} for a more detailed discussion.

\sect{Joint Probability Density of the Eigenvalues}
\label{sec:jpdf}

The main goal of this section is to derive the results~\eqref{eq:jpdfdef}-\eqref{jpdf-const}, but before doing so we show that  the ensemble~\eqref{eq:Yaltdef}-\eqref{eq:ZNdef-b} produces the same spectral statistics as the original random matrix model~\eqref{Jdef}-\eqref{eq:ZNdef}, see Subsection~\ref{sec:3matrix}. Thereafter, we compute the jpdf of the eigenvalues of $J$ in subsection~\ref{sec:jpdf.der}.

\subsection{Equivalence with a Three-Matrix Model}\label{sec:3matrix}

Let us spell out the anti-symmetric Hermitian matrix $Y=iH$ in block form:
\begin{align}
  Y=i\left(\begin{array}{cc}
  \widetilde{A} & V\\
  -V^{T} & \tilde{B}
 \end{array}
\right)\ \ \Rightarrow\ \ 
  J=Y+X=i\left(\begin{array}{cc}
  \tilde{A} & V+W\\
  -V^T-W^{T} & \widetilde{B}
 \end{array}
\right).
\label{eq:Jdef}
\end{align}
Here, $\widetilde{A}$ and $\widetilde{B}$ are real antisymmetric of dimensions $n$ and $n+\nu$, respectively, and $V$ and $W$ are  real rectangular $n\times(n+\nu)$ matrices. 
In terms of these matrices the probability density~\eqref{eq:ZNdef} reads
\begin{align}
 P(Y,X)=
 \left(\frac{\pi a^2}{2}\right)^{-\frac{N(N-1)}{4}}
 \left(\frac{\pi (1-a^2)}{2}\right)^{-\frac{n(n+\nu)}{2}}
  e^{\frac{1}{a^2}(\text{Tr}\,\widetilde{A}^2+\text{Tr}\,\widetilde{B}^2)-\frac{2}{a^{2}}\text{Tr}\,VV^T
 -\frac{2}{1-a^{2}}\text{Tr}\,WW^T}.
 \label{eq:ZN2}
\end{align}
With a slight abuse of notation regarding the labelling of the probability distributions for $J$ in terms of $Y$ and $X$, and of $A$, $B$, and $\widetilde{W}$, cf. Eqs.~\eqref{Jdef} and~\eqref{eq:Yaltdef}, we can identify 
\begin{align}
J=i\left(\begin{array}{cc}
  \tilde{A} & V+W\\
  -V^T-W^{T} & \widetilde{B}
 \end{array}
\right)=i\left(\begin{array}{cc}
 a A & \widetilde{W}\\
  -\widetilde{W}^{T} & a B
 \end{array}
\right).
\end{align}
Hence the two distributions are related as
\begin{align} \label{eq:ZN3}
\begin{split}
P(A,B,\widetilde{W})=&\int \delta(A-\widetilde{A}/a)\delta(B-\widetilde{B}/a)\delta(\widetilde{W}-V-W)P(Y,X)[d\widetilde{A}][d\widetilde{B}][dV][dW]\\
=&\left(\frac{\pi a^2}{2}\right)^{-\frac{N(N-1)}{4}}
 \left(\frac{\pi (1-a^2)}{2}\right)^{-\frac{n(n+\nu)}{2}}a^{\frac{n(n-1)}{2}+\frac{(n+\nu)(n+\nu-1)}{2}}\\
 &\times\int  e^{\text{Tr}\,A^2+\text{Tr}\,B^2-\frac{2}{a^{2}}\text{Tr}\,VV^T
 -\frac{2}{1-a^{2}}\text{Tr}\,(\widetilde{W}-V)(\widetilde{W}-V)^T}[dV]\\
=&\left(\frac{\pi a^2}{2}\right)^{-\frac{N(N-1)}{4}}
 \left(\frac{\pi (1-a^2)}{2}\right)^{-\frac{n(n+\nu)}{2}}a^{\frac{n(n-1)}{2}+\frac{(n+\nu)(n+\nu-1)}{2}}\left(\frac{\pi a^2(1-a^2)}{2}\right)^{\frac{n(n+\nu)}{2}}\\
 &\times e^{\text{Tr}\,A^2+\text{Tr}\,B^2-2\text{Tr}\,\widetilde{W}\widetilde{W}^T}.
\end{split}
\end{align}
After evaluating the Dirac delta-functions, we shifted the remaining integral over $V$ by $a^2\widetilde{W}$, which leads to a decoupling with the matrix $\widetilde{W}$. The integral over $V$ is then a centered Gaussian integral yielding the additional constant. The result~\eqref{eq:ZNdef-b} is obtained when the constant is simplified even more.

The random matrix~\eqref{eq:Yaltdef} thus constitutes an equivalent representation of the real two-matrix model. 
The advantage of the new representation~\eqref{eq:Yaltdef} is that the parameter $a$ can be extended to arbitrary real positive numbers instead of the open unit interval, which is of particular importance when applying the model to the physical system of Majorana modes in quantum wires~\cite{BA,Dumitrescu,Neven,Kitaev-Majorana}.

\subsection{Derivation of the JPDF}\label{sec:jpdf.der}

We want to note that both even and odd $n$ can be dealt simultaneously in the following. Only at the end, when we state the explicit result we have to distinguish between them.

We start from~\eqref{eq:ZNdef} and change variables $Y\to J=Y+X$, while keeping the matrix $X$ unchanged,
\begin{align}
 P(J,X) &= \left(\frac{2}{\pi a^2}\right)^{(2n+\nu)(2n+\nu-1)/4}
\left(\frac{2}{\pi (1-a^2)}\right)^{n(n+\nu)/2}\label{eq:ZJX}\\
&\times\exp\left[
 -\frac{1}{a^{2}}\text{Tr}\,J^{2}
 -\frac{1}{a^{2}(1-a^{2})}\text{Tr}\,X^{2}
 +\frac{2}{a^{2}}\text{Tr}\,JX
\right]\ .\nonumber
\end{align}
Once again the linear transformation only yields a Jacobian equal to unity. 
For the computation of the jpdf we have to proceed in two steps. First, we have to block-diagonalise the matrices $J$ and $X$ which is standard, yielding well-known Jacobians, and afterwards we integrate out the angular degrees resulting from the diagonalisations. As the group integral does not drop out in  the coupling term ${\rm Tr}\,JX$, its integration is performed with the help of the Harish-Chandra integral for the orthogonal group. In the second step we integrate over the eigenvalues of $X$ to obtain the jpdf of the eigenvalues of $J$.

Following Cartan's Theorem, the antisymmetric purely imaginary $N\times N$ matrix $J$ can be brought to the following block-diagonal form, using an orthogonal transformation $\mathcal{O}$
\begin{align}
J=i\mathcal{O}\Lambda_\lambda \mathcal{O}^T\ ,
\label{eq:Jdiag}
\end{align}
where the block-diagonal $N\times N$ matrix is
\begin{align}
\Lambda_\lambda=\begin{cases}
\text{diag}\left(\lambda_1 i\tau_2,\ldots,\lambda_n i\tau_2\right) \  ,&\mbox{for} \ \ N=2n\ \ (\nu=0)\ ,\\
\text{diag}\left(\lambda_1 i\tau_2,\ldots,\lambda_n i\tau_2,0\right)\ , &\mbox{for} \ \ N=2n+1\ \ (\nu=1)\ .
\end{cases}
\label{eq:blockdiag}
\end{align}
The matrix $\Lambda_\lambda$ comprises all eigenvalue pairs $\pm\lambda_{j=1,\ldots,n}$, with $\lambda_j>0$ being the singular values of $J$. The subscript of $\Lambda_\lambda$ indicates the singular values which we introduced, as we use the same notation for the matrix $X$ below. The matrix $\tau_2$ is the second Pauli matrix. It is clear that every individual subblock  $\lambda_j i\tau_2$ is invariant under the orthogonal group $O(2)$. Therefore, the orthogonal matrix in~\eqref{eq:Jdiag} belongs to the coset $\mathcal{O}\in O(N)/O(2)^n$.

The Jacobian for the transformation \eqref{eq:Jdiag} is known~\cite{EdelmanRao} and contains the Vandermonde determinant \eqref{eq:Vander} squared
\begin{align}
[dJ] &= \frac{2^n\pi^{n(n+\nu-\frac12)}}{n!\prod_{j=0}^{n-1}\Gamma(j+1)\Gamma\left(j+\nu+\frac12\right)} [d\mathcal{O}] \prod_{j=1}^n d\lambda_j\lambda_j^{2\nu}\ \Delta_n(\{\lambda^2\})^2\ ,
\label{eq:JacobiJ}
\end{align}
where $[d\mathcal{O}]$ denotes the normalised Haar measure on the orthogonal group, $\int[d\mathcal{O}]=1$.
The constant on the right-hand side of~\eqref{eq:JacobiJ} is equal to the following quotient of integrals\footnote{For its computation we can set $a=1$ here.}
\begin{align}
\frac{\int [dJ]\exp[-\Tr J^2]}{\prod_{k=1}^n\int_0^\infty d\lambda_k\lambda_k^{2\nu}\Delta_n(\{\lambda^2\})^2\exp[-2\sum_{j=1}^n\lambda_j^2]} = \frac{\left(\pi/2\right)^{\frac{(2n+\nu)(2n+\nu-1)}{4}}2^{n(n+\nu+\frac12)}}{\prod_{j=0}^{n-1}\Gamma(j+2)\Gamma\left(j+\nu+\frac12\right)}\ ,
\label{const1}
\end{align}
where we have used that $J$ has $N(N-1)/2$ independent real matrix elements as well as 
\begin{align}
\text{Tr}\,J^2=2\sum_{j=1}^n\lambda_j^2 \ .
\label{eq:TrJ2}
\end{align}
The integral in the denominator of \eqref{const1} is a Selberg integral, 
\begin{align}
\prod_{i=1}^n\int_0^\infty dx_ix_i^{\kappa}\ |\Delta_n(\{x\})|^\beta e^{-\frac{\beta}{2}\sum_{j=1}^nx_j}
=\left(\frac{2}{\beta}\right)^{n(\kappa+1)+\frac{\beta}{2}n(n-1)}
\prod_{j=0}^{n-1}\frac{\Gamma\left(1+\frac{\beta}{2}(j+1)\right)\Gamma\left( 1+\kappa+\frac{\beta}{2}j \right)}{\Gamma\left( 1+\frac{\beta}{2}\right)}
\ ,
\label{Selberg}
\end{align}
for $\beta>0$ and $\kappa>-1$, see e.g. \cite[Chapter 17]{Mehta}.
Here and in the following we use that $\nu(\nu-1)=0$ for $\nu=0,1$.

The non-zero eigenvalues of the matrix $X$ from \eqref{eq:Xdef} are given by the pairs $\pm x_{j=1,\ldots,n}$, where $x_{j}\geq 0$ are the singular values of the matrix $W$. Its  singular value composition reads $W=P\diag(x_1,\ldots,x_n)Q^T$, where $P,Q\in O(n)$ for $\nu=0$, and $P\in O(n)$ and $Q^TQ=\eins_n$ with $Q$ of size $(n+1)\times n$  for $\nu=1$. This 
leads to 
\begin{align}
X=i\left(\begin{array}{cc}
  P & 0\\
0 & Q
 \end{array}\right)
\left(\begin{array}{cc}
  0 & \diag(x_1,\ldots,x_n)\\
  -\diag(x_1,\ldots,x_n) & 0
 \end{array}\right)
\left(\begin{array}{cc}
  P^T & 0\\
0 & Q^T
 \end{array}\right).
\label{eq:X2}
\end{align}
By a linear transformation, in fact by a permutation of rows and columns, we 
can find a representation similar to \eqref{eq:Jdiag}, 
\begin{align}
X=i\widetilde{\mathcal{O}}\Lambda_x\widetilde{\mathcal{O}}^T
\label{eq:Xdiag}
\end{align}
using the notation~\eqref{eq:blockdiag} for $\Lambda_x$. The matrix $\widetilde{\mathcal{O}}$ is also orthogonal, i.e., $\widetilde{\mathcal{O}}\in O(2n+\nu)$, although it has some substructure. However the explicit form of this structure is not important, as can be seen below.
The corresponding Jacobian for the diagonalisation of $X$ is also known~\cite{VerbNc2} (and different from that of the matrix $J$), being given by
\begin{align}
[dX] = \frac{\pi^{\frac{n}{2}(n+\nu+1)}}{\prod_{j=0}^{n-1}\Gamma\left(\frac{j+3}{2}\right)\Gamma\left(\frac{j+\nu+1}{2}\right)} [d\widetilde{\mathcal{O}}] \prod_{j=1}^n dx_jx_j^{\nu}\ |\Delta_n(\{x^2\})|\ .
\label{eq:JacobiX}
\end{align} 
The normalised Haar measure on the corresponding coset is denoted by $[d\widetilde{\mathcal{O}}]$. Once more the quotient of integrals
\begin{align}
\frac{\int [dX]\exp[-\Tr X^2]}{\prod_{k=1}^n\int_0^\infty dx_kx_k^{\nu}|\Delta_n(\{x^2\})|\exp[-2\sum_{j=1}^nx_j^2]} = \frac{\left(\frac{\pi}{2}\right)^{\frac{n}{2}(n+\nu)}2^{\frac{n}{2}(n+\nu)}}{\pi^{-\frac{n}{2}}\prod_{j=0}^{n-1}\Gamma\left(\frac{j+3}{2}\right)\Gamma\left(\frac{j+\nu+1}{2}\right)}
\end{align}
fixes the normalisation constant; the denominator follows from  the Selberg integral~\eqref{Selberg} again.

Collecting all constants, we obtain
\begin{align}
\begin{split}
&P^{(\nu)}_{n}(\lambda_1,\dots,\lambda_n)\label{eq:Zev}\\
=& \left(\frac{2}{\pi a^2}\right)^{^{n(n+\nu-\frac12)}}
\left(\frac{2}{\pi(1-a^2)}\right)^{\frac{n(n+\nu)}{2}}\frac{2^n\pi^{\frac{3n}{2}(n+\nu)} }{n!\prod_{j=0}^{n-1}\Gamma(j+1)\Gamma\left(j+\nu+\frac12\right)\Gamma\left(\frac{j+3}{2}\right)\Gamma\left(\frac{j+\nu+1}{2}\right)}\\
&\times\int [d\widetilde{\mathcal{O}}][d\mathcal{O}] \prod_{j=1}^n \left(
\int_0^\infty dx_jx_j^{\nu}\lambda_j^{2\nu}\right) |\Delta_n(\{x^2\})| \Delta_n(\{\lambda^2\})^2\\
&\times \exp\left[ -\frac{2}{a^2}\sum_{j=1}^n\lambda_j^2 -\frac{2}{a^2(1-a^2)}\sum_{j=1}^n x_j^2 -\frac{2}{a^2}\Tr\,\mathcal{O}\Lambda_\lambda \mathcal{O}^T\widetilde{\mathcal{O}}\Lambda_x\widetilde{\mathcal{O}}^T\right],
\end{split}
\end{align}
Owing to the invariance of the Haar measures  $d[\mathcal{O}]$ and $d[\widetilde{\mathcal{O}}]$, in the last term in the exponential the conjugation by the group element
$\widetilde{\mathcal{O}}$ can be absorbed by $\mathcal{O}\in O(2n+\nu)/O(2)^n$, and we are left with the Harish-Chandra integral~\cite{HC} over the orthogonal group,
\begin{align}
\int [d\mathcal{O}]  \exp\left[-\frac{2}{a^2}\Tr\,\mathcal{O}\Lambda_\lambda \mathcal{O}^T\Lambda_x\right]= \left(\prod_{j=0}^{n-1}(2j+\nu)!\left(\frac{a^2}{4}\right)^{2j+\nu}\right) \frac{\det\left[f_\nu(x_i\lambda_j)\right]_{i,j=1}^n}{\Delta_n(\{x^2\}) \Delta_n(\{\lambda^2\})\prod_{k=1}^n(x_k\lambda_k)^\nu},
\label{eq:HC-On}
\end{align}
see~\cite{FEFZ} for details of its derivation.
Our normalisation is chosen such that at $\lambda_1,\dots,\lambda_n=0$ the integral is unity. Following \cite{FEFZ}, we define 
\begin{align}
f_\nu(x)= \left\{
\begin{array}{ll}
\cosh\left[\frac{4}{a^2}x\right]\ , & \mbox{for} \ \ \nu=0\ ,\\
&\\
\sinh\left[\frac{4}{a^2}x\right]\ , & \mbox{for} \ \ \nu=1\ .\\
\end{array}
\right.
\label{eq:fnudef}
\end{align}
Inserting this into \eqref{eq:Zev} we arrive at
\begin{align}
\begin{split}
&P^{(\nu)}_{n}(\lambda_1,\dots,\lambda_n) \label{eq:jpdf-xz}\\
=& 
\frac{\pi^{\frac{n}{2}} 2^{\frac{n}{2}(5-n-\nu)} }{a^n (1-a^2)^{\frac{n}{2}(n+\nu)}}
\frac{1}{n!}
\prod_{j=0}^{n-1}\frac{\Gamma(2j+\nu+1)}{\Gamma(j+1)\Gamma\left(j+\nu+\frac12\right)\Gamma\left(\frac{j+3}{2}\right)\Gamma\left(\frac{j+\nu+1}{2}\right)}\\
&\times\prod_{j=1}^n \left(
\lambda_j^{\nu}e^{-\frac{2}{a^2}\lambda_j^2} \int_0^\infty dx_j e^{-\frac{2}{a^2(1-a^2)}x_j^2}\right) 
 \frac{|\Delta_n(\{x^2\})|}{\Delta_n(\{x^2\})} \Delta_n(\{\lambda^2\})\det\left[f_\nu(x_k\lambda_l)\right]_{k,l=1}^n\ .
 \end{split}
\end{align}
The remaining integral can be brought into a standard form for random matrix ensembles yielding Pfaffian point processes, see~\cite{KieburgGuhrb}, because of the sign of the Vandermonde determinant $\Delta_n(\{x^2\})$. The sign of the Vandermonde determinant has the form~\cite{deBruijn}
\begin{align}
\begin{split}
 \frac{|\Delta_n(\{x^2\})|}{\Delta_n(\{x^2\})} =&\prod_{i<j}^n \mbox{sign}(x_j^2-x_i^2)= \prod_{i<j}^n \mbox{sign}(x_j-x_i)\\
 =&
\left\{
\begin{array}{ll}
\mbox{Pf}\left[{\rm sign}(x_j-x_i)\right]_{i,j=1}^n\ , & \mbox{for}\ \ n=2m\,,\\
&\\
\mbox{Pf}\left[
\begin{array}{c|c}
{\rm sign}(x_j-x_i)& \vec{1}\\ \hline
\overset{\ }{-\vec{1}^T}     & 0\\
\end{array}
\right]_{i,j=1}^n\ , &\mbox{for}\ \ n=2m-1\,,\\
\end{array}
\right.
\end{split}
\end{align}
where $\vec{1}$ is an $n$-dimensional vector whose entries are all equal to $1$. Note that we now have to distinguish between even and odd $n$. This identity is the asymptotics of the Schur-Pfaffian identity~\cite{Schur} in the limit of large distance of its arguments.
We apply the integration theorem of de Bruijn~\cite[Section 4]{deBruijn} which in its most general form reads
\begin{align}
\int dx_1\ldots dx_{n} \frac{|\Delta_n(\{x^2\})|}{\Delta_n(\{x^2\})} \det[\varphi_i(x_j)]_{i,j=1}^n=
\left\{
\begin{array}{ll}
\mbox{Pf}\left[a_{i,j}\right]_{i,j=1}^n\ , & \mbox{for}\ \ n=2m\,,\\
&\\
\mbox{Pf}\left[
\begin{array}{c|c}
a_{i,j}& b_i\\ \hline
-b_j     & 0\\
\end{array}
\right]_{i,j=1}^n\ , &\mbox{for}\ \ n=2m+1\,,\\
\end{array}
\right.
\label{eq:dB}
\end{align}
with 
\begin{eqnarray}
a_{i,j}=\int_0^\infty dx\int_0^\infty dy\ \mbox{sign}(y-x)\varphi_i(x)\varphi_j(y)\quad{\rm and}\quad b_j=\int_0^\infty dx\ \varphi_i(x)\,.
\label{eq:abdef}
\end{eqnarray}
Finally, the weights $\lambda_j^{\nu}e^{-\frac{2}{a^2}\lambda_j^2} $ 
in \eqref{eq:jpdf-xz} can be pulled into the rows and columns of the Pfaffian determinant so that we finally arrive at
\begin{align}
P^{(\nu)}_{n}(\lambda_1,\dots,\lambda_n)=C_{n,\nu}\ \Delta_n(\{\lambda^2\})
\left\{
\begin{array}{ll}
\mbox{Pf}\left[G_\nu(\lambda_i,\lambda_j)\right]_{i,j=1}^n\ , & \mbox{for}\ \ n=2m\,,\\
&\\
\mbox{Pf}\left[
\begin{array}{c|c}
G_\nu(\lambda_i,\lambda_j)& g_\nu(\lambda_i)\\ \hline
-g_\nu(\lambda_j)     & 0\\
\end{array}
\right]_{i,j=1}^n\ , &\mbox{for}\ \ n=2m+1\,,\\
\end{array}
\right.
\label{eq:ZNfinal}
\end{align}
with 
\begin{align}
G_\nu(\lambda,u)=&(\lambda u)^\nu e^{-\frac{2}{a^2}(\lambda^2+u^2)} \int_0^\infty dx\int_0^\infty dy\ \mbox{sign}(y-x)e^{-\frac{2}{a^2(1-a^2)}(x^2+y^2)}f_\nu(x\lambda)f_\nu(yu)\,,
\label{eq:Gnudef}\\
g_\nu(\lambda)=& \lambda^\nu e^{-\frac{2}{a^2}\lambda^2}\int_0^\infty dx\ e^{-\frac{2}{a^2(1-a^2)}x^2} f_\nu(x\lambda)\,,\label{eq:gnudef}\\
C_{n,\nu} =& 
\frac{2^{\frac{n}{2}(3+n+\nu)} }{a^n (1-a^2)^{\frac{n}{2}(n+\nu)}}
\prod_{j=0}^{n-1}\frac{1}{\Gamma\left(\frac{j+3}{2}\right)\Gamma\left(\frac{j+\nu+1}{2}\right)}
\ .
\label{jpdf-const1}
\end{align}
The antisymmetry $G_\nu(\lambda,u)=-G_\nu(u,\lambda)$ is obvious, due to the antisymmetry of the integrand under interchange of integration variables $x$ and $y$. It also follows from the definition~\eqref{eq:fnudef} of the function $f_\nu(x)$ that $x^\nu f_\nu(x)$ is an even function, and that therefore both $G_\nu(\lambda,u)$ and $g_\nu(\lambda)$ are even functions in their arguments $\lambda$ and $u$, separately.

The simplification leading from~\eqref{eq:Gnudef} and~\eqref{eq:gnudef} to~\eqref{eq:G01} and~\eqref{eq:g01} are provided in Appendix~\ref{sec:AppendixA}.
We also want to underline that the normalisation constant $C_{n,\nu}$ in~\eqref{jpdf-const1} is equally valid for even and odd $n$.

\sect{Skew-Orthogonal Polynomials}\label{sec:sOP}

The goal of this section is to derive the explicit results~\eqref{eq:sOPMain} for the sOP, as well as some equivalent expressions. Together with the normalisation constants, which are derived in Subsection~\ref{sec:norms}, they determine the three kernels and thus all $k$-point correlation functions in the respective cases of an even or odd dimension $n$.
Let us begin by recalling the Heine-like formulas~\cite{MarioTakuya,AKP,Eynard01} that we briefly rederive in Appendix~\ref{sec:HeineFormulas},
\begin{align}
\begin{split}
p_j^{(\nu)} (x) =& x^{-\nu}\langle \det(x \eins_{2j+\nu} - J)\rangle_{j,\nu},\\
q_{j}^{(\nu)} (x) =& x^{-\nu}\left< \det(x \eins_{2j+\nu} - J) \left(x^2 + \frac{1}{2}\Tr [J^2] + c_j^{(\nu)}(a)\right)\right>_{j,\nu}\ .\label{eq:HeineLike}
\end{split}
\end{align}

Our strategy is as follows. In Subsection~\ref{sec:contour}, a generating function is defined from which both expectation values \eqref{eq:HeineLike} follow. We compute this generating function by first integrating out the matrix $X$. Expressing the determinant over $J$ inside the expectation value as a Grassmann integral, we are able to perform the remaining Gaussian integrals over $J$. Using bosonisation, the resulting expression is then mapped to a double contour integral. In this form we can show that the polynomials $q_j^{(\nu)}(x)$ directly follow by applying a differential operator in $x$ and $a$ acting on the polynomials $p_j^{(\nu)}(x)$. 
In the following Subsection~\ref{sec:sOP:OtherRep}, we rewrite these contour integrals in terms of the Gaussian integrals quoted in~\eqref{eq:sOPMain}, 
as well as in terms of classical Hermite or Laguerre polynomials. 

\subsection{Derivation of Contour Integral Representations}
\label{sec:contour}

Let us define the following generating function
\begin{align} 
Q_j^{(\nu)}(x;s)
=
D_{j,\nu}(a)
\int [dJ] [dX] \det(x \eins_{2j+\nu} - J) 
\exp\left[
 -\frac{s}{a^{2}}\text{Tr}\left[J^{2}\right]
 -\frac{1}{a^{2}(1-a^{2})}\text{Tr}\left[X^{2}\right]
 +\frac{2}{a^{2}}\text{Tr}\left[JX\right]
\right],
\label{Qdef}
\end{align}
where we average over the matrices $J$ and $X$ of dimensions $2j+\nu$. Compared to the probability density~\eqref{eq:ZJX}, we have introduced an extra parameter $s$ in front of the term ${\rm Tr}\,J^2$.  The constant is
\begin{align}
D_{j,\nu}(a)= \left(\frac{2}{\pi a^2}\right)^{(2j+\nu)(2j+\nu-1)/4}
\left(\frac{2}{\pi (1-a^2)}\right)^{j(j+\nu)/2}
\label{Ddef}
\end{align}
and depends on $j,\nu$ and $a$, but not on $s$. The generating function~\eqref{Qdef} can be used to find the averages~\eqref{eq:HeineLike} in the following way,
\begin{align}
x^\nu p_j^{(\nu)}(x) =Q_j^{(\nu)}(x;s=1)\quad{\rm and}\quad x^{\nu} q_j^{(\nu)}(x) =
\left. 
\left(x^2 - \frac{a^2}{2}\frac{\partial}{\partial s} + c_j^{(\nu)}(a)\right) 
Q_j^{(\nu)}(x;s)\right|_{s=1}\ .
\label{pqQrel}
\end{align}

To evaluate the integrals in~\eqref{Qdef}, we first parametrise our matrices as follows,
\begin{eqnarray}
J = i\left(\begin{matrix}
	A & V\\
	-V^T & B
\end{matrix}\right)\quad{\rm and}\quad 
X = i\left(\begin{matrix}
	0 & W\\
	-W^T & 0
\end{matrix}\right)\ ,
\label{JXpara}
\end{eqnarray}
as in Section~\ref{sec:jpdf}, with $A$ and $B$ real antisymmetric matrices of dimensions $j$ and $j+\nu$, and $V$ as well as $W$ of dimensions $j\times(j+\nu)$, respectively. This leads to the expression
\begin{align}
\begin{split}
Q_j^{(\nu)}(x;s)=&
D_{j,\nu}(a)
 \int [dA][dB][dV][dW]\det
\left[\begin{matrix}
	x\eins_j-iA & -iV\\
	iV^T & x\eins_{j+\nu}-iB
	\end{matrix}
\right] 
e^{-\frac{s}{a^2}\left(2\text{Tr}\,VV^T-\text{Tr}\,A^2-\text{Tr}\,B^2\right)}\\
&\times  e^{ -\frac{2}{a^{2}(1-a^2)}\text{Tr}\,{W}{W}^T
+\frac{2}{a^{2}}\text{Tr}[{W}V^T+V{W}^T]}\\
=& D_{j,\nu}(a)\left(\frac{\pi a^2(1-a^2)}{2}\right)^{\frac{n(n+\nu)}{2}}
 \int [dA][dB][dV]\det
\left[\begin{matrix}
	x\eins_j-iA & -iV\\
	iV^T & x\eins_{j+\nu}-iB
	\end{matrix}
\right] \\
&\times e^{\frac{s}{a^2}(\text{Tr}\,A^2+\text{Tr}\,B^2)-\frac{2(s-1+a^2)}{a^2}\text{Tr}\,VV^T}.
\end{split}
\label{preQ}
\end{align}
In the second step we have integrated over the matrix $W$, yielding an extra $s$-independent constant. From now on we  disregard $s$-independent normalisation constants in~\eqref{preQ} and subsequent equations. The overall normalisation of the final form of the polynomials $p_j^{(\nu)}(x)$ and $q_j^{(\nu)}(x)$ is fixed by making them monic.

Next, we express the determinant as a Grassmann integral over two complex anti-commuting vectors $\psi_L$ and $\psi_R$ of dimensions $j$ and $j+\nu$, respectively. We refer to~\cite{Berezin} for some introduction into superalgebra and superanalysis. Denoting by $[d\psi]$ the product of differentials over all independent Grassmann variables $(\psi_L)_l,(\psi^*_L)_l,(\psi_R)_k,(\psi_R^*)_k$, we have 
\begin{align}
\begin{split}
Q_{j}^{(\nu)}(x,s) \propto &
\int [dA][dB][dV] [d\psi] 
\exp\left[ \frac{s}{a^2}\left(\Tr\,A^2 + \Tr\,B^2\right) - \frac{2(a^2 + s - 1)}{a^2}\Tr\,VV^T
\right]\\
&\hspace*{-1cm}\times \exp\left[x(\psi_L\hc \psi_L + \psi_R\hc\psi_R) + i\Tr\,A\psi_L \psi_L\hc+ i\Tr\,B\psi_R \psi_R\hc
-i\Tr[V^T\psi_L\psi_R^\dag-V\psi_R\psi_L^\dag]
\right].
\end{split}
\end{align}
The trace is projective so the antisymmetry of $A$ and $B$ is imposed on the terms $\psi_L\hc \psi_L$ and $\psi_R\hc \psi_R$, which therefore can be antisymmetrised as well. In doing so we have to take into account that Grassmann variables anti-commute. Introducing the $j\times 2$ and $(j+\nu)\times2$ dimensional matrices 
$\phi_L = (\psi_L , \psi_L^*)$ and $\phi_R = (\psi_R , \psi_R^*)$, which now contain all independent Grassmann variables, we rewrite the above equation as follows by using the Pauli matrices $\sigma_1$ and $\sigma_2$,
\begin{align}
Q_{j}^{(\nu)}(x,s) \propto & \int [dA][dB][dV] [d\psi] \exp\left[\frac{s}{a^2}\left(\Tr\,A^2 + \Tr\,B^2\right) - \frac{2(a^2 + s - 1)}{a^2}\Tr\,VV^T \right]\nn\\
&\times\exp\left[\frac{x}{2}\Tr\,i\sigma_2(\phi_L^T \phi_L + \phi_R^T\phi_R)+ \frac{i}{2}\Tr\,A\phi_L \sigma_1 \phi_L^T+ \frac{i}{2}\Tr\,B\phi_R \sigma_1 \phi_R^T\right]\nn\\
&\times \exp\left[\frac{i}{2}\Tr\,V\phi_R\sigma_1\phi_L^T - \frac{i}{2}\Tr\,V^T\phi_L\sigma_1\phi_R^T \right]\nn\\ 
\propto & \left(\frac{\pi a^2}{s}\right)^{\frac{n(n+\nu-1)}{2}}
\left(\frac{\pi a^2}{2(s+a^2-1)}\right)^{\frac{n(n+\nu)}{2}}\nn\\
&\times\int [d\psi] \exp\left[-\frac{a^2}{16s^2}\Tr\left(\sigma_1 \phi_L^T\phi_L\right)^2 - \frac{a^2}{16s^2}\Tr\left(\sigma_1 \phi_R^T\phi_R\right)^2 \right]\nn\\
&\times \exp\left[\frac{x}{2}\Tr\,i\sigma_2(\phi_L^T \phi_L + \phi_R^T\phi_R) - \frac{a^2}{8(a^2 + s - 1)}\Tr\,\sigma_1\phi_L^T\phi_L\sigma_1\phi_R^T\phi_R\right]\ .
\label{preBoson}
\end{align}
In the last step we have integrated out the Gaussian matrices $A,B$ and $V$, leading to extra normalisation factors in front that now depend on $s$.  For the polynomial $p_j^{(\nu)}(x)$ this is immaterial, as the monic normalisation can be fixed at the end. However, from~\eqref{pqQrel} we see that for the determination of $q_j^{(\nu)}(x)$ the differentiation with respect to $s$ also acts on this $s$-dependent prefactor as well as on the integrand. But as the differentiation of the prefactor only yields a term proportional to the polynomial $p_j^{(\nu)}(x)$, it just contributes to the constant $c_j^{(\nu)}(a)$ in~\eqref{eq:HeineLike}. We denote this modification by shifting $c_j^{(\nu)}(a)\to \tilde{c}_j^{(\nu)}(a)$. Because the previous constant has been arbitrary, we do not need to compute the precise value of this shift.
With this modification we can also drop the $s$-dependent prefactors in~\eqref{preBoson} in the ensuing computations.

Next, we perform the bosonisation~\cite{Sommers,LSZ,KSG} which allows us to write the Grassmann integrals as contour integrals. This step is possible since the right-hand side of~\eqref{preBoson} only depends on the combinations $\phi_L^T \phi_L$ and $\phi_R^T \phi_R$. Both matrices are two-dimensional and antisymmetric and their only non-zero entries are two nilpotent scalar variables which can be represented by $\Tr\,\sigma_2 \phi_L^T \phi_L$ and $\Tr\,\sigma_2 \phi_R^T \phi_R$. We may expand the function in a finite Taylor series of these two variables and, as a result of the integral over the Grassmann variables, we are only interested in the highest order of this expansion since it involves the product over all Grassmann variables. Exactly this term can also be obtained via a contour integral over two phases, in particular we replace $\phi_L^T \phi_L\to i e^{i\varphi_L}\sigma_2$ and $\phi_R^T \phi_R\to i e^{i\varphi_R}\sigma_2$ with $\varphi_L,\varphi_R\in[0,2\pi]$ and pick out the $j$'th and $(j+\nu)$'th power of these phases, respectively. This approach is exactly at the heart of bosonisation~\cite{Sommers,LSZ,KSG}. This procedure  leaves us with
\begin{align}
\begin{split}
Q_{j}^{(\nu)}(x,s) \propto& \int_{0}^{2\pi} \frac{d\varphi_L}{2\pi} e^{-i j\varphi_L} \int_{0}^{2\pi} \frac{d\varphi_R}{2\pi} e^{-i (j+\nu)\varphi_R} \exp\left[-\frac{a^2}{8s}(e^{2i\varphi_L} + e^{2i\varphi_R})\right]\\
&\times\exp\left[ - x(e^{i\varphi_L} + e^{i\varphi_R}) - \frac{a^2}{4(a^2 + s - 1)}e^{i(\varphi_L + \varphi_R)}\right].
\end{split}
\label{Qfinal}
\end{align}
Employing~\eqref{pqQrel}, we arrive at the first polynomial by setting $s=1$,
\begin{align}
\begin{split}
p_j^{(\nu)}(x) =& \frac{j!(j+\nu)!}{(-x)^\nu} \int_{0}^{2\pi} \frac{d\varphi_L}{2\pi} e^{-i j\varphi_L} \int_{0}^{2\pi} \frac{d\varphi_R}{2\pi} e^{-i (j+\nu)\varphi_R}\\
&\times\exp\left[-\frac{a^2}{8}(e^{2i\varphi_L} + e^{2i\varphi_R}) - x(e^{i\varphi_L} + e^{i\varphi_R}) - \frac{1}{4} e^{i(\varphi_L + \varphi_R)}\right],
\end{split}
\label{pcontour}
\end{align}
where we have already divided by $x^\nu$ and given the correct monic normalisation. The normalisation follows from an expansion of the two $x$-dependent exponential factors $-xe^{i\varphi_L}$ and $-xe^{i\varphi_R}$ in two Taylor series. The highest powers in $x^2$ that contribute to the angular integrals are of the orders $j$ and $j+\nu$, respectively. 
The other angle dependent terms in the second line of~\eqref{pcontour} only contribute with unity.
The Taylor coefficients cancel the factorials in~\eqref{pcontour} and we obtain
\begin{align}
p_j^{(\nu)}(x) = x^{2j}+ O(x^{2j-1})\ .
\label{pmonic}
\end{align}

For the polynomial $q_j^{(\nu)}(x)$ we have to differentiate~\eqref{Qfinal} and subsequently set $s=1$. As explained above, this yields the following answer, with the modified constant $\tilde{c}_j^{(\nu)}(a)$,
\begin{align}
q_{j}^{(\nu)}(x) =& \frac{j!(j+\nu)!}{(-x)^\nu} \int_{0}^{2\pi} \frac{d\varphi_L}{2\pi} e^{-i j\varphi_L} \int_{0}^{2\pi} \frac{d\varphi_R}{2\pi} e^{-i (j+\nu)\varphi_R} 
\exp\left[-\frac{a^2}{8}(e^{2i\varphi_L} + e^{2i\varphi_R}) - x(e^{i\varphi_L} + e^{i\varphi_R})\right]\nonumber\\
&\times\exp\left[ - \frac{1}{4} e^{i(\varphi_L + \varphi_R)}\right]\left(x^2 - \frac{a^4}{16}(e^{2i\varphi_L} + e^{2i\varphi_R}) -\frac{1}{8}e^{i(\varphi_L + \varphi_R)} + \tilde{c}_j^{(\nu)}(a)\right)\ .
\label{qcontour}
\end{align}
In this form the relation between $p_j^{(\nu)}(x)$ and $q_{j}^{(\nu)}(x)$ from~\eqref{eq:sOPMain} becomes more transparent, where $q_{j}^{(\nu)}(x)$ is not generated by differentiating with respect to an auxiliary variable like $s$. Namely, we can generate~\eqref{qcontour} by application of a second order differential operator in $x$ and $a$, acting directly on \eqref{pcontour}, i.e.
\begin{align}
q_{j}^{(\nu)}(x) &= x^{-\nu}\left(x^2 - \frac{1}{16}\partial_x^2 + \frac{a^4-1}{2}\partial_{a^2} + \tilde{c}^{(\nu)}_j(a)\right)\left(x^\nu p_{j}^{(\nu)}(x)\right). \label{eq:qjOp}
\end{align}
The fact that~\eqref{qcontour} is also monic, of degree $j+1$ in $x^2$, easily follows from~\eqref{eq:qjOp} and~\eqref{pmonic} as only the multiplication by $x^2$ contributes to the highest power.
Equations~\eqref{pcontour},~\eqref{qcontour} and the relation~\eqref{eq:qjOp} constitute the main results of this subsection.

\subsection{Equivalent Representations of sOP}\label{sec:sOP:OtherRep}

The representations of the polynomials $p_j^{(\nu)}(x)$ and $q_j^{(\nu)}(x)$ in terms of angular integrals will be complemented by three further equivalent representations. We derive an integral representation in terms of two Gaussian integrals as well as expressions yielding sums or integrals over classical Hermite and Laguerre polynomials, which are extremely helpful when taking limits.

\subsubsection{Representation as Gaussian Integrals}

Starting from the angular integral representation~\eqref{pcontour}, we apply two Hubbard-Stratonovich transformations in order to linearise the angular dependence in the exponent,
\begin{align}
\begin{split} 
p_j^{(\nu)}(x) =& \frac{j!(j+\nu)!}{(-x)^\nu} \frac{4}{\pi\sqrt{1-a^4}} \int_{-\infty}^{\infty}dy \int_{-\infty}^{\infty} d\lambda \int_{0}^{2\pi} \frac{d\varphi_L}{2\pi} e^{-i j\varphi_L} \int_{0}^{2\pi} \frac{d\varphi_R}{2\pi} e^{-i (j+\nu)\varphi_R}\\
&\times\exp\left[-\frac{4}{1+a^2}y^2-\frac{4}{1-a^2}\lambda^2 - (iy + \lambda + x) e^{i\varphi_L} - (iy - \lambda + x) e^{i\varphi_R}\right]\ .
\end{split}
\end{align}
The angular integrals can now be performed, leading to 
\begin{align}
p_j^{(\nu)}(x) = x^{-\nu} \frac{4}{\pi\sqrt{1-a^4}} \int_{-\infty}^{\infty}dy \int_{-\infty}^{\infty}d\lambda\, e^{-\frac{4}{1+a^2}y^2-\frac{4}{1-a^2}\lambda^2} (iy + \lambda + x)^j (iy - \lambda + x)^{j+\nu}\ .
\label{pGauss}
\end{align}
This is the form stated in~\eqref{eq:sOPMain} and the monic normalisation can be easily checked by looking at the limit for large $x$. We may then find $q_j^{(\nu)}(x)$ via the relation~\eqref{eq:qjOp}.

\subsubsection{Representation as Hermite Polynomials}

A Taylor expansion of the term coupling the two angles in the second line of~\eqref{pcontour} decouples the two angular integrals. We represent them as complex contour integrals, integrating counter-clockwise around the origin,
\begin{align}
\begin{split}
p_j^{(\nu)}(x) =& \frac{j!(j+\nu)!}{(-x)^\nu}\sum_{k=0}^{\infty}\frac{1}{(-4)^k k!} \oint \frac{dz_L}{2\pi i}\frac{1}{z_L^{j-k+1}} \oint \frac{dz_R}{2\pi i}\frac{1} {z_R^{j+\nu-k+1}}
e^{-\frac{a^2}{8}({z_L}^2 + {z_R}^2)- x(z_L + z_R)}\\
=&\left(\frac{a^2}{8}\right)^{j+\nu/2}x^{-\nu}\sum_{k=0}^{j}\frac{j!(j+\nu)!}{k!(j-k)!(j-k+\nu)!}\left(-\frac{2}{a^2}\right)^{k}H_{j-k}\left(\sqrt{\frac{2}{a^2}}x\right)H_{j-k+\nu}\left(\sqrt{\frac{2}{a^2}}x\right).
\end{split}\label{psum}
\end{align}
Note that the sum terminates at $k=j$ because of the orders of the poles at the origin. The second step is the result after identifying the contour representation of the Hermite polynomials and cancelling some signs for $\nu=1$. From here $q_j^{(\nu)}(x)$ may be found through the relation~\eqref{eq:qjOp}. The representation derived here is particularly useful for an explicit study of the polynomials at low degree.

\subsubsection{Representation as Laguerre Polynomials}

We start from the Gaussian representation \eqref{pGauss},
\begin{align}
\begin{split}
p_j^{(\nu)}(x) =& \frac{4x^{-\nu}}{\pi\sqrt{1-a^4}} \int_{-\infty}^{\infty}dy\  e^{-\frac{4}{1+a^2}y^2} \int_{-\infty}^{\infty} d\lambda\ e^{-\frac{4}{1-a^2}\lambda^2} ((iy+x)^2 - \lambda^2)^j (iy+x-\lambda)^{\nu}\\
=& \frac{4x^{-\nu}}{\pi\sqrt{1-a^4}} \int_{-\infty}^{\infty}dy\int_{-\infty}^{\infty}d\lambda\  e^{-\frac{4}{1+a^2}y^2-\frac{4}{1-a^2}\lambda^2} \sum_{k=0}^{j}\binom{j}{k} (iy+x)^{2(j-k)+\nu} (i\lambda)^{2k}\\
=& \frac{2x^{-\nu}}{\sqrt{\pi}\sqrt{1-a^2}} \int_{-\infty}^{\infty}d\lambda\  e^{-\frac{4}{1-a^2}\lambda^2}\sum_{k=0}^{j}\binom{j}{k}\left(\frac{\sqrt{1+a^2}}{4}\right)^{2j-2k+\nu}  H_{2(j-k)+\nu}\left(\frac{2x}{\sqrt{1+a^2}}\right) (i\lambda)^{2k}\ .
\end{split}
\label{prepL}
\end{align}
For $\nu=1$, the term $(iy+x-\lambda)^{\nu}$ in the first line can be replaced by $(iy+x)^{\nu}$ due to parity of the remaining integrand in $\lambda$. 
In the second line we have  made a binomial expansion
and in the last line we have used the integral representation of the Hermite polynomials, which are orthogonal with respect to $\exp[-x^2]$, i.e.
\begin{align}
H_n(x)=\frac{2^n}{\sqrt{\pi}}\int_{-\infty}^\infty dt (it+x)^n e^{-t^2}\ .
\label{Hint}
\end{align}
At this point we exploit the following identity
\begin{align}
(2i)^{2n}n!\int_{-\infty}^{\infty}dy\,e^{-(y-\lambda)^2}  L_n(x^2+y^2)  &= \int_{-\infty}^{\infty}dy\,e^{-(y-\lambda)^2} \sum_{m=0}^{n}\binom{n}{m}H_{2(n-m)}(x)H_{2m}(y) \nonumber\\
&= \sqrt{\pi}\sum_{m=0}^{n}\binom{n}{m}(2\lambda)^{2m}H_{2(n-m)}(x)\ ,
\label{Hid}
\end{align}
where the first line is given in~\cite[Eq. 18.18.40]{NIST} and the second line follows from~\cite[Sec. 7.374]{Gradshteyn}.
Inserting this into~\eqref{prepL} at $\nu=0$, we end up with
\begin{align}
\begin{split}
p_j^{(0)}(x) =& \frac{(-1)^{j}j!(1+a^2)^{j}}{2^{2j-1}\pi\sqrt{1-a^2}} \int_{-\infty}^{\infty} dy\int_{-\infty}^{\infty}d\lambda\  L_j\left(\frac{4x^2}{1+a^2}+y^2\right) e^{-\left(y-\frac{2i\lambda}{\sqrt{1+a^2}}\right)^2-\frac{4}{1-a^2}\lambda^2}\\
=& \frac{j!(1+a^2)^{j+1/2}}{(-4)^{j}\sqrt{2\pi} a}\int_{-\infty}^{\infty} dy\  L_j\left(\frac{4x^2}{1+a^2}+y^2\right) e^{-\frac{1+a^2}{2a^2}y^2}\ .
\end{split}
\label{pL0}
\end{align}
In the last line we have completed the square in $\lambda$ and integrated it out. 

For $\nu=1$, we modify the identity~\eqref{Hid}. Using the well-known relations for Hermite and Laguerre polynomials for $k>0$, 
\begin{eqnarray}
\frac{\partial H_{k}(x)}{\partial x}= 2k H_{k-1}(x)\quad{\rm and}\quad\frac{\partial L_k^{(0)}(x)}{\partial x} = - L_{k-1}^{(1)}(x),
\end{eqnarray}
we may differentiate \eqref{Hid} with respect to $x$, and then shift $n-1\to n$ to obtain
\begin{align}
2x(2i)^{2n}n!\int_{-\infty}^{\infty}dy\,e^{-(y-\lambda)^2}  L_n^{(1)}(x^2+y^2) =\sqrt{\pi}\sum_{m=0}^{n}\binom{n}{m}(2\lambda)^{2m}H_{2(n-m)+1}(x)\ .
\end{align}
Inserting this into~\eqref{prepL} we obtain the following polynomials for $\nu=1$,
\begin{align}
p_j^{(1)}(x) = 
\frac{(-1)^{j}j!(1+a^2)^{j}}{2^{2j-1}\pi\sqrt{1-a^2}} \int_{-\infty}^{\infty} dy\int_{-\infty}^{\infty}d\lambda\  L_j^{(1)}\left(\frac{4x^2}{1+a^2}+y^2\right) e^{-\left(y-\frac{2i\lambda}{\sqrt{1+a^2}}\right)^2-\frac{4}{1-a^2}\lambda^2}.
\end{align}
Hence we can write the polynomials in closed form for both values of $\nu$ and, after integrating over $\lambda$, it reads
\begin{align}
p_j^{(\nu)}(x)=\frac{j!(1+a^2)^{j+1/2}}{(-4)^j\sqrt{2\pi} a}\int_{-\infty}^{\infty} dy\ L_{j}^{(\nu)}\left(\frac{4x^2}{1+a^2}+y^2\right) 
e^{-\frac{1+a^2}{2a^2}y^2}\ .
\label{pjnu}
\end{align}
It can be easily checked via the leading order coefficient of the generalised Laguerre polynomial, given by $L^{(\alpha)}_j(x)=\frac{(-x)^j}{j!}+ O(x^{j-1})$ that the normalisation is indeed monic for both values of $\nu=0,1$.

For later convenience we also explicitly give the polynomials $q_j^{(\nu)}(x)$ expressed in terms of Laguerre polynomials. They are give by the relation~\eqref{eq:qjOp}, which leads to
\begin{align}
\begin{split}
q_j^{(\nu)}(x) =& \frac{j!(1+a^2)^{j+1/2}}{(-4)^j\sqrt{2\pi} a}\int_{-\infty}^{\infty} dy\ 
e^{-\frac{1+a^2}{2a^2}y^2}
\left\{\frac{-4x^2}{(1+a^2)^2} L_{j-2}^{(\nu+2)}\left(\frac{4x^2}{1+a^2} + y^2\right) \right.\\
&+\frac{1}{2(1+a^2)}\left(2\nu-1-4(1-a^2)x^2 \right)L_{j-1}^{(\nu+1)}\left(\frac{4x^2}{1+a^2} + y^2\right)\\
&+\left. \left[x^2-\frac{(1-a^2)}{4a^2}(2ja^2-1) -\frac{(1-a^4)}{4a^4} y^2+\tilde{c}_j^{(\nu)}(a)\right]L_{j}^{(\nu)}\left(\frac{4x^2}{1+a^2} + y^2\right)\right\}\ .
\end{split}
\label{qjexplicit}
\end{align}
For $j=1,2$, the Laguerre polynomials with negative subscript are absent, formally setting $L_{-k}^{(\nu)}(z)=0$ for $k>0$. We note that the term $(1-a^2)(2ja^2-1)/4a^2$ can alternatively be absorbed in the constant $\tilde{c}_j^{(\nu)}(a)$, slightly simplifying the expression.

\subsection{Determination of the Normalisation}\label{sec:norms}

After having determined the sOP \eqref{eq:sOPMain} and various equivalent representations thereof, we still need to compute their normalisation constants $h_j^{(\nu)}$ for $j$ even and odd in order to fully determine the three different kernels~\eqref{eq:3kernels} and~\eqref{eq:3kernelsOdd}.  
It turns out that a direct computation by inserting~\eqref{eq:sOPMain} into the respective skew-symmetric products~\eqref{eq:sOPprodEven} and~\eqref{eq:sOPprodOdd}, and then evaluating the various integrals, is very cumbersome. In fact we have not managed to simplify these integrals and thus have chosen a rather different route. We  exploit the fact that, on the one hand, the integrated unnormalised jpdf is proportional to the product of the normalisation constants and, on the other hand, we can compare with the known normalisation constant of the jpdf, see~\eqref{jpdf-const}.

We begin with the case of an even dimension $n=2m$. It is a well-known fact~\cite{Mehta} that the integral of the unnormalised jpdf yields the product of the normalisation constants of the monic sOP, i.e.
\begin{align}
C_{2m,\nu}^{-1} =\int_0^\infty d\lambda_1\ldots \int_0^\infty d\lambda_{2m} \Delta_{2m}\left(\{\lambda^2\}\right)
\text{Pf}\left[\ G_{\nu}(\lambda_j,\lambda_k) \ \right]_{j,k= 1,\ldots,2m} =(2m)! \prod_{k=0}^{m-1}h_{2k}^{(\nu)}\ .
\end{align}
This relation can be readily inverted and we obtain
\begin{align}
h_{2m}^{(\nu)}=&\frac{(2m)!}{(2m+2)!} \frac{C_{2m,\nu}}{C_{2m+2,\nu}}\nn\\
=& \frac{a^2(1-a^2)^{4m+2+\nu}}{(2m+2)(2m+1)2^{4m+5+\nu}}    \Gamma\left(\frac{2m+3}{2}\right)\Gamma\left(\frac{2m+\nu+1}{2}\right)
\Gamma\left(\frac{2m+4}{2}\right)\Gamma\left(\frac{2m+\nu+2}{2}\right)\nn\\
=& \frac{\pi a^2 (1-a^2)^{4m+2+\nu}}{2^{8m+2\nu+7}} (2m)! (2m+\nu)!\ .
\label{h2m}
\end{align}
In the first step we have inserted \eqref{jpdf-const} and in the second line the doubling formula for the Gamma-function has been exploited twice, leading to the final answer for the normalisation with $n=2m$.

Let us turn to $n=2m+1$ odd. In that case the integral over the unnormalised jpdf yields an additional contribution, the integral over the extra row and column denoted by $\bar{g}_\nu$ in~\eqref{bargdef},
\begin{align}
C_{2m+1,\nu}^{-1}&= 
 \int_0^\infty d\lambda_1\ldots\int_0^\infty  d\lambda_{2m+1}
\Delta_{2m+1}\left(\{\lambda^2\}\right)
\text{Pf}\left[
\begin{array}{c|c}
H_{\nu}(\lambda_{j},\lambda_{k}) & {g}_\nu(\lambda_{j}) \\ \hline
-{g}_\nu(\lambda_{k}) & 0 
\end{array}
\right]_{j,k= 1,\ldots,2m+1} \nn\\
&= (2m+1)!\ \bar{g}_\nu\prod_{k=1}^{m} h_{2k-1}^{(\nu)}
\end{align}
Because $\bar{g}_\nu$ is independent of $m$, it drops out when considering the ratio $C_{2m+1,\nu}/C_{2m+3,\nu}$, i.e.
\begin{align}
h_{2m+1}^{(\nu)}=\frac{(2m+1)!}{(2m+3)!} \frac{C_{2m+1,\nu}}{C_{2m+3,\nu}}=\frac{\pi a^2 (1-a^2)^{4m+4+\nu}}{2^{8m+2\nu+11}} (2m+1)! (2m+\nu+1)!\ ,
\end{align}
following the same calculation as for $n=2m$. As a result the normalisation constants can be written in the following unified closed form, valid for even and odd index $j$,
\begin{align}
h_{j}^{(\nu)}=
\frac{\pi a^2 (1-a^2)^{2j+2+\nu}}{2^{4j+2\nu+7}} j! (j+\nu)!\ .
\label{finalhj}
\end{align}

\sect{Spectral density and distribution of the smallest eigenvalue}
\label{rhop1}

\begin{figure}
		\centerline{\includegraphics[width=0.49\linewidth,angle=0]{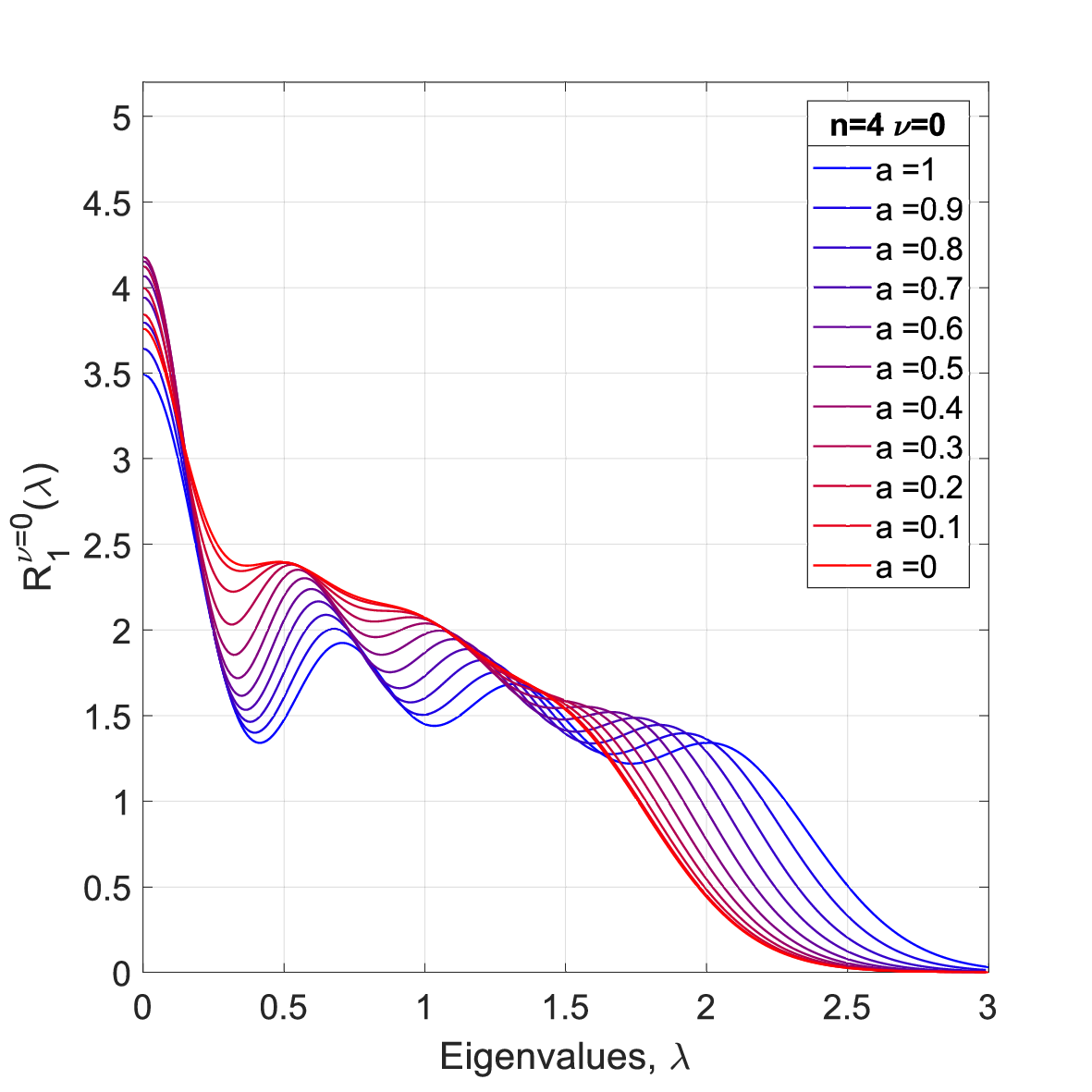}
		\includegraphics[width=0.49\linewidth,angle=0]{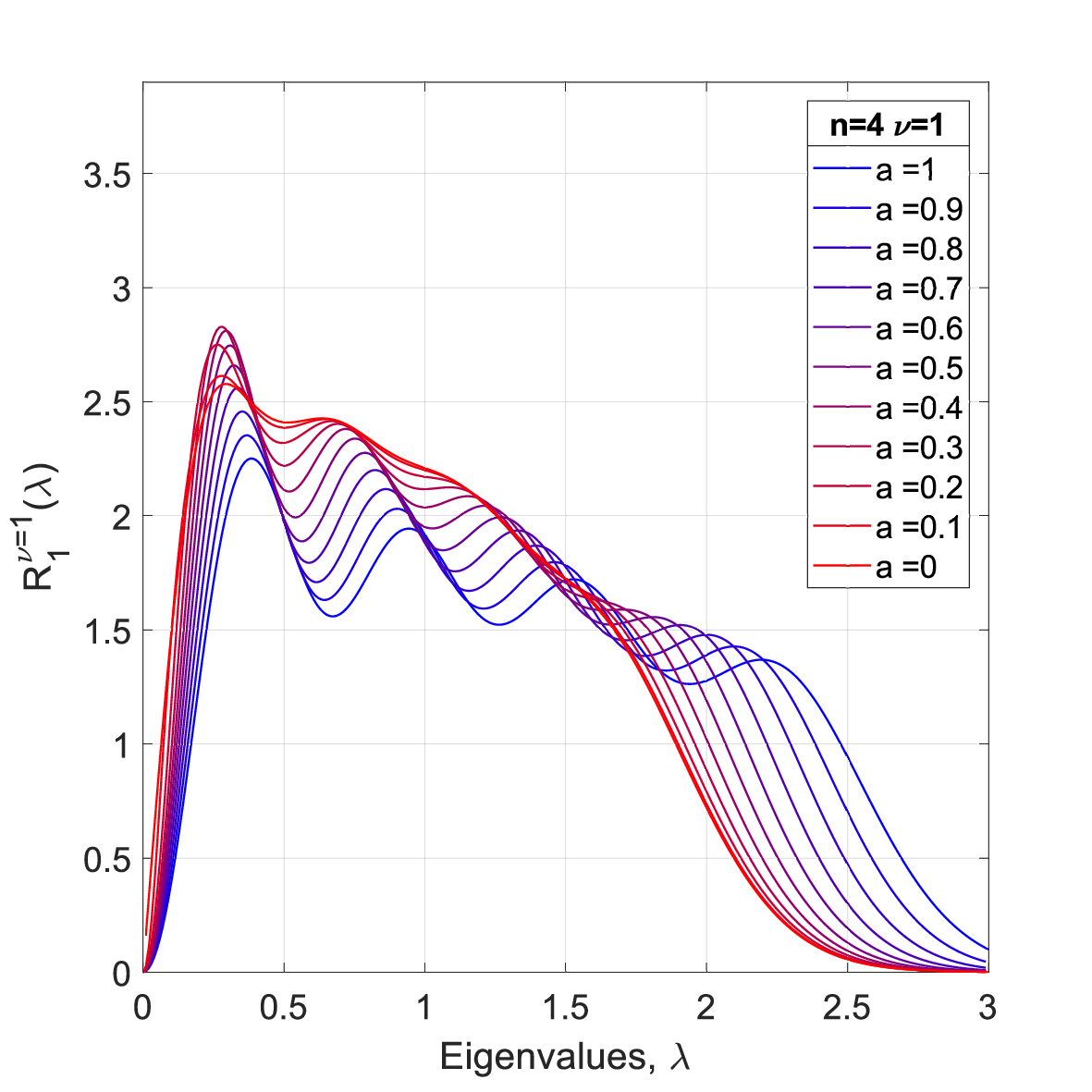}}
		\caption{
		The spectral density~\eqref{eq:DensEven1} is shown for $n=4$ with $\nu=0$ (left) and $\nu=1$ (right). 
		The parameter $a$ increases from the most narrow distribution at $a=0$ (red), corresponding to the density of the chGOE in~\eqref{R1Jac}, to the broadest distribution at $a=1$ (blue), corresponding to the density of the  GAOE in~\eqref{R1Mehta}.
		}
		\label{fig:cohort-even}
	\end{figure}

In this section we discuss the dependence of the spectral density and the distribution of the smallest eigenvalue on the symmetry transition parameter $a\in[0,1]$, including the limiting cases $a=0$ and $a=1$.  
Because the spectral density has been tested against Monte Carlo simulations of our random two-matrix model in Figures~\ref{fig:even} and~\ref{fig:odd} we shall not repeat that here. Instead, in Figure~\ref{fig:cohort-even} we show the parameter dependence of the spectral density~\eqref{eq:DensEven} for even $n=2m$ at $m=2$ and $\nu=0,1$ given by 
\begin{align}
R_1^\nu(\lambda)=S_{2m}(\lambda,\lambda)\ ,
\label{eq:DensEven1}
\end{align}
with the polynomials~\eqref{eq:sOPMain} and their integral transforms~\eqref{bpq-def}. The local maxima of the density correspond to the individual eigenvalues and are close to their average positions, e.g., in Figure~\ref{fig:cohort-even} there are $n=4$ eigenvalues. The density at $a=1$, corresponding to the Dyson index $\beta=2$, has the strongest level repulsion amongst the eigenvalues, as its local maxima and minima are most pronounced. Decreasing $a$ they flatten out till the density at $a=0$ is approached, corresponding to the Dyson index $\beta=1$ that exhibits the weakest level repulsion. 
Additionally, the extra zero eigenvalue for $\nu=1$ pushes the non-zero eigenvalues away from the origin.

For completeness we also give the
spectral density of the chGOE that we obtain in the limit $a\to0$,
\begin{align}
\begin{split}
\left.R_1^\nu(\lambda)\right|_{a=0}=&
\sum_{j=0}^{m-1} \frac{2^{2\nu+2}(2j)!}{(2j+\nu)!} \int_{0}^{\infty}du(\lambda u)^\nu e^{-2u^2-2\lambda^2}\sign(\lambda-u)
\\
&\hspace*{-1.5cm}\times\left[L^{(\nu)}_{2j}\left(4\lambda^2\right)\left((2j+1)L^{(\nu)}_{2j+1}\left(4u^2\right)-(2j+\nu)\left(L^{(\nu)}_{2j}\left(4u^2\right)+L^{(\nu)}_{2j-1}\left(4u^2\right)\right)\right) - (\lambda\leftrightarrow u)\right].
\label{R1Jac}
\end{split}
\end{align}
Apparently, this formula represents only the case with even $n=2m$, following~\cite{VerbNc2} (rescaling $\lambda\to2\lambda^2$ therein), cf. our limiting sOP in~\eqref{pja0},~\eqref{qja0}, and~\eqref{Lid}.
The spectral density of the GAOE corresponding to the limit $a=1$ is equal to
\begin{align}
\left.R_1^\nu(\lambda)\right|_{a=1}=\sum_{j=0}^{n-1}\frac{1}{\sqrt{\pi}2^{2j+\nu-3/2}(2j+\nu)!}e^{-2\lambda^2}H_{2j+\nu}\left(\sqrt{2}\ \lambda\right)^2\ ,
\label{R1Mehta}
\end{align}
where we follow~\cite{Mehta} (rescaling $\lambda\to\sqrt{2}\, \lambda$ therein), see also our limiting polynomials in \eqref{pja1}.

\begin{figure}
		\centerline{\includegraphics[width=0.49\linewidth,angle=0]{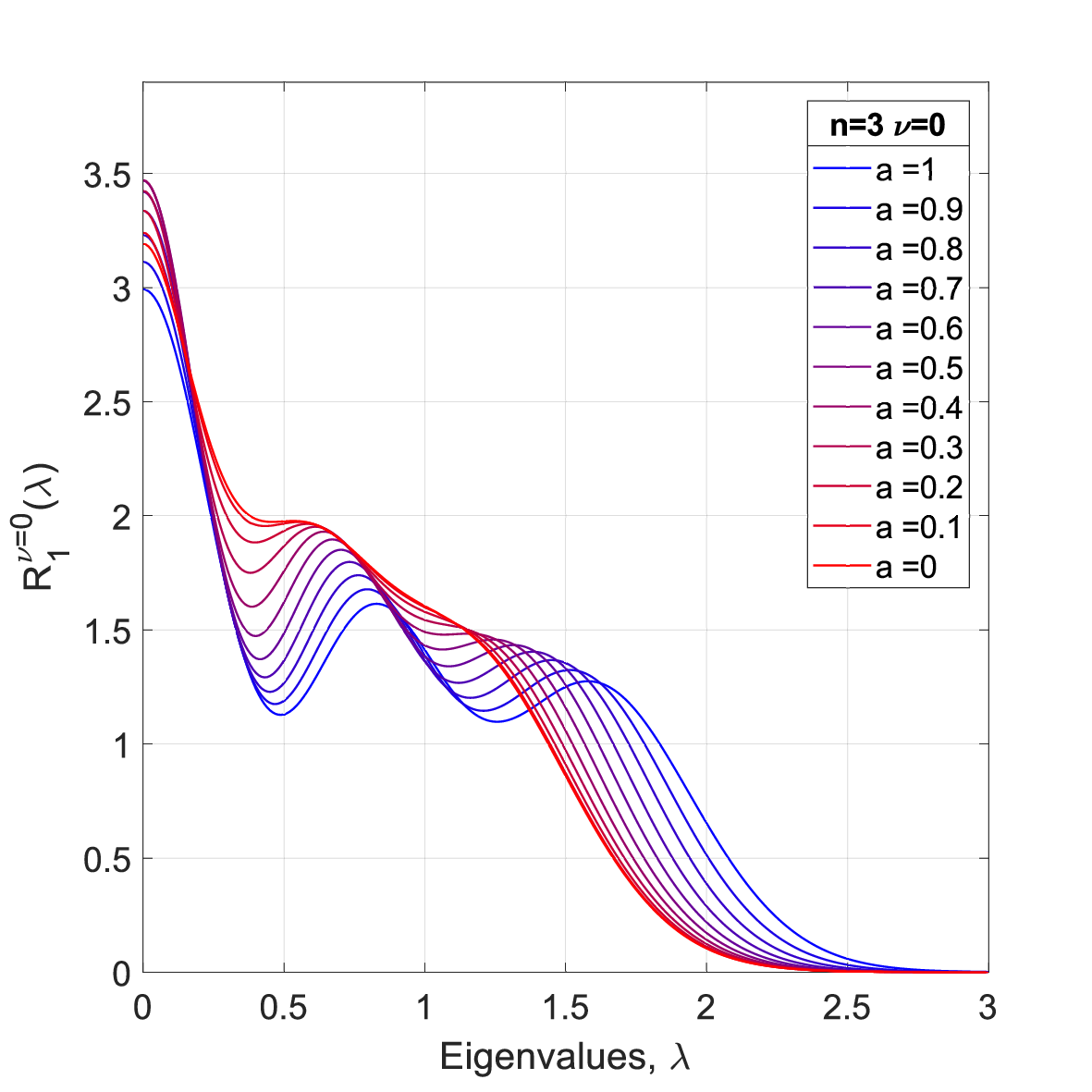}
		\includegraphics[width=0.49\linewidth,angle=0]{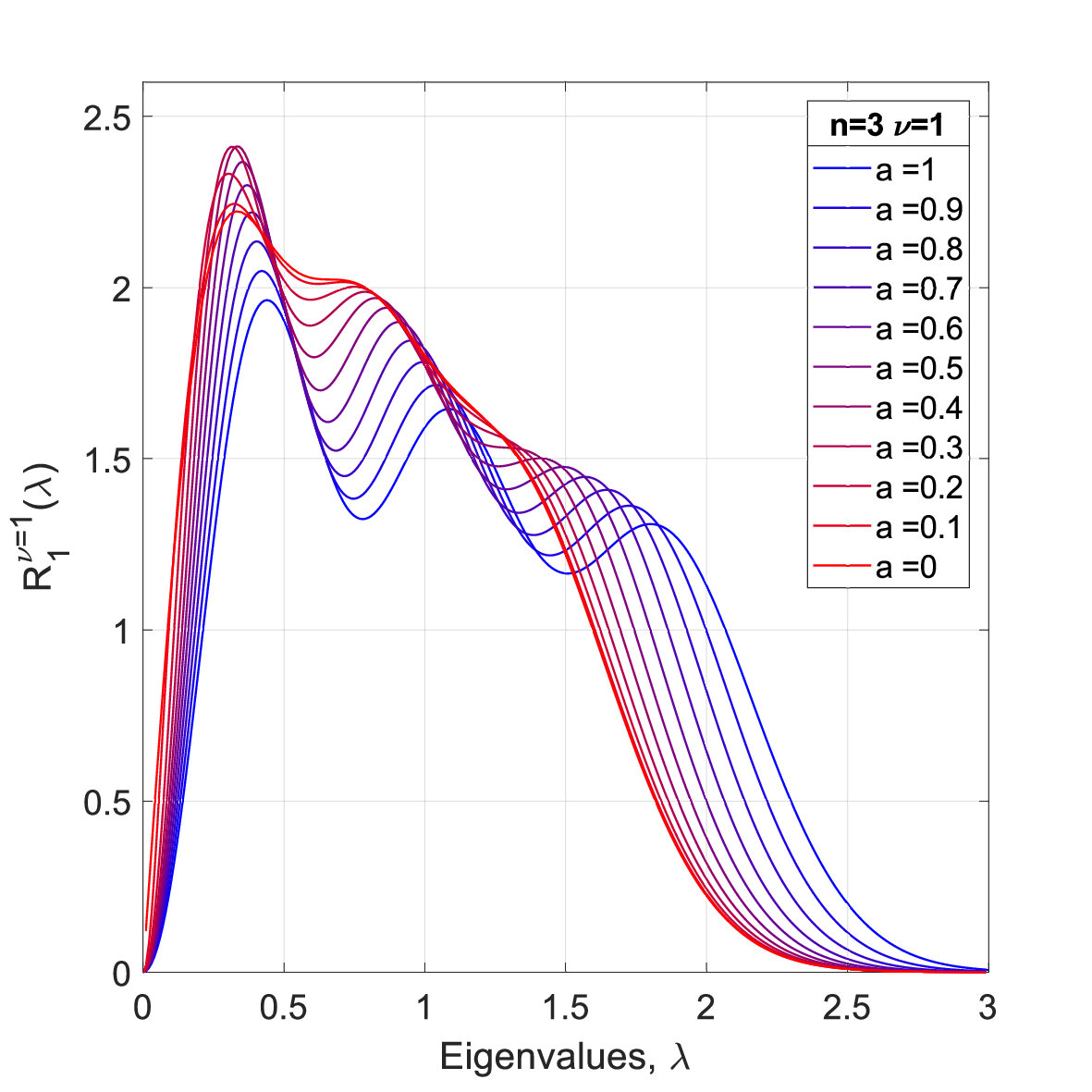}}
		\caption{
		The spectral density for odd $n=3$  from~\eqref{eq:DensOdd1} is plotted for $\nu=0$ (left) and $\nu=1$ (right). 
		As in Figure~\ref{fig:cohort-even}, the parameter $a$ varies from a narrow distribution at $a=0$ (red, chGOE) to the broad distribution which corresponds to $a=1$ (blue, GAOE, see~\eqref{R1Mehta}).
		}
		\label{fig:cohort-odd}
	\end{figure}

A similar parameter dependent plot for the spectral density \eqref{eq:DensOdd}, with odd $n=2m-1 $ at $m=2$ and $\nu=0,1$, given by
\begin{align}
R_1^\nu(\lambda)=S_{2m-1}(\lambda,\lambda)\ ,
\label{eq:DensOdd1}
\end{align}
is shown in Figure~\ref{fig:cohort-odd}.

Both Figures \ref{fig:cohort-even} and \ref{fig:cohort-odd} show a peculiar behaviour for the smallest eigenvalue, given by the leftmost peak: It first increases from $a=0$ to reach its maximal height as a function of $a$, to decrease again to its lowest value for $a=1$.
For that reason we have investigated the smallest eigenvalue distribution separately below, also because of the apparently strong overlap with the second largest eigenvalues close to $a=0$. It has to be said that the densities we plot in Figures~\ref{fig:cohort-even} and~\ref{fig:cohort-odd} are at finite $n$ and thus not universal. 
For example, there is a competition between the level repulsion varying from $\beta=2$ for $a=1$ to $\beta=1$ for $a=0$, and the fact that the overall support of the density narrows with decreasing $a$, thus pushing the eigenvalues closer together. Therefore it is difficult to decide what the ``true" impact of $a$ is when sorting out the scaling effects.
It remains to be seen how these features carry over to  the microscopic large-$n$ limit at the origin, when e.g. effects of the $a$-dependent edge of the finite-$n$ densities in Figures~\ref{fig:cohort-even} and \ref{fig:cohort-odd} are no longer seen. This investigation is left for future work.

There is a second motivation to study the distribution of smallest eigenvalues, apart from isolating its behaviour as a function of $a$. 
As it is true for any determinantal or Pfaffian point process, the $k$-th gap probability at the origin and the resulting distribution of the $k$-th smallest eigenvalue can be expanded in terms of the $l$-point density correlation functions. Referring e.g.~to \cite{AD03} for a derivation we only display it for the smallest eigenvalue with $k=1$,
\begin{align}
p_1^\nu(s)= \sum_{l=1}^n\frac{(-1)^{l-1}}{(l-1)!}\int_0^s dx_1\ldots dx_{l-1} R_l^\nu(s,x_1,\ldots,x_{l-1})= R_1^\nu(s)-\int_0^s dx_1 R_2^\nu(s,x_1)+\ldots\ ,
\label{p1exp}
\end{align}
where for $l=1$ we have no integral in the sum. 
It was found in \cite{AD03} for a different symmetry class that this expansion may converge remarkably well. 
For that reason we compare the truncated expansion to Monte-Carlo simulations, keeping only the first two terms in~\eqref{p1exp}.

	\begin{figure}
		\centerline{\includegraphics[width=0.49\linewidth,angle=0]{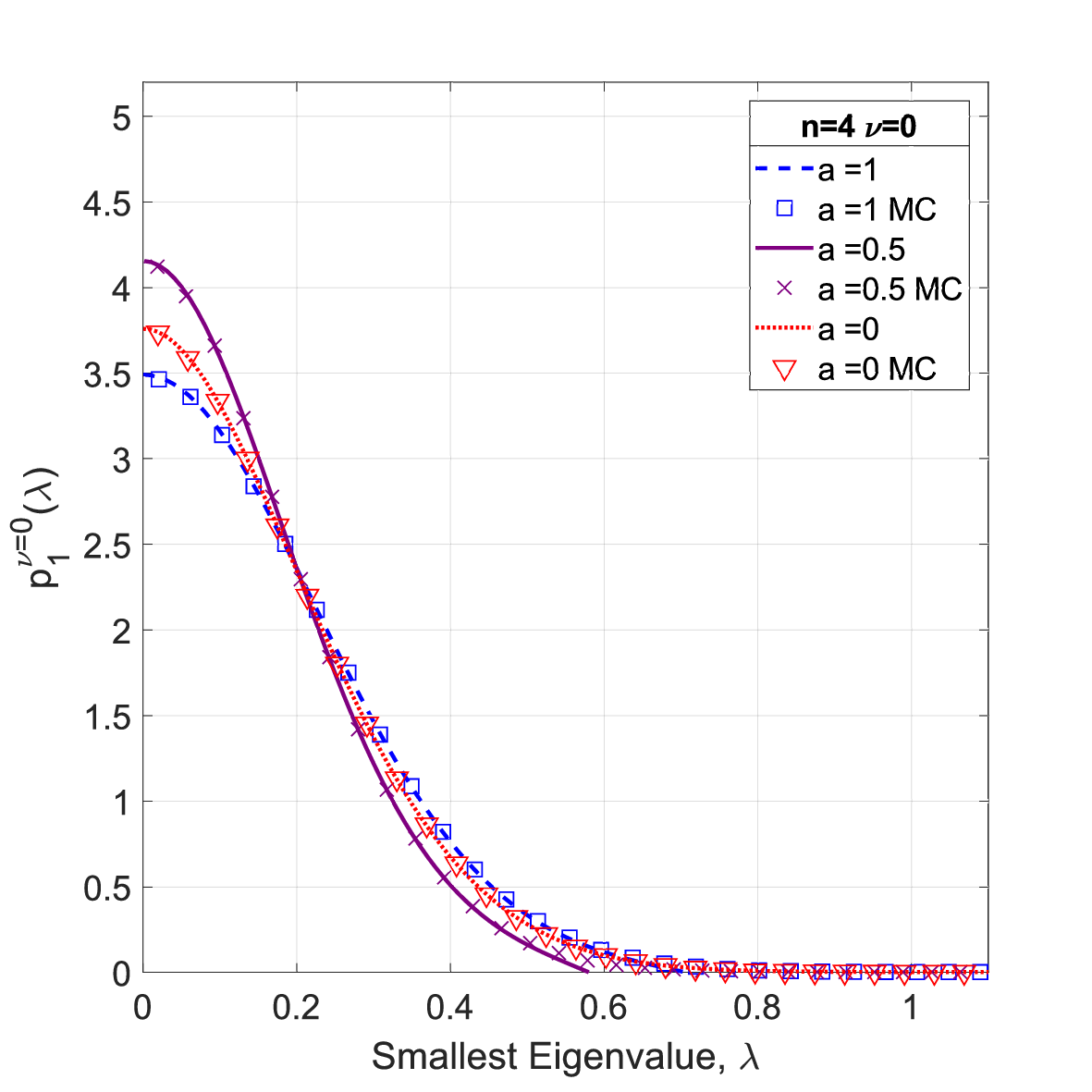}
		\includegraphics[width=0.49\linewidth,angle=0]{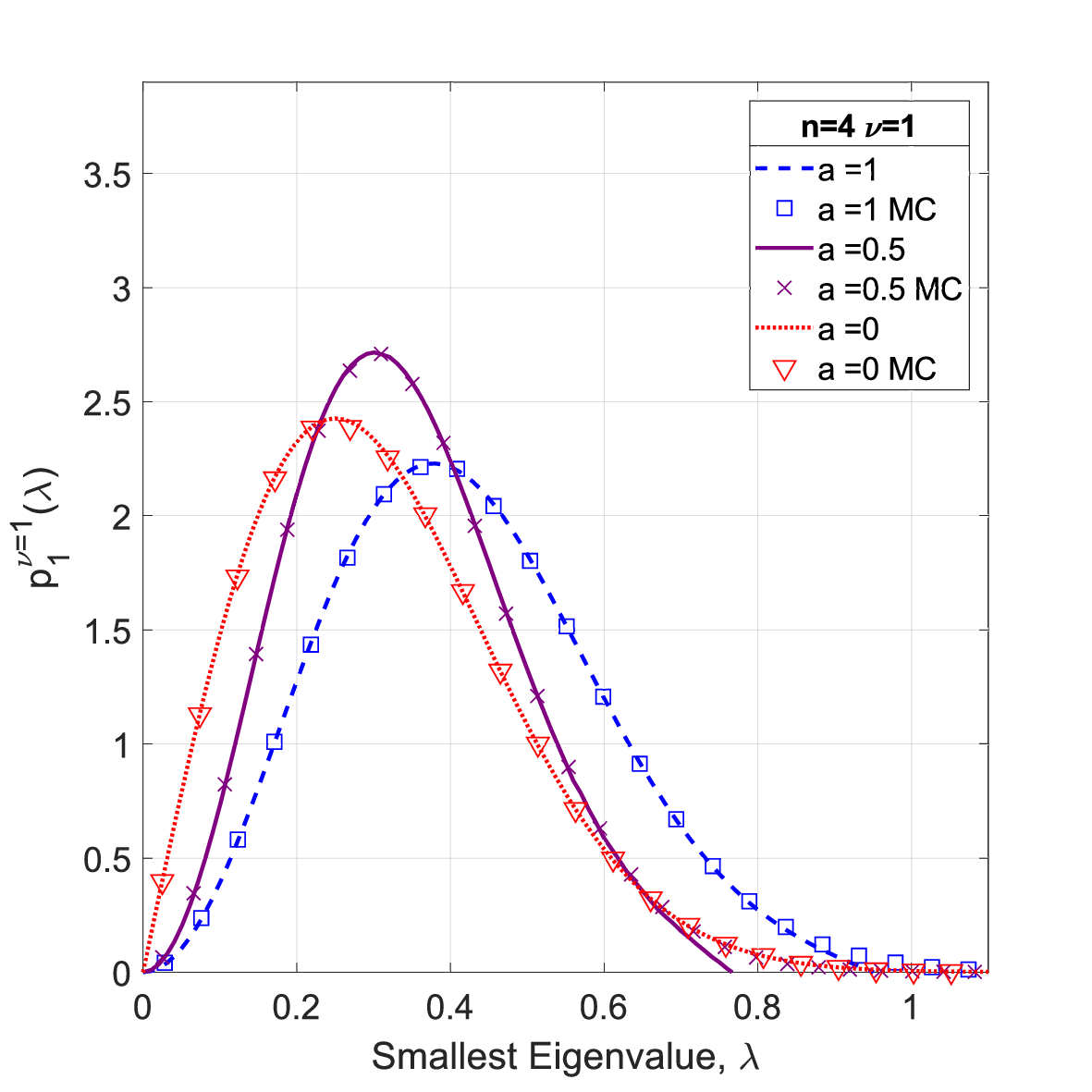}}
		\caption{
		The approximate distribution of the smallest eigenvalue $p_1^\nu(s)$, see~\eqref{p1exp}, (curves) is compared to Monte-Carlo (MC) simulations (symbols). Here, we have plotted the results for $n=4$ with $\nu=0$ (left) and $\nu=1$ (right). The ensemble comprises the generation of $10^{6}$ matrices and the events have been collected in bins of an approximate size of $0.05$. 
		For $a=0$ the exact curves for $p_1^\nu(s)$ are drawn, see~\eqref{p1a0nu1} and~\eqref{p1a0nu0}.
		}
		\label{fig:p1-even}
	\end{figure}

As an extra benefit, the expression~\eqref{p1exp} is sensitive not only to the density, but also to the $2$-point and in principle also to all higher $k$-point functions, although they have a weaker impact. The higher order correlation functions also depend on the off-diagonal elements of the matrix valued kernel in~\eqref{eq:RkPf}. When comparing the analytical expansion to Monte Carlo simulations that a priori yield the full distribution of the smallest eigenvalue, we can test the convergence of our expansion and at the same time cross-check the validity of the off-diagonal kernel elements $I_n^\nu(x,y)$ and $D_n^\nu(x,y)$, as well as $S_n^\nu(x,y)$ at unequal arguments, i.e.
\begin{align}
R_2^\nu(x,y)=S_n^\nu(x,x)S_n^\nu(y,y)-I_n^\nu(x,y)D_n^\nu(x,y)-S_n^\nu(x,y)^2\ .
\label{R2}
\end{align}
Certainly, other, more sophisticated methods 
exist for a controlled approximation of the Fredholm expansion, see e.g.~\cite{Nishigaki} for the distribution of the smallest eigenvalues of a random two-matrix model that describes the chGUE-chGSE transition. Because we deal with quantities at finite (and small) $n$ we have not aimed at a better precision. 

	\begin{figure}
		\centerline{\includegraphics[width=0.49\linewidth,angle=0]{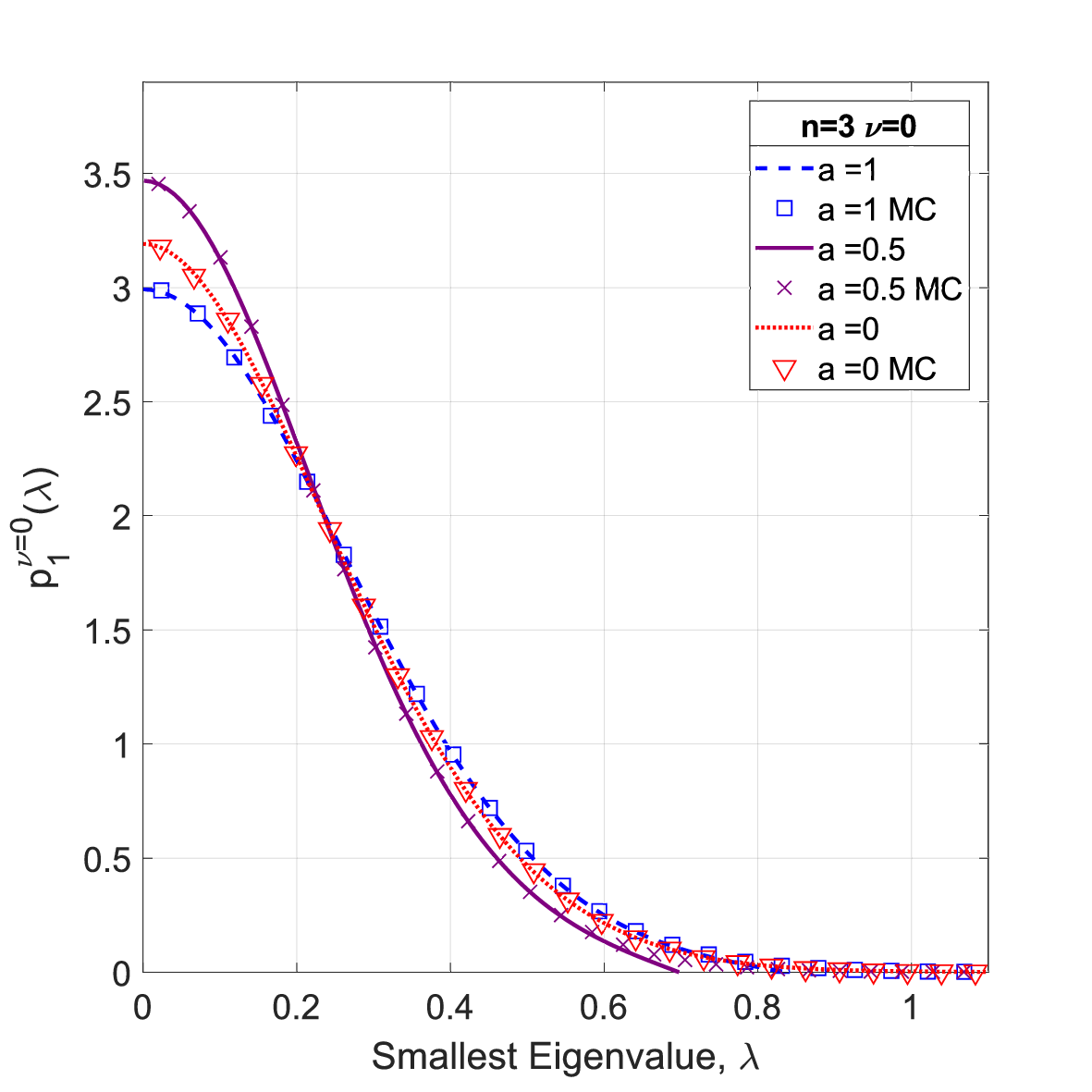}
		\includegraphics[width=0.49\linewidth,angle=0]{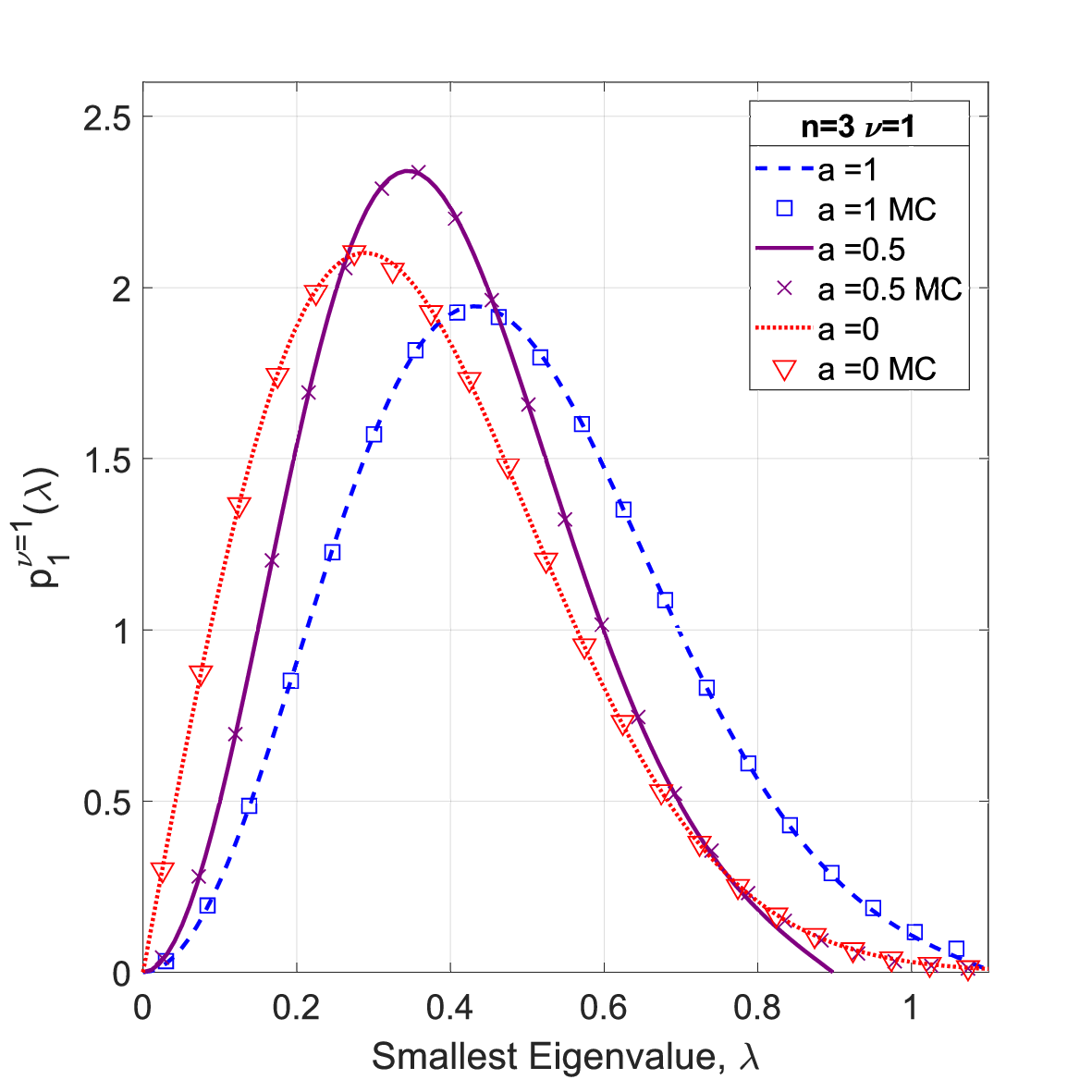}}
		\caption{
	The analytical approximation~\eqref{p1exp} ($a>0$)  and the exact 
distributions~\eqref{p1a0nu1} and~\eqref{p1a0nu0} ($a=0$) (curves) as well as Monte-Carlo simulations (symbols)
		of the smallest eigenvalue 
	$p_1^\nu(s)$ 
		are shown  for $n=3$ with $\nu=0$ (left) and $\nu=1$ (right). As before we have generated $10^{6}$ matrices to keep the statistical error very low and the bin size has been chosen to be $0.05$. 
		}
		\label{fig:p1-odd}
	\end{figure}

In Figures~\ref{fig:p1-even} and \ref{fig:p1-odd} the expansion~\eqref{p1exp} given by solid curves  is compared to Monte Carlo simulations which are represented by symbols. 
The non-monotonous behaviour of the maximum becomes particularly transparent, but it remains to be seen whether this behaviour will 
carry over to the large-$n$ limit.
Clearly,  when the truncated expansion~\eqref{p1exp} of the smallest eigenvalue density becomes negative, the analytical approximation keeping only the first two terms  breaks down at latest. Nevertheless, this  approximation works remarkably well, almost all the way down to vanishing density, as this truncated sum is smoothly approached by the symbols. 

For the chGOE with $a=0$ the distribution of the smallest eigenvalue is known exactly for finite $n$, where we use the expressions from~\cite{Edelman,DN} for $\nu=0,1$. We start with $\nu=1$ which has a particularly simple form,
\begin{align}
\left.p_1^{\nu=1}(s)\right|_{a=0}=4ns\,e^{-2ns^2}\ .
\label{p1a0nu1}
\end{align}
For $\nu=0$ it contains Tricomi's confluent hypergeometric function $U$,
\begin{align}
\left.p_1^{\nu=0}(s)\right|_{a=0}=n \sqrt{\frac{8}{\pi}}\ \Gamma\left(\frac{n+1}{2}\right) e^{-2n s^2} U\left(\frac{n-1}{2},-\frac{1}{2},2s^2\right).
\label{p1a0nu0}
\end{align}
Both expressions are valid for even and odd $n$ alike.
We do not expect that such simple expressions exist for our one-parameter family of real random two-matrix models. At $a=1$ it is known for the GAOE that the distribution of the smallest eigenvalue  is proportional to the expectation value of a characteristic polynomial to half-integer power, see e.g. \cite{Timetal} for the corresponding expression in the chGOE. This is an open problem in itself, which is why we also included the expansion from~\eqref{p1exp} in our plot for $a=1$.

\sect{Conclusion}\label{sec:Conc}

In the present work we have introduced and solved a parameter-dependent family of random two-matrix models with real matrix elements drawn from Gaussian distributions. They describe the symmetry transition between the following two ensembles:
the chiral Gaussian orthogonal ensemble (chGOE), which is also called real Wishart-Laguerre ensemble belonging to the Cartan class B$|$DI,  and the ensemble of Gaussian antisymmetric Hermitian random matrices (GAOE), which are denoted by B$|$D in the Cartan classification scheme.
Both ensembles are invariant under the action of subgroups of orthogonal groups, although those groups are not the same. Since the resulting group integral is of the real type one would presume that it represents a particular challenge. Fortunately, in our case the corresponding group integral has been computed by Harish-Chandra. 

On the physics side, our main motivation has originated from topological insulators, in particular from the disordered system of a quantum wire with two Majorana modes, one at each end. This system satisfies several symmetry constraints. One of them is that the Hamiltonian is antisymmetric and real. Another condition is that, in the ideal, unperturbed system, the Hamiltonian splits into a direct sum of two identical Hamiltonians that have one generic zero eigenvalue each. Including the perturbations in the system, the two Hamiltonians couple and the pair of former zero modes broadens. We suppose that the main features of this situation can be realized by our model, choosing the matrix dimension to be $n=2m-1$ odd and $\nu=0$ in the regime $1/a^2\propto1/n$.

It is very likely that our model enjoys other applications as well due to the ubiquity of random matrices in general. Especially the realisation that our model protects the topology $\nu=0,1$ while global symmetries change could be of physical interest. For example, the transition of lattice QCD to continuum QCD may exhibit such a transition for a particular dimension, gauge group, and discretisation. 
There are also other field theories involving unpaired Majorana modes which could be topologically protected as in our random matrix model for $0\leq a\leq 1$.

From a mathematical perspective, we have shown that our ensemble belongs to the class of Pfaffian point processes which can be solved using skew-orthogonal polynomials. We have explicitly constructed these polynomials for arbitrary matrix dimension and topological index $\nu=0,1$. They interpolate between the skew-orthogonal Laguerre polynomials of the chGOE and the orthogonal Hermite polynomials with parity $\nu$ of the GAOE.
The resulting matrix kernels that determine all $k$-point density correlation functions depend on the matrix dimension $N=2n+\nu$, and take different forms when $n$ is even or odd. 
Our analytical results have been confirmed by performing the limits $a\to0,1$, and $\infty$. This yields the known results for the chGOE and the GAOE, and for $a\to\infty$ for the direct sum of two GAOEs. Furthermore, we have successfully compared our results to Monte-Carlo simulations for the spectral density and the distribution of the smallest eigenvalue, for all parities of $n$ and $\nu$. For the latter we found an interesting non-monotonous behaviour of its height throughout the transition. It has to be seen whether this behaviour carries over to the large-$n$ limit when the spectrum is properly scaled. While for the chGOE the distribution of the smallest eigenvalue is known analytically for finite $n$, we used a truncated Fredholm expansion in terms of the density and two-point density correlation function for values $a>0$ all the way up to $a=1$ (GAOE), where no closed form expressions exist either.

The microscopic large-$n$ limit that we expect to be universal is left for future work. A new family of transition kernels is most likely to be found in the vicinity of the chGOE ($a\approx0$) as well as in the vicinity where the random matrix splits into a direct sum ($1/a\approx0$), whereas we do not expect any
deviations from the GAOE in between. Our prediction is based on the experience that the less symmetric ensemble rapidly dominates on the local scale of the spectrum.
A deformation of the Hamiltonian with more symmetries can only be observed in a small vicinity, with a very restricted parameter range. To make these regimes more precise, we conjecture that the deformed version of the chGOE kernels can be found in the scaling regime, where $na^2$ is fixed in the limit $n\to\infty$ and $a\to0$. Similarly, we would expect deformed kernels of the direct sum of two GAOEs for $n/a^2$ fixed, when $n\to\infty$ and $a\to\infty$. The latter limit may be of interest in the study of quantum wires with Majorana modes in the gapless phase, as already mentioned above.

\ \\

{\it Acknowledgements:} 

We would like to thank Lara Benfatto for an early discussion initiating this study, Alexander Altland for suggesting this symmetry transition, and Taro Nagao for a discussion about skew-orthogonal polynomials.
Support by the German research council
DFG through grant AK35/2-1 "Products of Random Matrices" of  (G.A. and M.K.), 
International Research Training Group 2235 Bielefeld-Seoul "Searching for the regular in the irregular: Analysis of singular and random systems" (A.M.), and
IGK 1132 ”Stochastics and Real World Models” Beijing-Bielefeld (P.V.) is kindly acknowledged.

\begin{appendix}

\sect{Simplification of the weight functions $G_\nu(x,y)$ and $g_\nu(x)$
}\label{sec:AppendixA}

We begin with the simplification of the one-point weight function $g_\nu(x)$ starting from the definition \eqref{eq:gnudef}, given by a single integral. Together with \eqref{eq:fnudef} it reads for $\nu=0\ (1)$ with upper (lower) signs
\begin{align}
\begin{split}
g_{\nu=0,1}(x)=& \frac12 x^\nu e^{-\frac{2}{a^2}x^2}\int_0^\infty dt \ e^{-\frac{2}{a^2(1-a^2)}t^2}\left( e^{\frac{4}{a^2}xt}\pm e^{-\frac{4}{a^2}xt}\right)\\
=&\frac12 x^\nu e^{-\frac{2}{a^2}x^2+\frac{2(1-a^2)}{a^2}x^2}\left(\int_0^\infty dt \ e^{-\frac{2(t-x(1-a^2))^2}{a^2(1-a^2)}}\pm \int_0^\infty dt \  e^{-\frac{2(t+x(1-a^2))^2}{a^2(1-a^2)}}\right)\\
=& \frac{\sqrt{\pi}}{4}\sqrt{\frac{a^2(1-a^2)}{2}}\,x^\nu e^{-2x^2}\left( 1+\erf\left[x\sqrt{\frac{2(1-a^2)}{a^2}} \right]\pm 1\mp\erf\left[x\sqrt{\frac{2(1-a^2)}{a^2}} \right]\right),
\end{split}
\label{eq:gnuerf}
\end{align}
where we have completed the squares in the second line. In the last step we have used the relation for the  complementary error function 
\begin{align}
\erfc(x)=\frac{2}{\sqrt{\pi}}\int_x^\infty dt\,e^{-t^2}=1-\erf(x) \ ,
\end{align}
 and the fact that the error function is odd, $\erf(-x)=-\erf(x)$. The last line of \eqref{eq:gnuerf} is equivalent to \eqref{eq:g01}, and it is also manifest in the simplified form \eqref{eq:g01} that $g_\nu(x)$ is an even function in $x$ for both values of $\nu=0,1$.

We turn to the simplification of the two-point weight function $G_\nu(z,u)$, given as a double integral in its definition \eqref{eq:Gnudef}. 
Let us start with $\nu=0$. Following from $\sign(y-x)=\sign(y^2-x^2)$ valid for $x,y>0$, we have an integrand that is even in both arguments, $x$ and $y$ separately, so we can extend both integrations to the entire real line,
\begin{align}
\begin{split}
G_0(s,t)=& \frac14 e^{-\frac{2}{a^2}(s^2+t^2)}\int_{-\infty}^\infty dx\int_{-\infty}^\infty dy\ \sign(y^2-x^2)e^{-\frac{2}{a^2(1-a^2)}(x^2+y^2)}\cosh\left(\frac{4xs}{a^2}\right)\cosh\left(\frac{4yt}{a^2}\right)\\
=&\frac12 e^{-\frac{2}{a^2}(s^2+t^2)}\int_{-\infty}^\infty du\int_{-\infty}^\infty dv\ \sign(u)\sign(v)e^{-\frac{4}{a^2(1-a^2)}(u^2+v^2)}\\
&\times\frac14\left( e^{\frac{4}{a^2}(v-u)s}+ e^{-\frac{4}{a^2}(v-u)s}\right)\left( e^{\frac{4}{a^2}(v+u)t}+ e^{-\frac{4}{a^2}(v+u)t}\right)\\
=&\frac{\pi a^2(1-a^2)}{8}\  e^{-2(s^2+t^2)}\erf\left[(t-s)\sqrt{\frac{(1-a^2)}{a^2}} \right]\erf\left[(t+s)\sqrt{\frac{(1-a^2)}{a^2}} \right]\ .
\label{eq:G0erf}
\end{split}
\end{align}
A change of variables $u=(y-x)/2$ and $v=(y+x)/2$ in the second line decouples the integrals, and the sign-function can be evaluated. Multiplying out and completing the squares as in \eqref{eq:gnuerf} leads to the last line, which is what was claimed in \eqref{eq:G01} for $\nu=0$. Also it is manifest in \eqref{eq:G0erf} that the function is an even function separately in both arguments $s$ and $t$, $G_0(-s,t)=G_0(s,t)=G_0(s,-t)$.

Turning to $\nu=1$, the definition \eqref{eq:gnudef} contains the function $\sinh$ instead of $\cosh$. The integrand thus has less symmetry and we use the addition theorem $\sinh(x)\sinh(y)=\cosh(x)\cosh(y)-\cosh(x-y)$ first. This leads to 
\begin{align}
G_1(s,t)=st G_0(s,t) - \widetilde{G}_1(s,t),\qquad {\rm for}\ a<1, \label{eq:GTdef}
\end{align}
where the latter function still has to be calculated,
\begin{align}
\begin{split}
\widetilde{G}_1(s,t)=& st e^{-\frac{2}{a^2}(s^2+t^2)} \int_0^\infty dx\int_0^\infty dy\ \mbox{sign}(y-x)e^{-\frac{2}{a^2(1-a^2)}(x^2+y^2)}
\cosh\left(\frac{4(xs-yt)}{a^2}\right)\\
=& st e^{-\frac{2}{a^2}(s^2+t^2)} \int_0^\infty dx\int_0^x dy \ e^{-\frac{2}{a^2(1-a^2)}(x^2+y^2)}\frac12 
\left\{-e^{\frac{4}{a^2}xs}e^{-\frac{4}{a^2}yt} - e^{-\frac{4}{a^2}xs}e^{\frac{4}{a^2}yt} 
\right.\\
& \quad\quad\quad\quad \quad\quad \quad\quad \quad\quad \quad\quad \quad\quad \quad\quad \quad\quad \quad\quad  \left. \ +\ e^{\frac{4}{a^2}ys}e^{-\frac{4}{a^2}xt}  + e^{-\frac{4}{a^2}ys}e^{\frac{4}{a^2}xt} \right\}.
\label{eq:GT2}
\end{split}
\end{align}
Here, we have explicitly evaluated the sign function, yielding two terms, and decomposed the $\cosh$ in its two exponential terms. In particular, the term for the region $y>x$ was rewritten with the aid of $\int_0^\infty dx\int_x^\infty dy=\int_0^\infty dy\int_0^y dx$ and, then, we exchanged the labelling of the variables $x\leftrightarrow y$ therein. The four terms in \eqref{eq:GT2} are integrated separately, where we first complete the squares and then shift the integration domains. Hence, we obtain
\begin{align}
\begin{split}
\widetilde{G}_1(s,t)=& -\frac12 st e^{-2(s^2+t^2)} \left\{
\int_{-s(1-a^2)}^\infty dx \int_{t(1-a^2)}^{x+(s+t)(1-a^2)}dy 
+\int_{s(1-a^2)}^\infty dx \int_{-t(1-a^2)}^{x-(s+t)(1-a^2)}dy 
\right.\\
&\left.
-\int_{t(1-a^2)}^\infty dx \int_{-s(1-a^2)}^{x-(s+t)(1-a^2)}dy 
-\int_{-t(1-a^2)}^\infty dx \int_{s(1-a^2)}^{x+(s+t)(1-a^2)}dy 
\right\}e^{-\frac{2}{a^2(1-a^2)}(x^2+y^2)}.
\end{split}
\label{eq:GT1}
\end{align}
Using 
\begin{align}
\frac{2}{\sqrt{\pi}}\int_b^a dy e^{-y^2}=\erf(a)-\erf(b),
\end{align}
all compact inner integrals over $y$ can be computed in \eqref{eq:GT1}. Notably, only half of the error functions obtained depend on $x$ and become significant. In contrast, the other half of the integrals factorise and all those contributions eliminate each other.
Defining
\begin{align}
\label{eq:ABdef}
A=s\sqrt{\frac{2(1-a^2)}{a^2}}\ \ ,\ B=t \sqrt{\frac{2(1-a^2)}{a^2}}\ ,
\end{align}
and rescaling $u=x\sqrt{2/(a^2(1-a^2))}$ we have for the remaining terms in \eqref{eq:GT1}
\begin{align}
\begin{split}
\widetilde{G}_1(s,t)=&-\frac{\sqrt{\pi}}{8}a^2(1-a^2) st e^{-2(s^2+t^2)}\left\{
\int_{-A}^\infty du \erf[u+A+B]+ \int_{A}^\infty du \erf[u-A-B]
\right.\\
&\left.-\int_{B}^\infty du \erf[u-A-B]- \int_{-B}^\infty du \erf[u+A+B]
\right\}e^{-u^2}\\
=& -\frac{\sqrt{\pi}}{8}a^2(1-a^2) st e^{-2(s^2+t^2)}\left\{
\int_A^{B} du \erf[u-A-B]
+\int_{-A}^{-B} du \erf[u+A+B]
\right\}e^{-u^2}\\
=& -\frac{\sqrt{\pi}}{4}a^2(1-a^2) st e^{-2(s^2+t^2)}
\int_{s\sqrt{\frac{2(1-a^2)}{a^2}}}^{t\sqrt{\frac{2(1-a^2)}{a^2}}} du \erf\left[u-(s+t)\sqrt{\frac{2(1-a^2)}{a^2}}\right]e^{-u^2}.
\end{split}
\label{eq:GTfinal}
\end{align}
In the first step, all parts of the integrals over $[0,\infty)$ cancel and, in the second step, the parity of the error function was exploited. The antisymmetry $\widetilde{G}_1(s,t)=-\widetilde{G}_1(t,s)$ is again manifest, due to the integration boundaries. Inserting \eqref{eq:GTfinal} and the result \eqref{eq:G0erf} for $\nu=0$ into  \eqref{eq:GTdef} we arrive at \eqref{eq:G01} as has been claimed for $\nu=1$. 

We now compute the integrals over the one- and two-point weight functions as they are needed in the modification of the scalar product for odd $n=2m+1$, see \eqref{Hdef}. We take up the calculation with the integral
\begin{align}
\begin{split}
\bar{G}_\nu(t)=&\int_0^\infty ds G_\nu(s,t).
\end{split}
\end{align}
Starting with the definition~\eqref{eq:Gnudef} (rather than with the results which we derived above), we can first perform the integral over $s$ by completing the square,
\begin{align}
\begin{split}
\bar{G}_\nu(t)=&\frac{1}{2}\int_{-\infty}^\infty ds (st)^\nu\int_0^\infty dx \int_0^\infty dy\sign(y-x)e^{-\frac{2}{a^{2}}(s^{2}+t^2) -\frac{2}{a^2(1-a^{2})}(x^{2}+y^2)+\frac{4}{a^2}sx}f_\nu(ty)\\
=&\sqrt{\frac{\pi a^{2}}{2^{3}}}\int_0^\infty dx \int_0^\infty dy\sign(y-x)(xt)^\nu e^{-\frac{2}{a^{2}}t^2-\frac{2}{1-a^{2}}x^2 -\frac{2}{a^2(1-a^{2})}y^2}f_\nu(ty)
\end{split}
\end{align}
In this form we can (partly) 
perform the integrals for $\nu=0,1$, separately.

First in order is the case $\nu=0$, where we integrate over $x$, leading to error functions, and afterwards rescale as $y=\sqrt{a^2(1-a^2)/2}u$, i.e.
\begin{align}
\begin{split}
\bar{G}_0(t)=&\sqrt{\frac{\pi a^2}{2^{3}}}e^{-\frac{2}{a^{2}}t^2}
\int_0^\infty dx \int_0^\infty dy\sign(y-x)e^{-\frac{2}{1-a^{2}}x^2 -\frac{2}{a^2(1-a^{2})}y^2}\cosh\left[\frac{4}{a^2}ty\right]\\
=&\frac{\pi }{2^{7/2}}a^2(1-a^2)e^{-\frac{2}{a^{2}}t^2}\int_0^\infty du(2\erf(au)-1)e^{ -u^2}\cosh\left[2\sqrt{\frac{2(1-a^2)}{a^2}}tu\right]\\
=&\frac{\pi }{2^{5/2}}a^2(1-a^2)e^{-\frac{2}{a^{2}}t^2}\int_0^\infty du\erf(au)e^{-u^2}\cosh\left[2\sqrt{\frac{2(1-a^2)}{a^2}}tu\right]\\
&-\frac{\pi }{2^{9/2}}a^2(1-a^2)e^{-\frac{2}{a^{2}}t^2}\int_{-\infty}^\infty du\,e^{-u^2+2\sqrt{\frac{2(1-a^2)}{a^2}}tu}\\
=& \frac{\pi a^2(1-a^2)}{2^{7/2}}e^{-2t^2}\int_{-\infty}^\infty du\erf(a|u|)
e^{-\left(u-\sqrt{\frac{2(1-a^2)}{a^2}}t\right)^2}-\frac{\pi^{3/2} a^2(1-a^2)}{2^{9/2}}e^{-2t^2}\ .
\end{split} \label{Gbar0.res}
\end{align}
For the second equality sign, we have extended the integration to the full real line for the second term in the initial integral, which is even in $u$, and thereupon performed the integral. 
The integral containing $\cosh$ appears to be non-elementary, nevertheless it can be simplified too by rewriting $\cosh$ into its to exponential terms and considering the two resulting terms as the two halves of an integration over the whole real line, and by eventually completing the squares.

Coming to the case $\nu=1$, we can perform all integrations as follows,
\begin{align}
\begin{split}
\bar{G}_1(t)=&\sqrt{\frac{\pi a^2}{2^{3}}}e^{-\frac{2}{a^{2}}t^2}\int_0^\infty dx \int_0^\infty dy\sign(y-x)x t\ e^{-\frac{2}{1-a^{2}}x^2 -\frac{2}{a^2(1-a^{2})}y^2}\sinh\left[\frac{4}{a^2}ty\right]\\
=&\frac{\sqrt{\pi}a(1-a^2)}{2^{7/2}} t\, e^{-\frac{2}{a^{2}}t^2}\int_0^\infty dy\left(e^{-\frac{2}{a^2(1-a^{2})}y^2}-2e^{ -\frac{2(1+a^2)}{a^2(1-a^{2})}y^2}\right)\sinh\left[\frac{4}{a^2}ty\right]\\
=&\frac{\pi a^2(1-a^2)^{3/2}}{2^{5}} t\,e^{-2t^2}\erf\left[\sqrt{\frac{2(1-a^2)}{a^2}}t\right]
-\frac{\pi a^2(1-a^2)^{3/2}}{2^{4}\sqrt{1+a^2}} t\, e^{-\frac{4}{1+a^{2}}t^2}\erf\left[\sqrt{\frac{2(1-a^2)}{a^2(1+a^2)}}t\right]\ .
\end{split}\label{Gbar1.res}
\end{align}
Evaluating the sign function in the integral over $x$ and using that the integrand is a total derivative, we obtain the second line. In the last step, we multiply out the parentheses with the two components of the $\sinh$-function and complete the squares in order to obtain the two error functions above.

The integral over the one-point function $g_\nu$ can be done in two alternative ways. First and foremost, the following direct computation can be pursued exploiting the definition~\eqref{eq:gnudef},
\begin{align}
\begin{split}
\bar{g}_\nu =& \int_0^\infty ds g_\nu(s) 
=\int_0^\infty ds\int_0^\infty dx\, s^\nu e^{-\frac{2s^2}{a^2}-\frac{2x^2}{a^2(1-a^2)}}\frac12 \left( e^{\frac{4sx}{a^2}}+(-1)^\nu e^{-\frac{4sx}{a^2}}\right)\\
=& \frac12 \int_{-\infty}^\infty ds\, s^\nu e^{-\frac{2s^2}{a^2}} \int_0^\infty dx\ e^{-\frac{2x^2}{a^2(1-a^2)}}e^{\frac{4sx}{a^2}}= \frac12 \int_0^\infty dx\ e^{-\frac{2x^2}{(1-a^2)}} \int_{-\infty}^\infty ds\, s^\nu e^{-\frac{2}{a^2}(s-x)^2}\\
=& \sqrt{\frac{\pi a^2}{8}} \int_0^\infty dx\, x^\nu e^{-\frac{2x^2}{(1-a^2)}}= \frac{\pi \sqrt{a^2(1-a^2)}}{8} \left( \frac{1-a^2}{2\pi}\right)^{\frac{\nu}{2}}.
\end{split}
\label{gbdirect}
\end{align}
We re-expressed the last term in the integral as an extension of the domain of $s$ over the full real line. After completing the square in the $s$-integral and the shift $s\to s+x$, we get the extra term $x^\nu$ (note that $\nu=0,1$) due to the symmetry of the Gaussian integral; in particular odd moments vanish. The remaining integral is elementary.

As a shortcut and cross-check we could have used the known normalisation of the jpdf 
\eqref{eq:jpdfdef} for $n=1$, which is just the integral over $g_\nu(s)$. This simply yields
\begin{align}
\bar{g}_\nu = C_{1,\nu}^{-1}=\frac{a(1-a^2)^{\frac12(1+\nu)}}{2^{\frac12(4+\nu)}}\Gamma\left(\frac32\right)\Gamma\left(\frac{1+\nu}{2}\right)\ ,
\label{gbcheck}
\end{align}
which agrees with \eqref{gbdirect}.

Before closing this chapter,  we make a consistency check for the result \eqref{eq:gnuerf} for $g_\nu(s)$, using the results for $G_\nu(s,t)$ that we just derived.
As mentioned in Subsection \ref{sec:Odd}, the joint density~\eqref{eq:jpdfdef} with $n=2m+1$ odd can also be derived from $n=2m+2$ even, by sending 
$\lambda_{2m+2}$ to infinity and factorising out its contribution,
\begin{align}
P^{(\nu)}_{2m+2}(\lambda_1,\ldots,\lambda_{2m+2})\overset{\lambda_{2m+2}\gg 1}{\approx} 
\eta_{2m+2}^{(\nu)}(\lambda_{2m+2})P^{(\nu)}_{2m+1}(\lambda_1,\ldots,\lambda_{2m+1})\ .
\end{align} 
Here $\eta_{2m+2}^{(\nu)}(\lambda_{2m+2})$ is some function that combines the leading power $\lambda_{2m+2}^{2m}$ of the Vandermonde determinant $\Delta_{2m+2}(\{\lambda^2\})$ with a factor coming from  the asymptotic limit of the two-point weight function $G_\nu(\lambda_j,\lambda_{2m+2})$.

Considering first $\nu=0$, equation~\eqref{eq:G0erf} can be approximated by
\begin{align}
G_0(s,z)\overset{z\gg 1}{\approx} \frac{\pi a^2(1-a^2)}{8}\  e^{-2(s^2+z^2)}= g_0(s)g_0(z) \ ,
\end{align}
with $g_0(z)$ from \eqref{eq:gnuerf} at $\nu=0$. We employed the asymptotic expansion $\erf(\lambda)\sim 1-e^{-\lambda^2}/(\lambda^2\sqrt{\pi})$ for $\lambda\gg 1$. 
Thus, $\eta_{2m+2}^{(0)}(z)=z^{2m}g_0(z)$ can be identified, after pulling out $g_0(z)$ from the Pfaffian~\eqref{eq:jpdfdef} for $n=2m+2$ (and keeping $g_0(s)$ inside), leading precisely to the expression for $n=2m+1$ at $\nu=0$.

Turning to $\nu=1$, we can use  the same asymptotics to replace the error function inside the integral~\eqref{eq:GTfinal} by unity. The remaining integral leads again to error functions,
\begin{align}
G_1(s,z)\overset{z\gg 1}{\approx}& \frac{\pi a^2(1-a^2)}{8}\  sz\ e^{-2(s^2+z^2)}\left\{ 1- \left( \erf\left[z\sqrt{\frac{(1-a^2)}{a^2}} \right]-\erf\left[s\sqrt{\frac{(1-a^2)}{a^2}} \right]\right)\right\}\nn\\
\overset{z\gg 1}{\approx}& \frac{\pi a^2(1-a^2)}{8}\  sz\ e^{-2(s^2+z^2)}\erf\left[s\sqrt{\frac{(1-a^2)}{a^2}} \right]=
zg_0(z) g_1(s)\ .
\end{align}
Once again we can pull out $g_0(z)$ from the Pfaffian~\eqref{eq:jpdfdef} for $n=2m+2$, with $\eta_{2m+2}^{(1)}(z)=z^{2m+1}g_0(z)$. This leads precisely to the expression for $n=2m+1$, with 
$g_1(s)$ from~\eqref{eq:gnuerf} at $\nu=1$ remaining in the extra row and column of the Pfaffian and shows our claim that the odd-dimensional case can be considered as a limit of the even-dimensional case.

\sect{Heine-like formulas for the skew-orthogonal polynomials}\label{sec:HeineFormulas}

In this appendix we recall a derivation for the following representation of the skew-orthogonal polynomials at given $N=2j+\nu$,
\begin{align}
p^{(\nu)}_{j}(x) &= x^{-\nu}\langle \det[x\eins_{2j+\nu}-J]\rangle_{j,\nu}
=\left\langle  \prod_{k=1}^{j}(x^2 - \lambda^2_k)\right\rangle_{j,\nu}
\ ,
\label{eq:pdef}\\
q^{(\nu)}_{j}(x) &= x^{-\nu}\Big\langle \det[x\eins_{2j+\nu}-J]\Big(x^2 +\frac12 \Tr J^2+c^{(\nu)}_j(a)\Big)\Big\rangle_{j,\nu} 
\nn\\
&= \left\langle  \prod_{k=1}^{j}(x^2 - \lambda^2_k)\left(x^2 +  \sum_{p=1}^{j}\lambda^2_p+ c^{(\nu)}_j(a)\right) 
\right\rangle_{j,\nu} 
\ .
\label{eq:qdef}
\end{align}
The averages $ \langle \ldots \rangle_{j,\nu}$ are taken 
over a random matrix $J$ of size $(2j+\nu)\times(2j+\nu)$, or over its $j$ singular values $\lambda_k$ in the second representation where the determinant and the trace are spelled out.

The term $c_j^{(\nu)}(a)$ is an arbitrary constant that may depend on $j,\nu$ and $a$. It reflects that the polynomials $q_j^{(\nu)}(x)$ are not uniquely defined. In this section we set all of these constants to zero, $c_j^{(\nu)}(a)=0$, although they are non-zero in the body of our work. The choice here is only for the sake of simplicity and clearness but has no further impact on the results. One can readily reintroduce those constant by adding a multiple of the polynomial $p^{(\nu)}_{j}(x)$ to the result of $q^{(\nu)}_{j}(x)$.

The relations~\eqref{eq:pdef} and \eqref{eq:qdef} are very much reminiscent to the form derived for general sOP in \cite{Eynard01} where orthogonal and symplectic ensembles with a general potential of the form $\prod_{k=1}^je^{-V(\lambda_k)}$ have been considered. One prominent difference from~\cite{Eynard01} is that we are dealing with polynomials in $x^2$, and hence in \eqref{eq:qdef} the second factor contains $x^2$ (and not $x$) as well as $\Tr J^2$ (as $J$ is traceless).
The calculation we draw here follows closely~\cite{AKP}, see also \cite{MarioTakuya}.  

As it is clear from taking large arguments, both polynomials are monic, i.e.
\begin{eqnarray}
p^{(\nu)}_{j}(x) = x^{2j}+O(x^{2j-2})\quad {\rm and}\quad q^{(\nu)}_{j}(x) = x^{2j+2}+O(x^{2j})\ ,
\label{pqmonic}
\end{eqnarray} 
respectively. What needs to be proven are the skew-orthogonality relations~\eqref{eq:sOPProdEven} and \eqref{eq:sOPProdOddI}, specifically that both
$p^{(\nu)}_{j}(x)$ and $q^{(\nu)}_{j}(x) $ are skew-orthogonal to all polynomials of degree up to $j-1$. Due to antisymmetry they are each skew-orthogonal to themselves and hence build a skew-orthogonal pair ($q^{(\nu)}_{j}(x)$ is the dual partner of $p^{(\nu)}_{j}(x)$ and vice versa). In other words we need to show that with $e_a(x)=x^{2a}$
\begin{align}
\langle {p}^{(\nu)}_{j},e_a\rangle_{e/o}&=0\ \ \mbox{for}\ \ a=0,1,\ldots, j\ ,\\
\langle   q^{(\nu)}_{j},e_a \rangle_{e/o} &=0 \ \ \mbox{for}\ \ a=0,1,\ldots, j-1\ ,
\end{align}
where the skew-symmetric product~\eqref{eq:sOPprodEven} labelled by ``$e$" corresponds to even $j$  and the skew-symmetric product~\eqref{eq:sOPprodOdd} denoted by ``$o$" relates to odd $j$.

\subsection{Even Dimension $j=2m$}
For completeness, we repeat here the skew-symmetric product~\eqref{eq:sOPprodEven}
\begin{align}
 \langle f_1,f_2\rangle_e=\int_0^\infty dx\int_0^\infty dy\,  G_\nu(x,y)f_1(x)f_2(y)
 \label{eq:Bskew}
\end{align}
for two functions $f_1,f_2$.
Taking up the representation~\eqref{eq:pdef} for $j=2m$ even with $m=0,1,\ldots$, we find
\begin{align}
\begin{split}
p^{(\nu)}_{2m}(x) =& C_{2m,\nu}
\int_0^\infty d\lambda_1\ldots \int_0^\infty d\lambda_{2m}\prod_{l=1}^{2m}(x^2-\lambda_l^2) 
\ \Delta_{2m}\left(\{\lambda^2\}\right)\text{Pf}\left[G_{\nu}(\lambda_a,\lambda_b)\right]_{a,b= 1}^{2m}\\
=& C_{2m,\nu}
\frac{(2m)!}{2^{m}m!}\int_0^\infty d\lambda_1\ldots \int_0^\infty d\lambda_{2m}  
\ \Delta_{2m+1}\left(\{\lambda^2\},x^2\right)\prod_{l=1}^m G_\nu(\lambda_{2l-1},\lambda_{2l})\\
=& 
C_{2m,\nu}
\frac{(2m)!}{2^{m}} \ 
\text{Pf}\left[
\begin{array}{cc}
\langle e_{a-1},e_{b-1}\rangle_e & e_{a-1}(x) \\
-e_{b-1}(x) & 0 
\end{array}
\right]_{a,b= 1}^{2m+1}.
\end{split}
\label{eq:Rj1}
\end{align}
In the first step we have spelled out the determinant. The product $\prod_{l=1}^{2m}(x^2-\lambda_l^2)$ and the Vandermonde determinant can be combined which, consequently, becomes a Vandermonde determinant of $2m+1$ variables, $x^2$ is the additional variable. Furthermore, the Pfaffian has been expanded where each of its terms yields the same contribution, namely the product $\prod_{l=1}^m G_\nu(\lambda_{2l-1},\lambda_{2l})$ times a combinatorial factor. In the last line we have applied a generalisation of the de Bruijn integral identity~\cite[Appendix~C.2]{Kieburg:2010}. The skew-symmetric product of the final result with the monomial $e_{c-1}(y)$ leads to
\begin{align}
\begin{split}
\langle p^{(\nu)}_{j}, e_{c-1}\rangle_e =& \int_0^\infty dx \int_0^\infty dy\ p^{(\nu)}_{2m}(x) G_\nu(x,y) e_{c-1}(y) \\
=& C_{2m,\nu}
\frac{(2m)!}{2^{m}} \ 
\text{Pf}\left[
\begin{array}{cc}
\langle e_{a-1},e_{b-1}\rangle_e & \langle e_{a-1},e_{c-1}\rangle_e \\
\langle e_{c-1},e_{b-1}\rangle_e & 0 
\end{array}
\right]_{a,b= 1}^{2m+1} = 0, \qquad\mbox{for}\ \ c=1,\ldots, 2m+1\ .
\end{split}
\label{eq:Rjproof}
\end{align}
The multi-linearity of the Pfaffian allows us to take the integral into the last row and column, and the antisymmetry of the Pfaffian leads to the vanishing of the right-hand side 
due to equal rows and columns, as claimed.

In the same fashion, the skew-orthogonality of the representation~\eqref{eq:qdef} can be shown. 
Equation~\eqref{eq:qdef} can be explicitly formulated as
\begin{align}
q^{(\nu)}_{2m}(x) =C_{2m,\nu}
\int_0^\infty d\lambda_1\ldots \int_0^\infty d\lambda_{2m}\  \prod_{l=1}^{2m}(x^2-\lambda_l^2) \left( x^2 +\sum_{p=1}^{2m}\lambda_p^2\right) 
\ \Delta_{2m}\left(\{\lambda^2\}\right)\text{Pf}\left[G_{\nu}(\lambda_a,\lambda_b)\right]_{a,b= 1}^{2m}\ .
\end{align}
Again the variable $x$ is regarded as an extra eigenvalue, $x=\lambda_{2m+1}$, enlarging the Vandermonde determinant to one of $2m+1$ variables,  
$\Delta_{2m+1}\left(\{\lambda^2\},x^2\right)=\Delta_{j+1}(\{\lambda^2\})$. We still have to deal with the sums. Here, the following identity \cite[Eq. (4.12)]{AKP} is particularly helpful,
\begin{align}
\sum_{a=1}^{j+1}\lambda_a^2\ \Delta_{j+1}(\{\lambda^2\})=\det
\left[
\begin{array}{ccccc}
1 & \lambda_1^2 &\cdots & \lambda_1^{2(j-1)} & \lambda_1^{2(j+1)} \\
\vdots & \vdots & & \vdots & \vdots \\
1 & \lambda_{j+1}^2 & \cdots & \lambda_{j+1}^{2(j-1)} &  \lambda_{j+1}^{2(j+1)}
\end{array}
\right]
=\widetilde{\Delta}_{j+1}(\{\lambda^2\})\ .
\label{eq:Deltaid}
\end{align}
For that reason, we can proceed as in the previous case,
\begin{align}
\begin{split}
q^{(\nu)}_{2m}(x)=& C_{2m,\nu}
\frac{(2m)!}{2^{m}m!}\int_0^\infty d\lambda_1\ldots \int_0^\infty d\lambda_{2m}  
\ \widetilde{\Delta}_{2m+1}\left(\{\lambda^2\},x^2\right)\prod_{l=1}^m G_\nu(\lambda_{2l-1},\lambda_{2l})\\
=& C_{2m,\nu}
\frac{(2m)!}{2^{m}} \ 
\text{Pf}\left[
\begin{array}{ccc}
\langle e_{a-1},e_{b-1}\rangle_e & \langle e_{a-1},e_{2m+1}\rangle_e &e_{a-1}(x) \\
\langle e_{2m+1},e_{b-1}\rangle_e & 0& e_{2m+1}(x)\\
- e_{b-1}(x) & - e_{2m+1}(x)&0 
\end{array}
\right]_{a,b= 1}^{2m}.
\end{split}
\label{eq:Rhatj1}
\end{align}
In the first line we have again replaced the Pfaffian by its diagonal and in the second step we repeated the application of de Bruijn's identity, this time to the modified Vandermonde determinant~\eqref{eq:Deltaid}. As before, a simple integration together with the Pfaffian's multi-linearity and antisymmetry yields
\begin{align}
\begin{split}
\langle q^{(\nu)}_{2m}, e_{c-1}\rangle_e = &\int_0^\infty dx \int_0^\infty dy \ q^{(\nu)}_{2m}(x) G_\nu(x,y) e_{c-1}(y)\\
=&C_{2m,\nu}
\frac{(2m)!}{2^{m}} \ 
\text{Pf}\left[
\begin{array}{ccc}
\langle e_{a-1},e_{b-1}\rangle_e & \langle e_{a-1},e_{2m+1}\rangle_e &\langle e_{a-1},e_{c-1}\rangle_e \\
\langle e_{2m+1},e_{b-1}\rangle_e & 0& \langle e_{2m+1},e_{c-1}\rangle_e \\
\langle e_{c-1}, e_{b-1}\rangle_e & \langle e_{c-1}e_{2m+1}\rangle_e &0 
\end{array}
\right]_{a,b= 1}^{2m}
\\
=& 0 \ \ \mbox{for}\ \ c=1,\ldots, 2m, 2m+2\ ,
\end{split}
\label{eq:Rhatjproof}
\end{align}
due to equal rows and columns. This finishes the proof for all vanishing skew-symmetric products when $j=2m$. Let us underline that the non-degeneracy of our skew-symmetric product is assumed and hence that all normalisations $h_{2j}^{(\nu)}=\langle p_{2j}^{(\nu)},q_{2j}^{(\nu)}\rangle_e \neq0$. The determination of the normalisation constants is done in Subsection \ref{sec:norms}.

\subsection{Odd Dimension $j=2m'+1$}
The skew-symmetric product for odd $n$ has a different weight function according to~\eqref{eq:sOPprodOdd},
\begin{eqnarray}
 \langle f_1,f_2\rangle_o=\int_0^\infty dx\int_0^\infty dy\  H_\nu(x,y)f_1(x)f_2(y)\ ,
\end{eqnarray}
with $f_1, f_2$ two suitably integrable functions and
\begin{eqnarray}
H_\nu(x,y) = G_\nu(x,y) - \frac{g_\nu(x)}{\bar{g}_\nu} \int_{0}^{\infty}dx'G_\nu(x',y) - \frac{g_\nu(y)}{\bar{g}_\nu} \int_{0}^{\infty}dy'G_\nu(x,y')\ .
\end{eqnarray}
The relation $H_\nu(x,y)=-H_\nu(y,x)$ is pellucid, due to the antisymmetry of the original two-point weight $G_\nu(x,y)$.
Additionally, the jpdf for odd $j=2m'+1$  with $m'=0,1,\ldots$ can be written in terms of this new two-point weight by using the invariance of the Pfaffian under simultaneous addition of rows and columns,
\begin{align}
\begin{split}
P_{2m'+1}^{(\nu)}(\lambda_1,\ldots,\lambda_n) =& C_{2m'+1,\nu}\Delta_{2m'+1}(\{\lambda^2\})\Pf\left[\begin{matrix}
G_\nu (\lambda_a,\lambda_b) & g_\nu(\lambda_a)\\
-g_\nu(\lambda_b) & 0
\end{matrix}\right]_{a,b=1}^{2m'+1}\\
=& C_{2m'+1,\nu}\Delta_{2m'+1}(\{\lambda^2\})\Pf\left[\begin{matrix}
	H_\nu (\lambda_a,\lambda_b) & g_\nu(\lambda_a)\\
	-g_\nu(\lambda_b) & 0
\end{matrix}\right]_{a,b=1}^{2m'+1}\ .
\end{split}
\end{align}
The skew-product with respect to the new two-point weight immediately satisfies that all monomials are skew-orthogonal to the zeroth order polynomial (unity), i.e. 

\begin{align}
\langle 1, e_a\rangle_o =& \int_0^\infty dx\int_0^\infty dy\  H_\nu(x,y) y^{2a}\nn\\
=& \int_0^\infty dx\int_0^\infty dy\  G_\nu(x,y)y^{2a} 
- \frac{\int_0^\infty dx  g_\nu(x)}{\bar{g}_\nu} \int_0^\infty dy\int_{0}^{\infty}dx'G_\nu(x',y) y^{2a}\nn\\
&- \frac{\int_0^\infty dyg_\nu(y)y^{2a}}{\bar{g}_\nu} \int_0^\infty dx \int_{0}^{\infty}dy'  G_\nu(x,y')\nn\\
=& 0\ .
\end{align}
The first two integrals cancel each other and the last integral vanishes due to the antisymmetry of $G_\nu(x,y')$.
Therefore the modified two-point weight $H_\nu(x,y)$ ensures that the lowest order polynomial is projected out in \eqref{skew1}. Consequently, we only need to prove skew-orthogonality for the remaining monomials, starting from degree one onwards with the polynomial $p_j^{(\nu)} (x)$ of odd degree $j=2m'+1$. The polynomial $p_{2m'+1}^{(\nu)} (x)$ reads, see~\eqref{eq:pdef},
\begin{align}
p_{2m'+1}^{(\nu)} (x) = C_{2m'+1,\nu}\int_{0}^{\infty}d\lambda_1\ldots \int_0^\infty d\lambda_{2m'+1}   
\Delta_{2m'+2}(\{\lambda^2\},x^2)\Pf\left[\begin{matrix}
	H_\nu (\lambda_a,\lambda_b) & g_\nu(\lambda_a)\\
	-g_\nu(\lambda_b) & 0
\end{matrix}\right]_{a,b=1}^{2m'+1},
\end{align}
where we have again combined the product $\prod_{k=1}^{2m'+1}(x^2 - \lambda^2_k)$ with the Vandermonde determinant $\Delta_{2m'+1}(\{\lambda^2\})$.
Exploiting the generalised de Bruijn integration identity~\cite[Appendix A.1]{KieburgMixing}, we may write
\begin{align}
p_{2m'+1}^{(\nu)} (x) = (2m'+1)! C_{2m'+1,\nu}\Pf\left[\begin{array}{c|c|c|c}
	0 & 0 & \bar{g}_{\nu} & 1\\ \hline
	0 & \langle e_{a}, e_{b}\rangle_o & \bar{g}_{a,\nu} & e_{a}(x)\\ \hline
	-\bar{g}_{\nu} & -\bar{g}_{b,\nu} & 0 & 0\\ \hline
	-1 & - e_{b}(x) & 0 & 0
\end{array}\right]_{a,b=1}^{2m'+1},
\label{pfinal}
\end{align}
where we have spelled out those terms involving the constant monomial $e_0(x)=1$, e.g. $\langle e_0,e_a\rangle_o = 0$ and have defined $\bar{g}_{a,\nu}=\int_0^\infty dz e_{a}(z) g_\nu(z)$ with $\bar{g}_{0,\nu}=\bar{g}_{\nu}$. The lines are meant as help for orientation. The skew-orthogonality readily follows,
\begin{align}
\langle p^{(\nu)}_{2m'+1}, e_k\rangle_o =& (2m'+1)! C_{2m'+1,\nu}\Pf\left[\begin{array}{c|c|c|c}
	0 & 0 & \bar{g}_{\nu} & 0\\ \hline
	0 & \langle e_{a}, e_{b}\rangle_o & \bar{g}_{a,\nu} & \langle e_{a}, e_k \rangle_o\\ \hline
	- \bar{g}_{\nu} & - \bar{g}_{b,\nu} & 0 & 0\\ \hline
	0 & \langle e_k, e_{b} \rangle_o & 0 & 0
\end{array}\right]_{a,b=1}^{2m'+1}\nn\\
=&(2m'+1)! C_{2m'+1,\nu}
\Pf\left[\begin{array}{c|c|c|c}
	0 & 0 & \bar{g}_{\nu} & 0\\ \hline
	0 & \langle e_{a}, e_{b}\rangle_o & 0 & \langle e_{a}, e_k \rangle_o\\ \hline
	- \bar{g}_{\nu} & 0 & 0 & 0\\ \hline
	0 & \langle e_k, e_{b} \rangle_o & 0 & 0
\end{array}\right]_{a,b=1}^{2m'+1}\nn\\
=& 0 ,\qquad\mbox{for}\ \ k=1,\ldots, 2m'+1\ .
\end{align}
The second equality is a consequence of the skew-symmetry of the Pfaffian, in particular the integrals $\int_0^\infty  e_{b}(z)g_\nu(z)dz$, especially $\int_0^\infty  g_\nu(z)dz\neq0$, are all constants and we can subtract any multiple of the first row and column from any entry in the second to last row and column. 
In this way we can read off for which values the Pfaffian vanishes, due to equal rows and columns.

It remains to show \eqref{eq:sOPProdOdd},
\begin{align}
\int_{0}^{\infty}dx\, p_{2j-1}^{(\nu)}(x)g_\nu(x)\ =0\ .
\end{align}
This can be easily seen by pulling the integral into the last row and column in \eqref{pfinal} so that the last two rows and columns agree and thus the Pfaffian vanishes.

For the polynomials $q_{j}^{(\nu)}(x) $ that are now of even degree $j+1 = 2m' + 2$ in $x^2$ the orthogonality follows in a  similar way. The polynomials $q_{j}^{(\nu)}(x) $, see~\eqref{eq:qdef}, are given by
\begin{align}
\begin{split}
q_{2m'+1}^{(\nu)} (x) =&
C_{2m'+1,\nu} 
\int_{0}^{\infty}d\lambda_1\ldots \int_0^\infty d\lambda_{2m'+1}   
\widetilde{\Delta}_{2m'+2}(\{\lambda^2\},x^2)\Pf\left[\begin{matrix}
	H_\nu (\lambda_a,\lambda_b) & g_\nu(\lambda_a)\\
	-g_\nu(\lambda_b) & 0
\end{matrix}\right]_{a,b=1}^{2m'+1}\\
=&(2m'+1)! C_{2m'+1,\nu}\Pf\left[\begin{array}{c|c|c|c|c}
	0 & 0 & 0 & \bar{g}_{\nu} & 1\\ \hline
	0 & \langle e_{a}, e_{b}\rangle_o & \langle e_{a}, e_{2m'+2}\rangle_o & \bar{g}_{a,\nu} & e_{a}(x)\\ \hline
	0 &  \langle e_{2m'+2},e_{b}\rangle_o & 0 & \bar{g}_{2m'+2,\nu} & e_{2m'+2}(x)\\ \hline
	- \bar{g}_{\nu} & - \bar{g}_{b,\nu} & - \bar{g}_{2m'+2,\nu} & 0 & 0\\ \hline
	-1 & - e_{b}(x) & - e_{2m'+2}(x) & 0 & 0
\end{array}\right]_{a,b=1}^{2m'}\ .
\end{split}
\label{qfinal}
\end{align}
In this calculation, we have again absorbed the product and the sum into a larger modified Vandermonde determinant, employing the identity \eqref{eq:Deltaid}, and then followed the same steps as above.
Thence, the skew-symmetric product is
\begin{align}
\begin{split}
\langle q^{(\nu)}_{2m'+1}, e_k\rangle_o =& (2m'+1)!C_{2m'+1,\nu}
\Pf\left[\begin{array}{c|c|c|c|c}
	0 & 0 & 0 & \bar{g}_\nu & 0\\ \hline
	0 & \langle e_{a}, e_{b}\rangle_o & \langle e_{a}, e_{2m'+2}\rangle_o & 0 & \langle e_{a}, e_k \rangle_o\\ \hline
	0 & \langle e_{2m'+2}, e_{b} \rangle_o & 0 & 0 &\langle e_{2m'+2}, e_k \rangle_o\\ \hline
	- \bar{g}_\nu & 0 & 0 & 0 & 0\\ \hline
	0 & \langle e_k, e_{b} \rangle_o & \langle e_k, e_{2m'+2} \rangle_o & 0 & 0
\end{array}\right]_{a,b=1}^{2m'}\\
=& 0, \qquad\mbox{for}\ \ k=1,\ldots, 2m',2m'+2\ .
\end{split}
\end{align}
where we have again removed the second two last column and row containing the constants with the help of the top row and column, respectively. 

We still need to show~\eqref{eq:sOPProdOdd},
\begin{align}
\int_{0}^{\infty}dx\,q_{2j-1}^{(\nu)}(x)g_\nu(x) =0\ .
\end{align}
Pulling the integration into the last row and column of \eqref{qfinal} leads again to a vanishing of the Pfaffian as the last two rows and columns agree thereafter. 

As in the even dimensional case, we only required that our skew-symmetric product is non-degenerate or, equivalently, $h_{2j-1}^{(\nu)}=\langle p_{2j-1}^{(\nu)},q_{2j-1}^{(\nu)}\rangle_o \neq0$ which are computed in Subsection~\ref{sec:norms}.


\sect{Limits $a\to 0,1,$ and $\infty$}\label{sec:Limits}

This appendix provides a few consistency checks of our results. By construction, our ensemble interpolates among the chGOE in the limit $a\to0$, the GAOE in the limit $a\to1$, and the direct sum of two GAOE's for $a\to\infty$. All three limits directly follow from Eq.~\eqref{eq:Yaltdef}, where we have to rescale the spectrum by $a$ when considering the limit $a\to\infty$. On the distributional level, see the initial probability density~\eqref{eq:ZNdef}, one needs the well-known relation
\begin{align}
\lim_{\epsilon\to 0} \sqrt{\frac{2}{\pi \epsilon^2}}\ e^{-\frac{2}{\epsilon^2}x^2} &= \delta(x)\ ;\label{eq:deltadef}
\end{align}
for instance to get the chGOE for $a\to0$ we have
\begin{align}
\begin{split}
\lim_{a\to0}\int[dY] P(Y,X)=&\lim_{a\to0} \prod_{i<k}^N\int_{-\infty}^\infty dH_{i,k} \left(\frac{\pi a^2}{2}\right)^{\frac12} e^{-\frac{2}{a^2}H_{i,k}^2}
 \left(\frac{\pi (1-a^2)}{2}\right)^{-n(n+\nu)/2}
e^{
 -\frac{1}{1-a^{2}}\text{Tr} \,X^{2}}\\
=& \left(\frac{\pi }{2}\right)^{-n(n+\nu)/2}
 \exp\left[
 -\text{Tr}\,X^{2}\right],
 \end{split}
\end{align}
Likewise, the limit $a\to1$ reproduces the GAOE. Only the limit $a\to\infty$ cannot be performed for~\eqref{eq:ZNdef} because it is not defined for $a\geq 1$, whereas the model~\eqref{eq:Yaltdef} is.

While the chGOE is a Pfaffian point process the GAOE is a determinatal point process, a situation similar to the classical interpolating ensembles of Mehta and Pandey~\cite{PandeyMehta,MehtaPandey}. Moreover, we have a factorisation of our model into two statistically independent spectra in the limit $a\to\infty$, at which each is a determinantal point process itself. Because our interpolating ensemble remains a Pfaffian point process for all parameter values $a\in(0,1)\cup(1,\infty)$, few quantities offer themselves for a consistency check in the limits $a\to 1,\infty$. Due to this problem, we only consider the jpdf  \eqref{eq:jpdfdef}, in Subsection \ref{jpdflim}, and the polynomials $p_j^{(\nu)}(x)$ (and $q_j^{(\nu)}(x)$), in Subsection \ref{sOPlim}. We show that these quantities reduce to the respective limiting results. 
Owing to the fact that the limiting polynomials constitute the corresponding limiting kernels, we will not further analyse the density or higher order correlation functions in these limits.

\subsection{Limiting JPDF}\label{jpdflim}

We begin with the limit $a\to0$. The constant $C_{n,\nu}$ in \eqref{jpdf-const} provides $n$ inverse powers in $a$ that we multiply into each row and column of the Pfaffian, for both $n$ even and odd. Note that for $n$ odd the last row and last column of the Pfaffian get multiplied only once with $a^{-1}$, though the other entries are multiplied by $a^{-2}$.
Using that $\gamma=\sqrt{(1-a^2)/a^2}$ diverges in the limit $a\to0$ and that 
\begin{eqnarray}
\lim_{\gamma\to\infty}\erf(\gamma z) = \sign(z)\ ,
\end{eqnarray}
we are lead to consider the following limits. For the two-point weight function we have 
\eqref{eq:G01}
\begin{align}
\lim_{a\to 0} \frac{G_\nu(x,y)}{a^2} 
= \frac{\pi}{8} (xy)^\nu e^{-2(x^2+y^2)}\sign(y^2-x^2)\ ,
\end{align}
since the second term in \eqref{eq:G01} vanishes as $a\exp[-4(\min\{x,y\})^2/a^2]$. The asymptotics for the one-point weight 
\eqref{eq:g01} can be obtained similarly,
\begin{align}
\lim_{a\to 0} \frac{g_\nu(y)}{a} =\sqrt{\frac{\pi}{8}} y^\nu e^{-2y^2}\ ,
\end{align}
where $\sign(y)=1$ due to $y>0$.
While all constants as well as the factors $y^\nu e^{-2y^2}$ can be pulled out of the rows and columns of the Pfaffian, we are left with
\begin{align}
\left. \begin{cases}
	\Pf[\sign(\lambda_b^2 - \lambda_a^2)]_{a,b=1}^n\,,\hspace{10pt}&\hspace{10pt} n\ \ \text{even}\\
	\Pf\left[\begin{array}{c|c}
		\sign(\lambda_b^2 - \lambda_a^2) & \vec{1}\\
		\hline
		-\vec{1}^T & 0
	\end{array}\right]_{a,b=1}^n,\hspace{10pt}&\hspace{10pt} n\ \ \text{odd}
\end{cases}\right\}
=\sign(\Delta_n(\{\lambda^2\}))
\end{align}
where we have used the Schur-Pfaffian identity~\cite{Schur} in the limit of large distances $|\lambda_b^2 - \lambda_a^2|\gg1$. Collecting everything, we find for the limiting jpdf
\begin{align}
\lim_{a\to 0} P_n^{(\nu)}\left(\lambda_1,\ldots,\lambda_n\right)
=&
2^{\frac{n}{2}(3+n+\nu)} 
\prod_{j=0}^{n-1}\frac{1}{\Gamma\left(\frac{j+3}{2}\right)\Gamma\left(\frac{j+\nu+1}{2}\right)}
\left(\frac{\pi}{8}\right)^{\frac{n}{2}}\prod_{j=1}^{n}\lambda_{j}^{\nu} e^{-2\lambda_j^2}\ 
\Delta_n(\{\lambda^2\}) 
\sign(\Delta_n(\lambda^2))\nonumber\\
=&
2^{\frac{n}{2}(n+\nu)} \pi^{\frac{n}{2}}
\prod_{j=0}^{n-1}\frac{1}{\Gamma\left(\frac{j+3}{2}\right)\Gamma\left(\frac{j+\nu+1}{2}\right)}
\prod_{j=1}^{n}\lambda_{j}^{\nu} e^{-2\lambda_j^2}\ 
|\Delta_n(\{\lambda^2\})|\ ,
\end{align}
for both even and odd $n$. This is the jpdf of the chGOE (in terms of squared singular values) for $\nu=0,1$.

The limit $a\to1$ is more involved, so we only sketch the derivation and omit the overall constants, knowing that the jpdf must be normalised. For this purpose, we split the weights $G_\nu(x,y)$ and $g_\nu(x)$ in terms which are independent of $a$ and expand the remainder in powers of $(1-a)$ leading to
\begin{align}
\begin{split}
G_\nu(x,y) 
\overset{a\approx1}{\approx}& (1-a)^{1+\nu}(xy)^{2\nu} e^{-2(x^2+y^2)}\sum_{k,l=0}^\infty c_{k,l}^{(\nu)}(1-a)^{k+l}x^{2k}y^{2l},\\
g_\nu(x)\overset{a\approx1}{\approx}&(1-a)^{(1+\nu)/2}x^{2\nu} e^{-2x^2}\sum_{k=0}^\infty d_{k}^{(\nu)}(1-a)^{k}x^{2k}.
\end{split}
\end{align}
Here, we have taken into account that both weights are even functions in all of their arguments, cf. the definitions~\eqref{eq:Gnudef} and~\eqref{eq:gnudef} which also can be exploited to explicitly calculate the Taylor coefficients. The coefficient $c_{k,l}^{(\nu)}=-c_{l,k}^{(\nu)}$ is  antisymmetric which is inherited from the antisymmetry of $G_\nu(x,y)=-G_\nu(y,x)$.

For even $n=2m$, this expansion implies for the Pfafffian that it is
\begin{align}
\Pf[G_\nu(\lambda_k,\lambda_l)]_{k,l=1}^{2m}=&(1-a)^{(1+\nu)m}\Pf\left[\sum_{r,t=0}^\infty c_{r,t}^{(\nu)}(1-a)^{r+t}\lambda_k^{2r}\lambda_l^{2t}\right]_{k,l=1}^{2m}\prod_{j=1}^{2m}\lambda_j^{2\nu}e^{-2\lambda_j^2}\nn\\
\overset{a\approx1}{\approx}&(1-a)^{(1+\nu)m}\Pf\left[\sum_{r,t=0}^{2m-1} c_{r,t}^{(\nu)}(1-a)^{r+t}\lambda_k^{2r}\lambda_l^{2t}\right]_{k,l=1}^{2m}\prod_{j=1}^{2m}\lambda_j^{2\nu}e^{-2\lambda_j^2}\nn\\
=&(1-a)^{(2m+\nu)m}\Pf\left[c_{r,t}^{(\nu)}\right]_{r,t=0}^{2m-1}\Delta_{2m}(\{\lambda^2\})\prod_{j=1}^{2m}\lambda_j^{2\nu}e^{-2\lambda_j^2}
\end{align}
in the lowest order in $(1-a)$. In the first line we have pulled out the $a$-independent factors while in the second line we have truncated the series since all other terms are of higher order in $(1-a)$. We cannot go below this truncation as the matrix inside the Pfaffian then becomes degenerate. The sum can be identified with a matrix product of the form $B^TAB$ where the matrices are $A=\{c_{r,t}^{(\nu)}\}_{r,t=0,\ldots,2m-1}$ and the Vandermonde matrix $B=\{(1-a)^t\lambda_l^{2t}\}_{\substack{t=0,\ldots,2m-1\\l=1,\ldots,2m}}$. Exploiting the identity $\Pf[B^TAB]=\Pf[A]\det[B]$, we end up with the last line.

In a similar way one can derive the case of odd $n=2m'+1$, i.e.
\begin{align}
\begin{split}
&\Pf\left[\begin{array}{cc} G_\nu(\lambda_k,\lambda_l) & g_\nu(\lambda_k) \\ -g_\nu(\lambda_l) & 0 \end{array}\right]_{k,l=1}^{2m'+1}\\
\overset{a\approx1}{\approx}&(1-a)^{(1+\nu)(2m'+1)/2}\Pf\left[\begin{array}{c|c} \sum_{r,t=0}^{2m'} c_{r,t}^{(\nu)}(1-a)^{r+t}\lambda_k^{2r}\lambda_l^{2t} & \sum_{r=0}^{2m'} d_{r}^{(\nu)}(1-a)^{r}\lambda_k^{2r} \\ \hline -\sum_{t=0}^{2m'} d_{t}^{(\nu)}(1-a)^{t}\lambda_l^{2t} & 0 \end{array}\right]_{r,t=1}^{2m'+1}\prod_{j=1}^{2m'+1}\lambda_j^{2\nu}e^{-2\lambda_j^2}\\
=&(1-a)^{(2m'+\nu)(2m'+1)/2}\Pf\left[\begin{array}{cc} c_{r,t}^{(\nu)} & d_{r}^{(\nu)} \\ - d_{t}^{(\nu)} & 0 \end{array}\right]_{r,t=0}^{2m'}\Delta_{2m'+1}(\{\lambda^2\})\prod_{j=1}^{2m'+1}\lambda_j^{2\nu}e^{-2\lambda_j^2}\ .
\end{split}
\end{align}
This time $A$ is the matrix inside the Pfaffian in the last line and $B$ is $\diag\left(\{(1-a)^t\lambda_l^{2t}\}_{\substack{t=0,\ldots,2m'\\l=1,\ldots,2m'+1}},1\right)$, meaning that it is a block diagonal matrix with a $(2m'+1)\times(2m'+1)$ block containing the Vandermonde matrix and a $1\times 1$ block being unity.

We combine the two asymptotics above with the remaining parts of the jpdf, in particular the Vandermonde determinant $\Delta_{n}(\{\lambda^2\})$ and the factor $(1-a)^{-n(n+\nu)/2}$ in the normalisation constant~\eqref{jpdf-const}, which cancels with the lowest order of the expansions. When suppressing all constants we obtain
\begin{align}
\lim_{a\to1}P^{(\nu)}_{n}(\lambda_1,\ldots,\lambda_{n}) &\propto  \prod_{j=1}^{n} \lambda_j^{2\nu}e^{-2\lambda_j^2}
\ \Delta_{n}(\{\lambda^2\})^2\ ,
\end{align}
agreeing with the jpdf of the GAOE for a matrix of size $N=2n+\nu$, cf.~\cite{MarioTim,Mehta}.

Finally, we want to study the limit $a\to\infty$. Beforehand, we have to rewrite the results~\eqref{eq:G01} and~\eqref{eq:g01} in their analytically continued forms for $a>1$ which are
\begin{align}
\begin{split}
G_\nu(x,y)=&\frac{\pi a^2(a^2-1)}{8}
(xy)^\nu e^{-2(x^2+y^2)}\Bigg({\rm erfi}\left[\sqrt{\frac{a^2-1}{a^2}}(y-x)\right] {\rm erfi}\left[\sqrt{\frac{a^2-1}{a^2}}(x+y)\right]
\\
&-\delta_{\nu,1}\frac{2}{\sqrt{\pi}}
\int_{\sqrt{2(a^2-1)/a^2} x}^{\sqrt{2(a^2-1)/a^2} y} du\, {\rm erfi}\left[\sqrt{\frac{2(a^2-1)}{a^2}}(x+y)-u\right]e^{u^2}
\Bigg),\\
g_\nu(y)=& i^{1+\nu}\sqrt{\frac{\pi a^2(a^2-1)}{8}}\exp\left[-2y^2\right]\left(y\, {\rm erfi} \left[\sqrt{\frac{2(a^2-1)}{a^2}}y\right]\right)^{\nu}.
\end{split}
\end{align}
For large positive real arguments $z$ the function ${\rm erfi} (z)$ satisfies the asymptotics
\begin{align}\label{erfi-lim}
{\rm erfi} (z) = \erf(iz)/i = \frac{2}{\sqrt{\pi}}\int_0^z e^{x^2}dx\overset{z\gg1}{\approx}\frac{1}{\sqrt{\pi}z}e^{z^2}.
\end{align}
In order to apply this asymptotics, we have to bear in mind that the spectrum scales with the coupling constant $a$. Thence, we rescale $x=a x'$ and $y=a y'$ and the weights have the asymptotics
\begin{align}
\begin{split}
G_\nu(a x',a y')\overset{a\gg1}{\approx}(-1)^\nu\frac{a^{2(1+\nu)}}{16}\frac{{y'}^{2\nu}+{x'}^{2\nu}}{{y'}^2-{x'}^2} e^{-2({x'}^2+{y'}^2)}\quad{\rm and}\quad g_\nu(y)\overset{a\gg1}{\approx}\frac{ i^{1+\nu}a^2}{(2\pi)^{\nu/2}}\sqrt{\frac{\pi}{8}}\exp\left[-2\frac{y^2}{a^{2(\nu-1)}}\right],
\end{split}\label{weight:ainfty}
\end{align}
which only holds for $G_\nu(a x',a y')$ when $x'\neq y'$, otherwise it vanishes.
The two-point weight function $G_\nu(a x',a y')$ results from the asymptotics~\eqref{erfi-lim} for the first term, keeping both orders in $\sqrt{2(a^2-1)}y'$ in the exponent, and the following Laplace approximation for the second term,
\begin{align}
&\int_{\sqrt{2(a^2-1)} x'}^{\sqrt{2(a^2-1)} y'} du\, {\rm erfi}\left[\sqrt{2(a^2-1)}(x'+y')-u\right]e^{u^2}\nn\\
=&\frac{4(a^2-1)}{\sqrt{\pi}}\int_{ x'}^{ y'} du'\int_0^{1} dv  \left(x'+y'-u'\right)\exp\left[2(a^2-1)[{u'}^2+(x'+y'-u')^2v^2]\right]\nn\\
\overset{a\gg1}{\approx}&\frac{1}{\sqrt{\pi}}\int_{ x'}^{ y'} \frac{du'}{x'+y'-u'}\exp\left[2(a^2-1)[{u'}^2+(x'+y'-u')^2]\right]\nn\\
\overset{a\gg1}{\approx}&\frac{1}{4\sqrt{\pi}(a^2-1)}\frac{y'+x'}{y'x'(y'-x')}e^{2(a^2-1)({x'}^2+{y'}^2)}.
\end{align}
In the first line, we have expressed the error function by its original definition as an integral and have rescaled $u=\sqrt{2(a^2-1)} u'$. In the second and third step we performed a Laplace approximation for $v$ and $u'$, respectively. The $v$-integral takes its maximum at $v=1$ whereas there are two maxima for $u'=x',y'$. The sum of the two terms in the weight $G_\nu(a x',a y')$ gives~\eqref{weight:ainfty}.

Next, we plug the asymptotics~\eqref{weight:ainfty} into the jpdf~\eqref{eq:jpdfdef}. For even $n=2m$ it reads
\begin{align}
\begin{split}
a^{2m}P^{(\nu)}_{2m}(a\lambda'_1,\ldots,a\lambda'_{2m})\overset{a\gg1}{\approx}&a^{2m}C_{2m,\nu}\,
\Delta_{2m}\left(\{a^2\lambda^2\}\right)\\
&\times
\text{Pf}\left[(1-\delta_{jk})(-1)^\nu\frac{a^{2(1+\nu)}}{16}\frac{{\lambda'_k}^{2\nu}+{\lambda'_j}^{2\nu}}{{\lambda'_k}^2-{\lambda'_j}^2} e^{-2({\lambda'_j}^2+{\lambda'_k}^2)}\right]_{j,k=1}^{2m}\\
\overset{a\gg1}{\approx}&
2^{m(2m+\nu-1)}\prod_{j=0}^{2m-1}\frac{1}{\Gamma\left(\frac{j+3}{2}\right)\Gamma\left(\frac{j+\nu+1}{2}\right)}\Delta_{2m}\left(\{\lambda^2\}\right)\prod_{j=1}^{2m}e^{-2{\lambda'_j}^2}\\
&\times\text{Pf}\left[\frac{{\lambda'_k}^{2\nu}+{\lambda'_j}^{2\nu}}{{\lambda'_k}^2-{\lambda'_j}^2} (1-\delta_{jk})\right]_{j,k=1}^{2m}\ .
\end{split}
\label{delta2}
\end{align}
The Kronecker delta inside the Pfaffian accounts to the vanishing of the two-point weight function $G_\nu(a x',a y')$ when both arguments coincide. The Pfaffian determinant can be expanded as follows
\begin{align}
\text{Pf}\left[\frac{{\lambda'_k}^{2\nu}+{\lambda'_j}^{2\nu}}{{\lambda'_k}^2-{\lambda'_j}^2} (1-\delta_{jk})\right]_{j,k=1}^{2m}=&\frac{1}{2^m m!}\sum_{\omega\in\mathbb{S}_{2m}}{\rm sign}(\omega)\prod_{j=1}^{m}\frac{(\lambda'_{\omega(2j)})^{2\nu}+(\lambda'_{\omega(2j-1)})^{2\nu}}{(\lambda'_{\omega(2j)})^2-(\lambda'_{\omega(2j-1)})^2}\nn\\
=&\frac{1}{(m!)^2}\sum_{\sigma\in\mathbb{S}_{m}}{\rm sign}(\sigma)\sum_{\omega\in\mathbb{S}_{2m}}{\rm sign}(\omega)\prod_{j=1}^{m}\frac{(\lambda'_{\omega(2j)})^{2\nu}}{(\lambda'_{\omega(2j)})^2-(\lambda'_{\omega(\sigma(2j-1))})^2}\nn\\
=&\frac{1}{(m!)^2}\sum_{\omega\in\mathbb{S}_{2m}}{\rm sign}(\omega) \det\left[\frac{(\lambda'_{\omega(2k)})^{2\nu}}{(\lambda'_{\omega(2k)})^2-(\lambda'_{\omega(2j-1)})^2}\right]_{j,k=1}^{m},
\label{Pfaffexpan}
\end{align}
where $\mathbb{S}_{2m}$ is the symmetric group permuting $2m$ elements, and ${\rm sign}(\omega)$ is $-1$ for odd permutations and unity for even ones. Here we have first employed the definition of the Pfaffian over the matrix $A_{j,k}$ on the left-hand side. In the second line we exploited the invariance under pairwise permutation of each pair $(\lambda'_{\omega(2j)},\lambda'_{\omega(2j-1)})$ and we have used the invariance under the permutation of the variables $\lambda'_{\omega(1)},\lambda'_{\omega(3)},\ldots,\lambda'_{\omega(2m-1)}$. The changes of the combinatorial prefactors reflect these transformations. At the end we identified the sum over the artificially introduced permutation $\sigma$ with the definition of the determinant.

Let us denote with $\lambda^{(\omega,o)}$ and $\lambda^{(\omega,e)}$ the set of odd and even eigenvalues $\lambda'_{\omega(1)},\ldots,\lambda'_{\omega(2m-1)}$ and $\lambda'_{\omega(2)},\ldots,\lambda'_{\omega(2m)}$, respectively, and pull the factor $(\lambda'_{\omega(2k)})^{2\nu}$ out of the determinant in the last line of~\eqref{Pfaffexpan}. Then, the resulting Cauchy determinant~\cite{Schechter} can be evaluated as
\begin{align}\label{Cauchydet}
\det\left[\frac{1}{(\lambda'_{\omega(2k)})^2-(\lambda'_{\omega(2j-1)})^2}\right]_{j,k=1}^{m}=\sign(\omega)\frac{\Delta_m^2(\{(\lambda^{(\omega,o)})^2\})\Delta_m^2(\{(\lambda^{(\omega,e)})^2\})}{\Delta_{2m}\left(\{\lambda'^2\}\right)}.
\end{align}
Note that the additional sign$(\omega)$ on the right-hand side originates from a reordering of the arguments of the larger Vandermonde determinant in the denominator to $\lambda'_1,\lambda'_2\ldots,,\lambda'_{2m}$.
Plugging this intermediate result into the jpdf we get the asymptotic formula
\begin{align}
\begin{split}
a^{2m}P^{(\nu)}_{2m}(a\lambda'_1,\ldots,a\lambda'_{2m})\overset{a\gg1}{\approx}&
\frac{1}{(2m)!}\sum_{\omega\in\mathbb{S}_{2m}}\frac{\Delta_m^2(\{(\lambda^{(\omega,o)})^2\})}{m!\prod_{j=0}^{m-1}\sqrt{\pi}2^{-4j-3/2}(2j)!}\\
&\hspace*{-1.5cm}\times\frac{\Delta_m^2(\{(\lambda^{(\omega,e)})^2\})}{m!\prod_{j=0}^{m-1}\sqrt{\pi}2^{-4j-2\nu-3/2}(2j+\nu)!}\prod_{j=1}^{m}(\lambda'_{\omega(2j)})^{2\nu}e^{-2((\lambda'_{\omega(2j)})^2+(\lambda'_{\omega(2j-1)})^2)},
\end{split}
\end{align}
where the factor $a^{2m}$ in front of the jpdf originates from the Jacobian of the rescaling $\lambda\to a\lambda'$.
The constant prefactor was simplified with the help of the doubling formula of the Gamma-function to easily identify the correctness of the normalising factors.
Without the sum the jpdf factorises into two jpdfs, one jpdf of a GAOE with dimension $2m$, and one of a GAOE with dimension $2m+\nu$. This is exactly what we have expected since for large coupling constant $a$ the original random matrix $J$, see~\eqref{eq:Yaltdef}, takes effectively the form $J=a\diag(A,B)$. The sum reflects only the fact that we cannot judge which eigenvalue belongs to which matrix.

A similar limit can be found for 
odd $n=2m'+1$. In this case, we need to consider the Pfaffian
\begin{align}
\label{Pfaff.1}
&\text{Pf}\left[\begin{array}{c|c}
\displaystyle(1-\delta_{jk})(-1)^\nu\frac{a^{2(1+\nu)}}{16}\frac{{\lambda'_k}^{2\nu}+{\lambda'_j}^{2\nu}}{{\lambda'_k}^2-{\lambda'_j}^2} e^{-2({\lambda'_j}^2+{\lambda'_k}^2)} & \displaystyle\frac{ i^{1+\nu}}{(2\pi)^{\nu/2}}\sqrt{\frac{\pi a^4}{8}}e^{-\frac{2}{a^{2(\nu-1)}}{\lambda'_j}^{2}} \\  \hline \displaystyle
-\frac{ i^{1+\nu}}{(2\pi)^{\nu/2}}\sqrt{\frac{\pi a^4}{8}}e^{-\frac{2}{a^{2(\nu-1)}}{\lambda'_k}^{2}}
& 0
\end{array}\right]_{j,k=1}^{2m'+1}\\
=&\frac{i^{1+\nu}(-1)^{m'\nu}}{(2\pi)^{\nu/2}}\sqrt{\frac{\pi a^4}{8}}\left(\frac{a^{2(1+\nu)}}{16}\right)^{m'}\frac{1}{(2m')!}\sum_{\sigma\in\mathbb{S}_{2m'+1}}{\rm sign}(\sigma)\text{Pf}\left[\frac{(\lambda'_{\sigma(k)})^{2\nu}+(\lambda'_{\sigma(j)})^{2\nu}}{(\lambda'_{\sigma(k)})^2-(\lambda'_{\sigma(j)})^2} (1-\delta_{jk})\right]_{j,k=1}^{2m'}\nn\\
&\times e^{-\frac{2}{a^{2(\nu-1)}}(\lambda'_{\sigma(2m'+1)})^{2}}\prod_{j=1}^{2m'}e^{-2(\lambda'_{\sigma(k)})^{2}}\nn
\end{align}

\begin{align}
=&\frac{i^{1+\nu}(-1)^{m'\nu}}{(2\pi)^{\nu/2}}\sqrt{\frac{\pi a^4}{8}}\left(\frac{a^{2(1+\nu)}}{16}\right)^{m'}\frac{1}{(m'!)^2}\sum_{\sigma\in\mathbb{S}_{2m'+1}}{\rm sign}(\sigma)\det\left[\frac{(\lambda'_{\sigma(2k)})^{2\nu}}{(\lambda'_{\sigma(2k)})^2-(\lambda'_{\sigma(2j-1)})^2}\right]_{j,k=1}^{m'}\nn\\
&\times e^{-\frac{2}{a^{2(\nu-1)}}(\lambda'_{\sigma(2m'+1)})^{2}}\prod_{j=1}^{2m'}e^{-2(\lambda'_{\sigma(k)})^{2}},\nn
\end{align}
where we first expanded the Pfaffian in the last row and column as well as symmetrised the expression with respect to the symmetric group $\mathbb{S}_{2m'+1}$ yielding the normalising factor $1/(2m')!$ and thereafter we used~\eqref{Pfaffexpan} again. The sum over the permutation $\omega\in\mathbb{S}_{2m'}$ has been absorbed in the permutation $\sigma$ which has produced a combinatorial factor of $(2m')!$ agreeing with the number of elements $\omega$ takes.

Before we exploit the relation~\eqref{Cauchydet} for the Cauchy determinant,  we have to case by case discuss $\nu=0$ and $\nu=1$ separately. Denoting again $\lambda^{(\sigma,o)}$ and $\lambda^{(\sigma,e)}$ as the set of eigenvalues $\lambda'_{\sigma(1)},\ldots,\lambda'_{\sigma(2m'-1)}$ and $\lambda'_{\sigma(2)},\ldots,\lambda'_{\sigma(2m')}$, respectively, we find for the jpdf~\eqref{eq:jpdfdef} in the case $\nu=0$
\begin{align}
\begin{split}
&a^{2m'+1} P^{(\nu=0)}_{2m'+1}(a\lambda'_1,\ldots,a\lambda'_{2m'+1})\\
\overset{a\gg1}{\approx}& \frac{1}{(2m'+1)!} \sum_{\sigma\in\mathbb{S}_{2m'+1}}\sqrt{\frac{8a^2}{\pi }}e^{-2a^{2}(\lambda'_{\sigma(2m'+1)})^{2}}\frac{\Delta_{m'}^2(\{(\lambda^{(\sigma,o)})^2\})\prod_{k=1}^{m'}((\lambda'_{\sigma(2k-1)})^2-(\lambda'_{\sigma(2m'+1)})^2)}{m'!\prod_{j=0}^{m'-1}\sqrt{\pi}2^{-4j-7/2}(2j+1)!}\\
&\times \frac{\Delta_{m'}^2(\{(\lambda^{(\sigma,e)})^2\})\prod_{k=1}^{m'}((\lambda'_{\sigma(2k)})^2-(\lambda'_{\sigma(2m'+1)})^2)}{m'!\prod_{j=0}^{m'-1}\sqrt{\pi}2^{-4j-7/2}(2j+1)!}\prod_{j=1}^{m'}e^{-2((\lambda'_{\sigma(2j)})^2+(\lambda'_{\sigma(2j-1)})^2)}\\
\overset{a\gg1}{\approx}&\frac{1}{(2m'+1)!}\sum_{\sigma\in\mathbb{S}_{2m'+1}}\delta(\lambda'_{\sigma(2m'+1)})\prod_{j=1}^{m'}(\lambda'_{\sigma(2j)}\lambda'_{\sigma(2j-1)})^2e^{-2((\lambda'_{\sigma(2j)})^2+(\lambda'_{\sigma(2j-1)})^2)}\\
&\times \frac{\Delta_{m'}^2(\{(\lambda^{(\sigma,o)})^2\})}{m'!\prod_{j=0}^{m'-1}\sqrt{\pi}2^{-4j-7/2}(2j+1)!}\frac{\Delta_{m'}^2(\{(\lambda^{(\sigma,e)})^2\})}{m'!\prod_{j=0}^{m'-1}\sqrt{\pi}2^{-4j-7/2}(2j+1)!}.
\end{split}
\end{align}
Note that the factorisation inside the sum resembles three random matrix ensembles, two GAOEs of dimension $(2m'+1)\times(2m'+1)$ and one two-dimensional GAOE with a spectrum on the scale of $1/a$. The latter yields a residual interaction between the two former ones via the products, reminiscent of the original Vandermonde determinant $\Delta_{2m'+1}(\{\lambda^2\})$, as long as $1/a$ is not too tiny. Decreasing $1/a$ to zero the two-dimensional GAOE is described by a Dirac delta function that is properly normalised to positive eigenvalues here. Consequently at non-zero $1/a$, its eigenvalue spectrum describes a broadening of the pair of zero modes, one for each of the $(2m'+1)$-dimensional GAOE.

Let us underline that this limit is the only situation where the topology (number of zero eigenvalues) is actually changing and it is, in our opinion, the most relevant situation of our model in view of physical applicability. The zero modes are expected to represent the pair of Majorana modes generated at the two opposite sides of the quantum wire~\cite{BA,Dumitrescu,Neven,Kitaev-Majorana}. Then, a $1/a$ expansion would describe the perturbations of the setting in the wire, which weakly couples the two subsystems given by two identical Bogoliubov-de-Genne Hamiltonians. Such perturbations may arise from the quasi-one-dimensionality of the wire, impurities, thermal fluctuations, or inhomogeneities and inaccuracies in the external fields like the magnetic field.

For the second case ($\nu=1$) we modify the Pfaffian~\eqref{Pfaff.1} even further to
\begin{align}\label{Pfaff.2}
&\text{Pf}\left[\begin{array}{c|c}
\displaystyle(\delta_{jk}-1)\frac{a^4}{16}\frac{{\lambda'_k}^{2}+{\lambda'_j}^{2}}{{\lambda'_k}^2-{\lambda'_j}^2} e^{-2({\lambda'_j}^2+{\lambda'_k}^2)} & \displaystyle-\frac{a^2}{4}e^{-2{\lambda'_j}^{2}} \\  \hline \displaystyle
\frac{a^2}{4}e^{-2{\lambda'_k}^{2}}
& 0 \end{array}\right]_{j,k=1}^{2m'+1}\\
=&\frac{(-1)^{m'+1}a^2}{4\,m'!(m'+1)!}\left(\frac{a}{2}\right)^{4m'}\sum_{\sigma\in\mathbb{S}_{2m'+1}}{\rm sign}(\sigma)\prod_{j=1}^{2m'+1}e^{-2(\lambda'_{\sigma(k)})^{2}}\det\left[\begin{array}{c|c} \displaystyle \frac{(\lambda'_{\sigma(2k)})^{2}}{(\lambda'_{\sigma(2k)})^2-(\lambda'_{\sigma(2j-1)})^2} & \vec{1} \end{array}\right]_{\substack{j=1,\ldots,m'+1\\k=1,\ldots,m'}}.\nn
\end{align}
The factors $(\lambda'_{\sigma(2k)})^{2}$ can now be pulled out of the determinant. The Cauchy determinant is this time replaced by a Cauchy-Vandermonde determinant, see~\cite{Kieburg:2010}, which is
\begin{align}
\det\left[\begin{array}{c|c} \displaystyle \frac{1}{(\lambda'_{\sigma(2k)})^2-(\lambda'_{\sigma(2j-1)})^2} & \vec{1} \end{array}\right]_{\substack{j=1,\ldots,m'+1\\k=1,\ldots,m'}}=\sign(\sigma)\frac{\Delta_{m'}^2(\{(\lambda^{(\sigma,e)})^2\})\Delta_{m'+1}^2(\{(\lambda^{(\sigma,o)})^2\})}{\Delta_{2m'+1}\left(\{\lambda^2\}\right)}
\end{align}
with $\lambda^{(\sigma,o)}=\diag(\lambda'_{\sigma(1)},\ldots,\lambda'_{\sigma(2m'+1)})$ and $\lambda^{(\sigma,e)}=\diag(\lambda'_{\sigma(2)},\ldots,\lambda'_{\sigma(2m')})$. The jpdf, hence, follows the asymptotics
\begin{align}
\begin{split}
a^{2m'+1} P^{(\nu=1)}_{2m'+1}(a\lambda'_1,\ldots,a\lambda'_{2m'+1})\overset{a\gg1}{\approx}&\frac{1}{(2m'+1)!}\sum_{\sigma\in\mathbb{S}_{2m'+1}}\left(\prod_{j=1}^{m'}(\lambda'_{\sigma(2j)})^2e^{-2(\lambda'_{\sigma(2j)})^2}\right)\!\!\left(\prod_{j=1}^{m'+1}e^{-2(\lambda'_{\sigma(2j-1)})^2}\right)\\
&\hspace*{-1cm}\times \frac{\Delta_{m'+1}^2(\{(\lambda^{(\sigma,e)})^2\})}{m'!\prod_{j=0}^{m'-1}\sqrt{\pi}2^{-4j-7/2}(2j+1)!}\frac{\Delta_{m'}^2(\{(\lambda^{(\sigma,o)})^2\})}{(m'+1)!\prod_{j=0}^{m'}\sqrt{\pi}2^{-4j-3/2}(2j)!}.
\end{split}
\end{align}
Without the symmetrising sum over $\sigma$ this is the factorising jpdf of a direct sum of an odd dimensional GAOE with size $(2m'+1)\times(2m'+1)$ and a $(2m'+2)\times(2m'+2)$ GAOE. Thus the zero eigenvalue of $J$ initially results from the matrix block $A$, cf.~\eqref{eq:Yaltdef}, when considering the random matrix $J$ as a perturbed model of the $a\to\infty$ limit.

\subsection{Limiting Polynomials}\label{sOPlim}

We again begin with the limit {$a\to 0$} to the chGOE. 
Using the representation~\eqref{pjnu} of $p_j^{(\nu)}(x)$ as an integral over a single Laguerre polynomial and the identity \eqref{eq:deltadef},
we find that
\begin{eqnarray}
\lim_{a\to 0} \ p_j^{(\nu)}(x) =\frac{j!}{(-4)^{j}}\int_{-\infty}^{\infty} dy\  L_j^{(\nu)}\left(4x^2+2y^2\right) \delta(y) = \frac{j!}{(-4)^{j}}\  L_j^{(\nu)}\left(4x^2\right)\ .
\label{pja0}
\end{eqnarray}
For $j=2m$ these polynomials agree with the even sOP polynomials in \cite{VerbNc2}, where only even $n$ was considered. 
Starting from \eqref{qjexplicit}  for $q_j^{(\nu)}(x)$, a tedious but straightforward calculation that we do not display here yields the following answer
\begin{align}
\lim_{a\to0}
q_j^{(\nu)}(x) &= \frac{j!}{(-4)^{j}}\left[(j+1)L_{j+1}^{(\nu)}\left(4x^2\right) 
- (j + \nu )L_{j}^{(\nu)}\left(4x^2\right) 
- (j+\nu)L_{j-1}^{(\nu)}\left(4x^2\right) 
\right.\nn\\
&\quad\quad\quad\quad\left.+ (\tilde{c}_j^{(\nu)}(0)-(j+1))L_{j}^{(\nu)}\left(4x^2\right) \right]\ .
\label{qja0}
\end{align}
The first three terms agree with the result for the odd sOP found in \cite{VerbNc2} (for even $j$), see \eqref{R1Jac}, after using the following identities for generalised Laguerre polynomials \cite{Gradshteyn},
\begin{eqnarray}
L_j^{(\nu-1)}(z)=L_j^{(\nu)}(z)-L_{j-1}^{(\nu)}(z)\quad{\rm and}\quad L_{j-1}^{(\nu)}(z)=(j+\nu)L_{j-1}^{(\nu)}(z)-jL_{j}^{(\nu)}(z)\ .
\label{Lid}
\end{eqnarray}
The last term in \eqref{qja0} can be set to zero by the appropriate choice of the constant as $\tilde{c}_j^{(\nu)}(0)=j+1$.

We turn to the limit {$a\to 1$} to the GAOE, where the jpdf becomes a determinantal point process, as we have seen above. Here, the corresponding single kernel and correlation functions are expressed in terms of orthogonal polynomials. Because these are given by the expectation value of a characteristic polynomial, we directly take the limit of the polynomials $x^\nu p_j^{(\nu)}(x)$ in \eqref{eq:sOPHeine}. In contrast to the polynomials $p_j^{(\nu)}(x)$, the expectation value that gives the polynomials $q_j^{(\nu)}(x)$ in \eqref{eq:sOPHeine} does not enjoy such a translucent interpretation for the GAOE. Even so, we can find one as follows. As it was shown in \cite{MarioPfaff}  each determinantal point process, especially the $\beta=2$ random matrix ensembles, can also be written as a Pfaffian point process in a non-trivial way. They can then be solved in terms of skew-orthogonal polynomials as well, given again by \eqref{eq:sOPHeine}. Because the corresponding solution of the GAOE has not been worked out in detail, we do not pursue this limit of the polynomials $q_j^{(\nu)}(x)$  further.

Returning to explicitly taking the limit, we employ the Gaussian integral representation~\eqref{eq:sOPMain}  for $p_j^{(\nu)}(x)$ and obtain
\begin{align}
\begin{split}
\lim_{a\to 1} \ x^\nu p_j^{(\nu)}(x) =& \frac{2}{\sqrt{2\pi}} \int_{-\infty}^{\infty}dy\int_{-\infty}^{\infty} dz\, e^{-2y^2} (iy + z + x)^j (iy - z + x)^{j+\nu}\delta(z)\\
=& \frac{1}{\sqrt{\pi}} \int_{-\infty}^{\infty}du\ e^{-u^2} \left(\frac{iu}{\sqrt{2}} + x\right)^{2j+\nu}\\
=& \frac{1}{2^{3(2j+\nu)/2}}
H_{2j+\nu}\left({\sqrt{2}}\ x\right).
\end{split}
\label{pja1}
\end{align}
The integral representation~\eqref{Hint} of the Hermite polynomials has been used for this result.
This result agrees with the orthogonal polynomials in \cite{Mehta}, finding only even or only odd Hermite polynomials for $\nu=0,1$, respectively. 

The $a\to\infty$ limit is as simple as the limit $a\to0$ though we start from a different representation of $p_j^{(\nu)}(x)$, namely, the second line of~\eqref{psum}. Yet, before we take the limit we need to rescale $x=a x'$ since the spectrum lives on this scale. This, thus, yields
\begin{align}\label{pliminfty}
\lim_{a\to\infty} a^{-2j}x^\nu p_j^{(\nu)}(ax) =  \frac{1}{2^{3(2j+\nu)/2}}H_{j}\left(\sqrt{2}x\right)H_{j+\nu}\left(\sqrt{2}x\right).
\end{align}
The overall prefactor $a^{-2j}$ corrects the scaling of the spectrum as well, considering that the determinant of $J$ scales with the inverse factor, see \eqref{eq:sOPHeine}. All other terms in the sum~\eqref{psum} are suppressed in $1/a^2$. Interpreting the result~\eqref{pliminfty}, it becomes immediate that the polynomials reflect the factorisation of the random matrix into two terms, where $H_{j}\left(\sqrt{2}x\right)$ and $H_{j+\nu}\left(\sqrt{2}x\right)$ correspond to the averages over $A$ and $B$ in the matrix $J$, respectively, cf.~\eqref{eq:Yaltdef}. Certainly these are the orthogonal polynomials for the two respective GAOEs. Also here we omit the discussion of the odd polynomials for the same reason as above.

This ends our consistency checks in the three limiting cases $a\to0,1$, and $\infty$.

\end{appendix}

\end{document}